\documentstyle[12pt]{article}
\begin{document}

\overfullrule 0 mm
\language 0
\centerline { \bf{ SOMMERFELD PARTICLE }}
\centerline { \bf{ IN STATIC MAGNETIC FIELD:}}
\centerline { \bf{TUNNELING AND DELAYED UNTWISTING IN CYCLOTRON}}
\vskip 0.5 cm \centerline {\bf{ Alexander A.  Vlasov}}
\vskip 0.3 cm \centerline {{  High Energy and Quantum Theory}}
\centerline {{  Department of Physics}} \centerline {{ Moscow State
University}} \centerline {{  Moscow, 119899}} \centerline {{
Russia}} \vskip 0.3 cm
 {\it Motion of a charged particle with finite size, described by
Sommerfeld model, in static magnetic field has two peculiar features:
1.) there is the effect of tunneling -  Sommerfeld particle overcomes
the  barrier and finds itself in the forbidden, from
classical point of view, area; 2.) the untwisting of trajectory in
cyclotron for Sommerfeld particle is strongly delayed compared to
that of a classical particle.}

03.50.De
\vskip 0.3 cm

Here we continue our investigation of peculiar features of motion
of Sommerfeld particle [1]. Let us remind that long time ago [2]
Sommerfeld proposed a model of a charged particle of finite size -
 sphere with uniform surface charge $Q$ and
mechanical mass $m$.  In nonrelativistic approximation such sphere
obeys the equation (see also [3]):
$$m\dot{\vec v} =\vec F_{ext}+  \eta\left[\vec v(t-2a/c) - \vec
v(t)\right] \eqno(1)$$ here $a$ - radius of the sphere,
 $\eta= {Q^2 \over 3  c a^2},\ \ \vec v= d \vec R /dt,\ \ \vec R$ -
coordinate of the center of the shell, $\vec F_{ext}$ - some external
force.

This model is a good tool to consider effects of radiation reaction
of a charged particle of finite size, free of problems of classical
point-like Lorentz-Dirac description.
\vskip 0.3 cm
{\bf {A.}}
\vskip 0.3 cm
   If Sommerfeld particle moves in the external static magnetic
field $\vec H$, the force $\vec F_{ext}=\int d \vec r \rho \cdot
[\dot {\vec R},\vec H]$   for $\rho=Q\delta( |\vec r - \vec R| -a)/4\pi a^2$
 has the form  $$ F_{ext}
={Q\over c} [\dot{\vec R},\vec H] $$

If magnetic field has non-zero values only in the shell of finite
size $S$ ( $0<Y<S$, $\vec H$ is parallel to $z$-axis,
$\vec R=(X,Y,0)$ ), then, as the particle has finite size $2a$, force
$\vec F_{ext}$ must be multiplied by the factor $f$:

$$f=\left\{\matrix{
0,& Y<-a; \cr {Y\over 2a} +{1\over 2},& -a<Y<a; \cr
1,& a<Y<S-a; \cr
{S-Y\over 2a} +{1\over 2},& S-a<Y<S+a; \cr
0,& S+a<y; \cr
}\right. \eqno(2)$$

For dimensionless variables $x=X/M,\ \ y=Y/M,\ \ \tau =ct/M$ ($M$
-scale factor) equation (1) takes the form $$\ddot y
=K\cdot\left[\dot y(\tau-d) -\dot y(\tau)\right]-\lambda \cdot \dot x
\cdot f,$$ $$\ddot x =K\cdot\left[\dot x(\tau-d) -\dot
x(\tau)\right]+\lambda \cdot \dot y \cdot f, \eqno (3)$$ here
$$f=\left\{\matrix{
0,& y<-{d\over 2}; \cr {y\over d} +{1\over 2},& -{d\over
2}<y<{d\over 2}; \cr 1,& {d\over 2}<y<L-{d\over 2}; \cr {L-y\over d}
+{1\over 2},& L-{d\over 2}<y<L+{d\over 2}; \cr 0,& L+{d\over 2}<y;
\cr }\right. \eqno(4)$$

and
$$K= { Q^2 M\over 3 a^2 m c^2},\ \ \lambda={Q H M\over mc^2},\ \
d={2a\over M},\ \ L={S\over M}.$$
Classical analog of equation (3) for point-like particle without
radiation reaction reads
$$\ddot y =-\lambda \cdot \dot x \cdot g,$$ $$\ddot x
=\lambda \cdot
\dot y \cdot g, \eqno (5)$$
here
$$ g= \left\{\matrix{ 0,& y<0; \cr 1,& 0<y<L; \cr 0,&
L<y; \cr }\right. \eqno(6)$$

For initial conditions $x(0)=0,\ \ y(0)=0,\ \ \dot x(0)=0,\ \ \dot
y(0)=v$ solution of (5) is
$$x=-{v\over \lambda}+{v\over \lambda}\cos{(\lambda\tau)},$$
$$y={v\over \lambda}\sin{(\lambda\tau)}\ \ (0<y<L) \eqno (7)$$

We see that for initial velocities $v$ smaller, then the critical
velocity $v_{cr}=\lambda L$, particle trajectory (half-circle) lies
inside the shell, i.e. particle cannot overcome the  barrier.  If
$L=10^4,\ \ \lambda=10^{-4}$ then $v_{cr}=1$.

We numerically investigated the particle motion governed by equation
(3) for the following values of initial velocity:
$$v=0.43,\ \ \ v=0.44$$
and for
$$L=10^4,\ \ \lambda=10^{-4}, \ \ d=1.0,\ \ K=4/(3d^2),$$
i.e. particle is of electron size and mass, magnetic field
approximately equals $10^{12}$ gauss and $S\approx 5,6 \cdot 10^{-9} $sm.

The result is shown on Fig. A, compared with classical
trajectory, governed by (7) with $v=0.44$. Horisontal axis is $x$ and
vertical axis is $y$.

The effect of tunneling for Sommerfeld particle is vividly seen:
velocity $v=0.44$ is smaller then the critical $v_{cr}=1$, but the
particle overcomes the barrier and finds itself in the forbidden from
classical point of view area $y>L=10^4$.
\vskip 0.3 cm
{\bf {B.}}
\vskip 0.3 cm

If magnetic field is parallel to $z$-axis for $y<0$ and
$y>L$ and equals to zero for $0<y<L$, and for  $0<y<L$ there is
static electric field $E$, parallel to $y$- axis in such
a way, that it is always collinear to $y$-component of particle
velocity (i.e.  particle is always accelerates in the clearance
$0<y<L$), then there is a model of cyclotron.

Equation of motion for Sommerfeld particle in cyclotron reads
$$\ddot y =K\cdot\left[\dot y(\tau-d) -\dot
y(\tau)\right]-\lambda \cdot \dot x \cdot f+\epsilon\cdot Sgn(\dot
y)\cdot(1-f),$$ $$\ddot x =K\cdot\left[\dot x(\tau-d) -\dot
x(\tau)\right]+\lambda \cdot \dot y \cdot f, \eqno (8)$$
here
$$\epsilon={QEM\over mc^2}$$
Classical analog of (8) one can construct replacing in (8) $K$ by
zero and $f$ by $g$ (6):
$$\ddot y =-\lambda \cdot \dot x \cdot g+\epsilon\cdot Sgn(\dot
y)\cdot(1-g),$$ $$\ddot x =\lambda \cdot \dot y \cdot g, \eqno (9)$$

Initial conditions are:
$$ x(0)=y(0)=\dot x(0)= \dot y(0) =0$$

Due to classical equation of motion without radiation reaction
(9) particle moves along untwisting trajectory. Total increase
of kinetic energy $W_c= (\dot x)^2/2 +(\dot y)^2/2$ of particle is
$N\cdot e \cdot L$:  $$W_c=N\cdot \epsilon \cdot L$$ where $N$ - is the
total number of passing of particle through the accelerating field
$E$.

If $N=10,\ \ \epsilon=\lambda=10^{-7},\ \ L=10^5$, then
$$W_c=10^{-1}.$$
We numerically calculated the particle motion governed by
equation (8) with zero initial conditions for the following values of
parameters:  $$L=10^5,\ \ \lambda=10^{-7}=\epsilon, \ \ d=0.3,\ \ K=2.0,$$
i.e. particle is of electron size and mass, magnetic field
approximately equals to $8.1\cdot 10^{7}$ gauss and electric field
produces in the clearance  potential difference  equal to $10^{4}$
eV.

The results of calculations are shown on Fig. B.1 - classical case
and on Fig. B.2 - case of Sommerfeld particle. Horisontal axis is
$x\cdot \lambda$ and vertical axis is $y\cdot \lambda$.

We see that for the same "time" $\tau \approx 10^{8}$ (i.e
$t \approx 10^{-4} sec $) classical particle (without radiation
reaction) made $N=10$ passings through the accelerating field $E$
with total energy increase $W_c=10^{-1}$, while Sommerfeld particle
made only $N=6$ passings with  total energy increase $W_s=0.0375$ (
$W_c$ for $N=6$ is equal to $0.06 $ ). Thus untwisting of trajectory
for Sommerfeld particle is strongly delayed compared to that of a
classical one.

Delay in energy increase falls mainly on the moments of passing
through the clearance. It can be  explained by difference in
accelerations in electric field (proportional to $\epsilon \approx
10^{-7}$) and in magnetic field (proportional to $v \cdot
\lambda \approx 10^{-8}$ ) as flux of radiating energy is
proportional to square of acceleration.

 \vskip 2 cm \centerline {\bf{REFERENCES}}

  \begin{enumerate}
\item Alexander A.Vlasov, physics/9905050, physics/9911059.

\item A.Sommerfeld, Gottingen Nachrichten, 29 (1904), 363 (1904), 201
  (1905).
\item L.Page, Phys.Rev., 11, 377 (1918).
 T.Erber, Fortschr. Phys., 9, 343 (1961).
 P.Pearle in "Electromagnetism",ed. D.Tepliz, (Plenum, N.Y.,
1982), p.211.
 A.Yaghjian, "Relativistic Dynamics of a Charged Sphere".
  Lecture Notes in Physics, 11 (Springer-Verlag, Berlin, 1992).

\end{enumerate}
\eject
\newcount\numpoint
\newcount\numpointo
\numpoint=1 \numpointo=1
\def\emmoveto#1#2{\offinterlineskip
\hbox to 0 true cm{\vbox to 0
true cm{\vskip - #2 true mm
\hskip #1 true mm \special{em:point
\the\numpoint}\vss}\hss}
\numpointo=\numpoint
\global\advance \numpoint by 1}
\def\emlineto#1#2{\offinterlineskip
\hbox to 0 true cm{\vbox to 0
true cm{\vskip - #2 true mm
\hskip #1 true mm \special{em:point
\the\numpoint}\vss}\hss}
\special{em:line
\the\numpointo,\the\numpoint}
\numpointo=\numpoint
\global\advance \numpoint by 1}
\def\emshow#1#2#3{\offinterlineskip
\hbox to 0 true cm{\vbox to 0
true cm{\vskip - #2 true mm
\hskip #1 true mm \vbox to 0
true cm{\vss\hbox{#3\hss
}}\vss}\hss}}
\special{em:linewidth 0.8pt}

\vrule width 0 mm height                0 mm depth 90.000 true mm

\special{em:linewidth 0.8pt}
\emmoveto{130.000}{10.000}
\emlineto{12.000}{10.000}
\emlineto{12.000}{80.000}
\emmoveto{71.000}{10.000}
\emlineto{71.000}{80.000}
\emmoveto{12.000}{45.000}
\emlineto{130.000}{45.000}
\emmoveto{130.000}{10.000}
\emlineto{130.000}{80.000}
\emlineto{12.000}{80.000}
\emlineto{12.000}{10.000}
\emlineto{130.000}{10.000}
\special{em:linewidth 0.4pt}
\emmoveto{12.000}{17.000}
\emlineto{130.000}{17.000}
\emmoveto{12.000}{24.000}
\emlineto{130.000}{24.000}
\emmoveto{12.000}{31.000}
\emlineto{130.000}{31.000}
\emmoveto{12.000}{38.000}
\emlineto{130.000}{38.000}
\emmoveto{12.000}{45.000}
\emlineto{130.000}{45.000}
\emmoveto{12.000}{52.000}
\emlineto{130.000}{52.000}
\emmoveto{12.000}{59.000}
\emlineto{130.000}{59.000}
\emmoveto{12.000}{66.000}
\emlineto{130.000}{66.000}
\emmoveto{12.000}{73.000}
\emlineto{130.000}{73.000}
\emmoveto{23.800}{10.000}
\emlineto{23.800}{80.000}
\emmoveto{35.600}{10.000}
\emlineto{35.600}{80.000}
\emmoveto{47.400}{10.000}
\emlineto{47.400}{80.000}
\emmoveto{59.200}{10.000}
\emlineto{59.200}{80.000}
\emmoveto{71.000}{10.000}
\emlineto{71.000}{80.000}
\emmoveto{82.800}{10.000}
\emlineto{82.800}{80.000}
\emmoveto{94.600}{10.000}
\emlineto{94.600}{80.000}
\emmoveto{106.400}{10.000}
\emlineto{106.400}{80.000}
\emmoveto{118.200}{10.000}
\emlineto{118.200}{80.000}
\special{em:linewidth 0.8pt}
\emmoveto{12.000}{10.013}
\emlineto{12.001}{10.174}
\emmoveto{12.002}{10.289}
\emlineto{12.004}{10.449}
\emmoveto{12.007}{10.565}
\emlineto{12.011}{10.725}
\emmoveto{12.016}{10.841}
\emlineto{12.022}{11.001}
\emmoveto{12.028}{11.117}
\emlineto{12.035}{11.277}
\emmoveto{12.043}{11.392}
\emlineto{12.052}{11.553}
\emmoveto{12.061}{11.668}
\emlineto{12.073}{11.829}
\emmoveto{12.083}{11.944}
\emlineto{12.096}{12.104}
\emmoveto{12.108}{12.219}
\emlineto{12.123}{12.380}
\emmoveto{12.136}{12.495}
\emlineto{12.153}{12.655}
\emmoveto{12.168}{12.770}
\emlineto{12.187}{12.930}
\emmoveto{12.203}{13.045}
\emlineto{12.224}{13.205}
\emmoveto{12.241}{13.320}
\emlineto{12.264}{13.480}
\emmoveto{12.283}{13.595}
\emlineto{12.307}{13.755}
\emmoveto{12.328}{13.870}
\emlineto{12.354}{14.030}
\emmoveto{12.376}{14.144}
\emlineto{12.404}{14.304}
\emmoveto{12.428}{14.419}
\emlineto{12.457}{14.578}
\emmoveto{12.482}{14.693}
\emlineto{12.513}{14.852}
\emmoveto{12.540}{14.966}
\emlineto{12.573}{15.126}
\emmoveto{12.602}{15.240}
\emlineto{12.636}{15.399}
\emmoveto{12.666}{15.513}
\emlineto{12.703}{15.672}
\emmoveto{12.734}{15.786}
\emlineto{12.772}{15.945}
\emmoveto{12.805}{16.059}
\emlineto{12.845}{16.218}
\emmoveto{12.880}{16.331}
\emlineto{12.921}{16.490}
\emmoveto{12.957}{16.603}
\emlineto{13.001}{16.762}
\emmoveto{13.038}{16.875}
\emlineto{13.084}{17.033}
\emmoveto{13.122}{17.147}
\emlineto{13.170}{17.305}
\emmoveto{13.210}{17.418}
\emlineto{13.259}{17.575}
\emmoveto{13.300}{17.688}
\emlineto{13.351}{17.846}
\emmoveto{13.394}{17.959}
\emlineto{13.447}{18.116}
\emmoveto{13.491}{18.228}
\emlineto{13.546}{18.386}
\emmoveto{13.592}{18.498}
\emlineto{13.648}{18.655}
\emmoveto{13.695}{18.767}
\emlineto{13.753}{18.924}
\emmoveto{13.802}{19.036}
\emlineto{13.862}{19.192}
\emmoveto{13.912}{19.304}
\emlineto{13.974}{19.460}
\emmoveto{14.026}{19.571}
\emlineto{14.089}{19.727}
\emmoveto{14.142}{19.838}
\emlineto{14.207}{19.994}
\emmoveto{14.262}{20.105}
\emlineto{14.328}{20.260}
\emmoveto{14.385}{20.371}
\emlineto{14.453}{20.526}
\emmoveto{14.511}{20.637}
\emlineto{14.581}{20.792}
\emmoveto{14.640}{20.902}
\emlineto{14.712}{21.056}
\emmoveto{14.772}{21.167}
\emlineto{14.846}{21.321}
\emmoveto{14.908}{21.431}
\emlineto{14.983}{21.584}
\emmoveto{15.047}{21.694}
\emlineto{15.124}{21.847}
\emmoveto{15.188}{21.957}
\emlineto{15.267}{22.110}
\emmoveto{15.333}{22.219}
\emlineto{15.414}{22.372}
\emmoveto{15.482}{22.480}
\emlineto{15.564}{22.633}
\emmoveto{15.619}{22.718}
\emlineto{15.689}{22.846}
\emmoveto{15.745}{22.931}
\emlineto{15.816}{23.059}
\emmoveto{15.873}{23.143}
\emlineto{15.945}{23.271}
\emmoveto{16.003}{23.356}
\emlineto{16.076}{23.483}
\emmoveto{16.135}{23.567}
\emlineto{16.209}{23.695}
\emmoveto{16.269}{23.779}
\emlineto{16.345}{23.906}
\emmoveto{16.406}{23.989}
\emlineto{16.482}{24.116}
\emmoveto{16.544}{24.200}
\emlineto{16.622}{24.326}
\emmoveto{16.684}{24.410}
\emlineto{16.763}{24.536}
\emmoveto{16.827}{24.619}
\emlineto{16.907}{24.745}
\emmoveto{16.971}{24.828}
\emlineto{17.052}{24.954}
\emmoveto{17.118}{25.036}
\emlineto{17.200}{25.162}
\emmoveto{17.266}{25.244}
\emlineto{17.350}{25.369}
\emmoveto{17.417}{25.451}
\emlineto{17.502}{25.576}
\emmoveto{17.570}{25.658}
\emlineto{17.655}{25.783}
\emmoveto{17.724}{25.865}
\emlineto{17.811}{25.989}
\emmoveto{17.881}{26.070}
\emlineto{17.969}{26.194}
\emmoveto{18.040}{26.276}
\emlineto{18.129}{26.399}
\emmoveto{18.201}{26.480}
\emlineto{18.291}{26.604}
\emmoveto{18.363}{26.684}
\emlineto{18.455}{26.807}
\emmoveto{18.528}{26.888}
\emlineto{18.620}{27.011}
\emmoveto{18.695}{27.091}
\emlineto{18.788}{27.213}
\emmoveto{18.864}{27.293}
\emlineto{18.958}{27.415}
\emmoveto{19.034}{27.495}
\emlineto{19.130}{27.617}
\emmoveto{19.207}{27.696}
\emlineto{19.304}{27.818}
\emmoveto{19.382}{27.897}
\emlineto{19.479}{28.018}
\emmoveto{19.558}{28.097}
\emlineto{19.657}{28.218}
\emmoveto{19.737}{28.296}
\emlineto{19.837}{28.417}
\emmoveto{19.917}{28.495}
\emlineto{20.018}{28.615}
\emmoveto{20.100}{28.693}
\emlineto{20.202}{28.813}
\emmoveto{20.284}{28.890}
\emlineto{20.387}{29.010}
\emmoveto{20.470}{29.087}
\emlineto{20.575}{29.206}
\emmoveto{20.659}{29.284}
\emlineto{20.764}{29.402}
\emmoveto{20.849}{29.479}
\emlineto{20.955}{29.597}
\emmoveto{21.041}{29.674}
\emlineto{21.149}{29.792}
\emmoveto{21.235}{29.868}
\emlineto{21.344}{29.986}
\emmoveto{21.431}{30.062}
\emlineto{21.541}{30.179}
\emmoveto{21.629}{30.255}
\emlineto{21.740}{30.372}
\emmoveto{21.829}{30.447}
\emlineto{21.940}{30.563}
\emmoveto{22.030}{30.638}
\emlineto{22.143}{30.754}
\emmoveto{22.234}{30.829}
\emlineto{22.347}{30.945}
\emmoveto{22.439}{31.019}
\emlineto{22.554}{31.135}
\emmoveto{22.646}{31.209}
\emlineto{22.762}{31.324}
\emmoveto{22.855}{31.397}
\emlineto{22.972}{31.512}
\emmoveto{23.066}{31.585}
\emlineto{23.184}{31.699}
\emmoveto{23.279}{31.772}
\emlineto{23.398}{31.886}
\emmoveto{23.493}{31.959}
\emlineto{23.613}{32.072}
\emmoveto{23.710}{32.145}
\emlineto{23.831}{32.258}
\emmoveto{23.928}{32.330}
\emlineto{24.050}{32.442}
\emmoveto{24.148}{32.514}
\emlineto{24.271}{32.626}
\emmoveto{24.370}{32.697}
\emlineto{24.494}{32.809}
\emmoveto{24.593}{32.880}
\emlineto{24.718}{32.991}
\emmoveto{24.819}{33.062}
\emlineto{24.945}{33.173}
\emmoveto{25.046}{33.243}
\emlineto{25.173}{33.353}
\emmoveto{25.275}{33.423}
\emlineto{25.403}{33.533}
\emmoveto{25.506}{33.603}
\emlineto{25.635}{33.712}
\emmoveto{25.738}{33.782}
\emlineto{25.868}{33.891}
\emmoveto{25.972}{33.960}
\emlineto{26.103}{34.068}
\emmoveto{26.208}{34.137}
\emlineto{26.340}{34.245}
\emmoveto{26.446}{34.313}
\emlineto{26.579}{34.421}
\emmoveto{26.685}{34.489}
\emlineto{26.819}{34.596}
\emmoveto{26.927}{34.663}
\emlineto{27.061}{34.770}
\emmoveto{27.169}{34.837}
\emlineto{27.305}{34.943}
\emmoveto{27.414}{35.010}
\emlineto{27.551}{35.116}
\emmoveto{27.660}{35.182}
\emlineto{27.798}{35.288}
\emmoveto{27.908}{35.354}
\emlineto{28.047}{35.458}
\emmoveto{28.158}{35.524}
\emlineto{28.297}{35.628}
\emmoveto{28.409}{35.694}
\emlineto{28.549}{35.797}
\emmoveto{28.662}{35.862}
\emlineto{28.803}{35.966}
\emmoveto{28.917}{36.030}
\emlineto{29.059}{36.133}
\emmoveto{29.173}{36.197}
\emlineto{29.316}{36.299}
\emmoveto{29.431}{36.363}
\emlineto{29.575}{36.465}
\emmoveto{29.690}{36.528}
\emlineto{29.835}{36.630}
\emmoveto{29.922}{36.674}
\emlineto{30.039}{36.757}
\emmoveto{30.126}{36.801}
\emlineto{30.243}{36.884}
\emmoveto{30.331}{36.928}
\emlineto{30.449}{37.010}
\emmoveto{30.537}{37.054}
\emlineto{30.655}{37.136}
\emmoveto{30.744}{37.180}
\emlineto{30.863}{37.261}
\emmoveto{30.952}{37.305}
\emlineto{31.071}{37.386}
\emmoveto{31.161}{37.429}
\emlineto{31.281}{37.510}
\emmoveto{31.371}{37.553}
\emlineto{31.491}{37.634}
\emmoveto{31.581}{37.676}
\emlineto{31.702}{37.757}
\emmoveto{31.793}{37.799}
\emlineto{31.914}{37.879}
\emmoveto{32.006}{37.921}
\emlineto{32.128}{38.001}
\emmoveto{32.219}{38.043}
\emlineto{32.342}{38.122}
\emmoveto{32.434}{38.164}
\emlineto{32.557}{38.243}
\emmoveto{32.649}{38.284}
\emlineto{32.772}{38.363}
\emmoveto{32.865}{38.404}
\emlineto{32.989}{38.483}
\emmoveto{33.083}{38.524}
\emlineto{33.207}{38.601}
\emmoveto{33.301}{38.642}
\emlineto{33.426}{38.720}
\emmoveto{33.520}{38.760}
\emlineto{33.645}{38.838}
\emmoveto{33.740}{38.878}
\emlineto{33.866}{38.955}
\emmoveto{33.960}{38.995}
\emlineto{34.087}{39.071}
\emmoveto{34.182}{39.111}
\emlineto{34.309}{39.187}
\emmoveto{34.405}{39.227}
\emlineto{34.532}{39.303}
\emmoveto{34.628}{39.342}
\emlineto{34.756}{39.418}
\emmoveto{34.853}{39.457}
\emlineto{34.981}{39.532}
\emmoveto{35.078}{39.571}
\emlineto{35.207}{39.645}
\emmoveto{35.304}{39.684}
\emlineto{35.433}{39.758}
\emmoveto{35.531}{39.797}
\emlineto{35.661}{39.871}
\emmoveto{35.759}{39.909}
\emlineto{35.889}{39.983}
\emmoveto{35.987}{40.020}
\emlineto{36.118}{40.094}
\emmoveto{36.217}{40.131}
\emlineto{36.348}{40.204}
\emmoveto{36.447}{40.241}
\emlineto{36.579}{40.314}
\emmoveto{36.678}{40.351}
\emlineto{36.811}{40.423}
\emmoveto{36.910}{40.460}
\emlineto{37.043}{40.532}
\emmoveto{37.143}{40.569}
\emlineto{37.277}{40.640}
\emmoveto{37.377}{40.676}
\emlineto{37.511}{40.748}
\emmoveto{37.611}{40.783}
\emlineto{37.746}{40.854}
\emmoveto{37.847}{40.890}
\emlineto{37.982}{40.961}
\emmoveto{38.083}{40.996}
\emlineto{38.218}{41.066}
\emmoveto{38.320}{41.101}
\emlineto{38.455}{41.171}
\emmoveto{38.557}{41.206}
\emlineto{38.694}{41.275}
\emmoveto{38.796}{41.310}
\emlineto{38.933}{41.379}
\emmoveto{39.035}{41.413}
\emlineto{39.172}{41.482}
\emmoveto{39.275}{41.516}
\emlineto{39.413}{41.584}
\emmoveto{39.516}{41.618}
\emlineto{39.654}{41.686}
\emmoveto{39.758}{41.719}
\emlineto{39.896}{41.787}
\emmoveto{40.000}{41.820}
\emlineto{40.139}{41.887}
\emmoveto{40.244}{41.920}
\emlineto{40.383}{41.987}
\emmoveto{40.488}{42.019}
\emlineto{40.627}{42.086}
\emmoveto{40.732}{42.118}
\emlineto{40.872}{42.184}
\emmoveto{40.978}{42.216}
\emlineto{41.118}{42.282}
\emmoveto{41.224}{42.314}
\emlineto{41.365}{42.379}
\emmoveto{41.471}{42.411}
\emlineto{41.612}{42.476}
\emmoveto{41.719}{42.507}
\emlineto{41.860}{42.571}
\emmoveto{41.967}{42.602}
\emlineto{42.109}{42.667}
\emmoveto{42.216}{42.697}
\emlineto{42.359}{42.761}
\emmoveto{42.466}{42.791}
\emlineto{42.609}{42.855}
\emmoveto{42.717}{42.885}
\emlineto{42.860}{42.948}
\emmoveto{42.968}{42.977}
\emlineto{43.112}{43.040}
\emmoveto{43.220}{43.070}
\emlineto{43.364}{43.132}
\emmoveto{43.473}{43.161}
\emlineto{43.617}{43.223}
\emmoveto{43.726}{43.252}
\emlineto{43.871}{43.313}
\emmoveto{43.980}{43.342}
\emlineto{44.126}{43.403}
\emmoveto{44.235}{43.431}
\emlineto{44.381}{43.492}
\emmoveto{44.490}{43.520}
\emlineto{44.637}{43.580}
\emmoveto{44.746}{43.608}
\emlineto{44.893}{43.668}
\emmoveto{45.003}{43.695}
\emlineto{45.150}{43.755}
\emmoveto{45.261}{43.782}
\emlineto{45.408}{43.841}
\emmoveto{45.519}{43.868}
\emlineto{45.666}{43.927}
\emmoveto{45.777}{43.953}
\emlineto{45.926}{44.012}
\emmoveto{46.037}{44.038}
\emlineto{46.185}{44.096}
\emmoveto{46.297}{44.122}
\emlineto{46.446}{44.179}
\emmoveto{46.558}{44.205}
\emlineto{46.707}{44.262}
\emmoveto{46.819}{44.287}
\emlineto{46.969}{44.344}
\emmoveto{47.081}{44.369}
\emlineto{47.231}{44.425}
\emmoveto{47.343}{44.450}
\emlineto{47.494}{44.506}
\emmoveto{47.607}{44.530}
\emlineto{47.757}{44.586}
\emmoveto{47.870}{44.610}
\emlineto{48.021}{44.665}
\emmoveto{48.135}{44.689}
\emlineto{48.286}{44.744}
\emmoveto{48.400}{44.767}
\emlineto{48.552}{44.822}
\emmoveto{48.665}{44.845}
\emlineto{48.817}{44.899}
\emmoveto{48.932}{44.922}
\emlineto{49.084}{44.975}
\emmoveto{49.198}{44.998}
\emlineto{49.351}{45.051}
\emmoveto{49.466}{45.073}
\emlineto{49.619}{45.126}
\emmoveto{49.734}{45.148}
\emlineto{49.887}{45.200}
\emmoveto{50.002}{45.222}
\emlineto{50.156}{45.273}
\emmoveto{50.271}{45.295}
\emlineto{50.425}{45.346}
\emmoveto{50.541}{45.367}
\emlineto{50.695}{45.418}
\emmoveto{50.811}{45.439}
\emlineto{50.966}{45.490}
\emmoveto{51.082}{45.510}
\emlineto{51.237}{45.560}
\emmoveto{51.353}{45.580}
\emlineto{51.508}{45.630}
\emmoveto{51.625}{45.650}
\emlineto{51.780}{45.699}
\emmoveto{51.897}{45.719}
\emlineto{52.053}{45.768}
\emmoveto{52.170}{45.787}
\emlineto{52.326}{45.835}
\emmoveto{52.443}{45.854}
\emlineto{52.599}{45.902}
\emmoveto{52.717}{45.921}
\emlineto{52.874}{45.969}
\emmoveto{52.991}{45.987}
\emlineto{53.148}{46.034}
\emmoveto{53.266}{46.052}
\emlineto{53.423}{46.099}
\emmoveto{53.541}{46.116}
\emlineto{53.699}{46.163}
\emmoveto{53.817}{46.180}
\emlineto{53.975}{46.226}
\emmoveto{54.094}{46.243}
\emlineto{54.252}{46.289}
\emmoveto{54.370}{46.305}
\emlineto{54.529}{46.350}
\emmoveto{54.648}{46.367}
\emlineto{54.806}{46.411}
\emmoveto{54.925}{46.427}
\emlineto{55.084}{46.472}
\emmoveto{55.204}{46.487}
\emlineto{55.363}{46.531}
\emmoveto{55.482}{46.547}
\emlineto{55.642}{46.590}
\emmoveto{55.761}{46.605}
\emlineto{55.921}{46.648}
\emmoveto{56.041}{46.663}
\emlineto{56.201}{46.705}
\emmoveto{56.321}{46.720}
\emlineto{56.481}{46.762}
\emmoveto{56.601}{46.776}
\emlineto{56.762}{46.818}
\emmoveto{56.882}{46.831}
\emlineto{57.043}{46.873}
\emmoveto{57.163}{46.886}
\emlineto{57.324}{46.927}
\emmoveto{57.445}{46.940}
\emlineto{57.606}{46.981}
\emmoveto{57.727}{46.993}
\emlineto{57.888}{47.033}
\emmoveto{58.009}{47.046}
\emlineto{58.171}{47.085}
\emmoveto{58.292}{47.097}
\emlineto{58.454}{47.137}
\emmoveto{58.576}{47.148}
\emlineto{58.738}{47.187}
\emmoveto{58.859}{47.199}
\emlineto{59.022}{47.237}
\emmoveto{59.143}{47.248}
\emlineto{59.306}{47.286}
\emmoveto{59.428}{47.297}
\emlineto{59.590}{47.334}
\emmoveto{59.712}{47.344}
\emlineto{59.875}{47.381}
\emmoveto{59.998}{47.392}
\emlineto{60.161}{47.428}
\emmoveto{60.283}{47.438}
\emlineto{60.446}{47.474}
\emmoveto{60.569}{47.483}
\emlineto{60.732}{47.519}
\emmoveto{60.855}{47.528}
\emlineto{61.019}{47.564}
\emmoveto{61.142}{47.572}
\emlineto{61.306}{47.607}
\emmoveto{61.429}{47.616}
\emlineto{61.593}{47.650}
\emmoveto{61.716}{47.658}
\emlineto{61.880}{47.692}
\emmoveto{62.003}{47.700}
\emlineto{62.168}{47.733}
\emmoveto{62.291}{47.741}
\emlineto{62.456}{47.774}
\emmoveto{62.579}{47.781}
\emlineto{62.744}{47.814}
\emmoveto{62.868}{47.820}
\emlineto{63.033}{47.853}
\emmoveto{63.156}{47.859}
\emlineto{63.321}{47.891}
\emmoveto{63.445}{47.897}
\emlineto{63.611}{47.928}
\emmoveto{63.735}{47.934}
\emlineto{63.900}{47.965}
\emmoveto{64.024}{47.970}
\emlineto{64.190}{48.001}
\emmoveto{64.314}{48.006}
\emlineto{64.480}{48.036}
\emmoveto{64.604}{48.040}
\emlineto{64.770}{48.070}
\emmoveto{64.895}{48.074}
\emlineto{65.061}{48.104}
\emmoveto{65.185}{48.108}
\emlineto{65.352}{48.136}
\emmoveto{65.476}{48.140}
\emlineto{65.643}{48.168}
\emmoveto{65.767}{48.172}
\emlineto{65.934}{48.199}
\emmoveto{66.059}{48.203}
\emlineto{66.225}{48.230}
\emmoveto{66.350}{48.233}
\emlineto{66.517}{48.259}
\emmoveto{66.642}{48.262}
\emlineto{66.809}{48.288}
\emmoveto{66.934}{48.290}
\emlineto{67.101}{48.316}
\emmoveto{67.227}{48.318}
\emlineto{67.394}{48.344}
\emmoveto{67.519}{48.345}
\emlineto{67.686}{48.370}
\emmoveto{67.812}{48.371}
\emlineto{67.979}{48.396}
\emmoveto{68.105}{48.396}
\emlineto{68.272}{48.421}
\emmoveto{68.398}{48.421}
\emlineto{68.565}{48.445}
\emmoveto{68.691}{48.445}
\emlineto{68.859}{48.468}
\emmoveto{68.984}{48.468}
\emlineto{69.152}{48.491}
\emmoveto{69.278}{48.490}
\emlineto{69.446}{48.512}
\emmoveto{69.572}{48.511}
\emlineto{69.740}{48.533}
\emmoveto{69.866}{48.532}
\emlineto{70.034}{48.553}
\emmoveto{70.160}{48.552}
\emlineto{70.328}{48.573}
\emmoveto{70.454}{48.571}
\emlineto{70.622}{48.591}
\emmoveto{70.748}{48.589}
\emlineto{70.917}{48.609}
\emmoveto{71.043}{48.607}
\emlineto{71.211}{48.626}
\emmoveto{71.338}{48.623}
\emlineto{71.506}{48.642}
\emmoveto{71.632}{48.639}
\emlineto{71.801}{48.658}
\emmoveto{71.927}{48.654}
\emlineto{72.096}{48.673}
\emmoveto{72.222}{48.669}
\emlineto{72.391}{48.686}
\emmoveto{72.517}{48.682}
\emlineto{72.686}{48.699}
\emmoveto{72.813}{48.695}
\emlineto{72.981}{48.712}
\emmoveto{73.108}{48.707}
\emlineto{73.277}{48.723}
\emmoveto{73.403}{48.718}
\emlineto{73.572}{48.734}
\emmoveto{73.699}{48.728}
\emlineto{73.868}{48.744}
\emmoveto{73.994}{48.738}
\emlineto{74.163}{48.753}
\emmoveto{74.290}{48.746}
\emlineto{74.459}{48.761}
\emmoveto{74.586}{48.754}
\emlineto{74.755}{48.769}
\emmoveto{74.881}{48.762}
\emlineto{75.050}{48.775}
\emmoveto{75.177}{48.768}
\emlineto{75.346}{48.781}
\emmoveto{75.473}{48.773}
\emlineto{75.642}{48.786}
\emmoveto{75.769}{48.778}
\emlineto{75.938}{48.791}
\emmoveto{76.065}{48.782}
\emlineto{76.234}{48.794}
\emmoveto{76.361}{48.785}
\emlineto{76.530}{48.797}
\emmoveto{76.657}{48.788}
\emlineto{76.826}{48.799}
\emmoveto{76.953}{48.789}
\emlineto{77.122}{48.800}
\emmoveto{77.249}{48.790}
\emlineto{77.418}{48.800}
\emmoveto{77.545}{48.790}
\emlineto{77.714}{48.800}
\emmoveto{77.841}{48.789}
\emlineto{78.010}{48.799}
\emmoveto{78.137}{48.788}
\emlineto{78.306}{48.797}
\emmoveto{78.432}{48.785}
\emlineto{78.602}{48.794}
\emmoveto{78.728}{48.782}
\emlineto{78.898}{48.790}
\emmoveto{79.024}{48.778}
\emlineto{79.193}{48.786}
\emmoveto{79.320}{48.774}
\emlineto{79.489}{48.780}
\emmoveto{79.616}{48.768}
\emlineto{79.785}{48.774}
\emmoveto{79.912}{48.762}
\emlineto{80.081}{48.768}
\emmoveto{80.208}{48.754}
\emlineto{80.377}{48.760}
\emmoveto{80.503}{48.747}
\emlineto{80.672}{48.752}
\emmoveto{80.799}{48.738}
\emlineto{80.968}{48.742}
\emmoveto{81.094}{48.728}
\emlineto{81.263}{48.732}
\emmoveto{81.390}{48.718}
\emlineto{81.559}{48.722}
\emmoveto{81.685}{48.707}
\emlineto{81.854}{48.710}
\emmoveto{81.981}{48.695}
\emlineto{82.149}{48.698}
\emmoveto{82.276}{48.682}
\emlineto{82.444}{48.685}
\emmoveto{82.571}{48.669}
\emlineto{82.739}{48.671}
\emmoveto{82.866}{48.654}
\emlineto{83.034}{48.656}
\emmoveto{83.161}{48.639}
\emlineto{83.329}{48.640}
\emmoveto{83.456}{48.624}
\emlineto{83.624}{48.624}
\emmoveto{83.750}{48.607}
\emlineto{83.919}{48.607}
\emmoveto{84.045}{48.589}
\emlineto{84.213}{48.589}
\emmoveto{84.339}{48.571}
\emlineto{84.507}{48.570}
\emmoveto{84.633}{48.552}
\emlineto{84.801}{48.551}
\emmoveto{84.927}{48.532}
\emlineto{85.095}{48.531}
\emmoveto{85.221}{48.512}
\emlineto{85.389}{48.510}
\emmoveto{85.515}{48.490}
\emlineto{85.683}{48.488}
\emmoveto{85.809}{48.468}
\emlineto{85.976}{48.465}
\emmoveto{86.102}{48.445}
\emlineto{86.270}{48.442}
\emmoveto{86.395}{48.421}
\emlineto{86.563}{48.417}
\emmoveto{86.689}{48.397}
\emlineto{86.856}{48.392}
\emmoveto{86.981}{48.372}
\emlineto{87.149}{48.367}
\emmoveto{87.274}{48.345}
\emlineto{87.441}{48.340}
\emmoveto{87.567}{48.318}
\emlineto{87.734}{48.313}
\emmoveto{87.859}{48.291}
\emlineto{88.026}{48.285}
\emmoveto{88.151}{48.262}
\emlineto{88.318}{48.256}
\emmoveto{88.443}{48.233}
\emlineto{88.609}{48.226}
\emmoveto{88.734}{48.203}
\emlineto{88.901}{48.195}
\emmoveto{89.026}{48.172}
\emlineto{89.192}{48.164}
\emmoveto{89.317}{48.141}
\emlineto{89.483}{48.132}
\emmoveto{89.608}{48.108}
\emlineto{89.774}{48.099}
\emmoveto{89.899}{48.075}
\emlineto{90.064}{48.066}
\emmoveto{90.189}{48.041}
\emlineto{90.355}{48.031}
\emmoveto{90.479}{48.006}
\emlineto{90.645}{47.996}
\emmoveto{90.769}{47.971}
\emlineto{90.934}{47.960}
\emmoveto{91.059}{47.935}
\emlineto{91.224}{47.923}
\emmoveto{91.348}{47.897}
\emlineto{91.513}{47.886}
\emmoveto{91.637}{47.860}
\emlineto{91.802}{47.848}
\emmoveto{91.926}{47.821}
\emlineto{92.090}{47.809}
\emmoveto{92.214}{47.782}
\emlineto{92.379}{47.769}
\emmoveto{92.502}{47.741}
\emlineto{92.667}{47.728}
\emmoveto{92.790}{47.700}
\emlineto{92.954}{47.687}
\emmoveto{93.078}{47.659}
\emlineto{93.242}{47.645}
\emmoveto{93.365}{47.616}
\emlineto{93.529}{47.602}
\emmoveto{93.652}{47.573}
\emlineto{93.815}{47.558}
\emmoveto{93.938}{47.529}
\emlineto{94.102}{47.513}
\emmoveto{94.224}{47.484}
\emlineto{94.388}{47.468}
\emmoveto{94.510}{47.439}
\emlineto{94.673}{47.422}
\emmoveto{94.796}{47.392}
\emlineto{94.959}{47.375}
\emmoveto{95.081}{47.345}
\emlineto{95.244}{47.328}
\emmoveto{95.366}{47.297}
\emlineto{95.528}{47.280}
\emmoveto{95.650}{47.249}
\emlineto{95.812}{47.231}
\emmoveto{95.934}{47.199}
\emlineto{96.096}{47.181}
\emmoveto{96.218}{47.149}
\emlineto{96.380}{47.130}
\emmoveto{96.501}{47.098}
\emlineto{96.663}{47.079}
\emmoveto{96.784}{47.047}
\emlineto{96.945}{47.027}
\emmoveto{97.066}{46.994}
\emlineto{97.228}{46.974}
\emmoveto{97.348}{46.941}
\emlineto{97.509}{46.920}
\emmoveto{97.630}{46.887}
\emlineto{97.791}{46.866}
\emmoveto{97.911}{46.832}
\emlineto{98.072}{46.811}
\emmoveto{98.192}{46.777}
\emlineto{98.352}{46.755}
\emmoveto{98.473}{46.721}
\emlineto{98.633}{46.698}
\emmoveto{98.753}{46.664}
\emlineto{98.912}{46.641}
\emmoveto{99.032}{46.606}
\emlineto{99.192}{46.583}
\emmoveto{99.311}{46.548}
\emlineto{99.470}{46.524}
\emmoveto{99.590}{46.488}
\emlineto{99.749}{46.464}
\emmoveto{99.868}{46.428}
\emlineto{100.027}{46.404}
\emmoveto{100.146}{46.368}
\emlineto{100.304}{46.343}
\emmoveto{100.423}{46.306}
\emlineto{100.581}{46.281}
\emmoveto{100.700}{46.244}
\emlineto{100.858}{46.218}
\emmoveto{100.976}{46.181}
\emlineto{101.134}{46.155}
\emmoveto{101.252}{46.117}
\emlineto{101.409}{46.091}
\emmoveto{101.527}{46.053}
\emlineto{101.684}{46.026}
\emmoveto{101.802}{45.988}
\emlineto{101.959}{45.960}
\emmoveto{102.077}{45.922}
\emlineto{102.233}{45.894}
\emmoveto{102.350}{45.855}
\emlineto{102.507}{45.827}
\emmoveto{102.624}{45.788}
\emlineto{102.780}{45.759}
\emmoveto{102.897}{45.720}
\emlineto{103.052}{45.691}
\emmoveto{103.169}{45.651}
\emlineto{103.324}{45.621}
\emmoveto{103.441}{45.581}
\emlineto{103.596}{45.551}
\emmoveto{103.712}{45.511}
\emlineto{103.867}{45.481}
\emmoveto{103.983}{45.440}
\emlineto{104.137}{45.409}
\emmoveto{104.253}{45.368}
\emlineto{104.407}{45.337}
\emmoveto{104.522}{45.296}
\emlineto{104.676}{45.264}
\emmoveto{104.791}{45.223}
\emlineto{104.945}{45.191}
\emmoveto{105.060}{45.149}
\emlineto{105.213}{45.116}
\emmoveto{105.328}{45.074}
\emlineto{105.481}{45.041}
\emmoveto{105.595}{44.999}
\emlineto{105.748}{44.965}
\emmoveto{105.862}{44.923}
\emlineto{106.014}{44.889}
\emmoveto{106.128}{44.846}
\emlineto{106.280}{44.812}
\emmoveto{106.394}{44.768}
\emlineto{106.545}{44.734}
\emmoveto{106.659}{44.690}
\emlineto{106.810}{44.655}
\emmoveto{106.923}{44.611}
\emlineto{107.074}{44.576}
\emmoveto{107.187}{44.532}
\emlineto{107.337}{44.496}
\emmoveto{107.450}{44.451}
\emlineto{107.600}{44.415}
\emmoveto{107.713}{44.370}
\emlineto{107.862}{44.334}
\emmoveto{107.975}{44.289}
\emlineto{108.124}{44.252}
\emmoveto{108.236}{44.206}
\emlineto{108.385}{44.169}
\emmoveto{108.497}{44.123}
\emlineto{108.645}{44.085}
\emmoveto{108.757}{44.039}
\emlineto{108.905}{44.001}
\emmoveto{109.016}{43.955}
\emlineto{109.164}{43.916}
\emmoveto{109.275}{43.869}
\emlineto{109.422}{43.830}
\emmoveto{109.533}{43.783}
\emlineto{109.680}{43.744}
\emmoveto{109.790}{43.697}
\emlineto{109.937}{43.657}
\emmoveto{110.047}{43.609}
\emlineto{110.194}{43.569}
\emmoveto{110.303}{43.521}
\emlineto{110.449}{43.481}
\emmoveto{110.559}{43.433}
\emlineto{110.704}{43.392}
\emmoveto{110.814}{43.343}
\emlineto{110.959}{43.302}
\emmoveto{111.068}{43.253}
\emlineto{111.213}{43.212}
\emmoveto{111.321}{43.163}
\emlineto{111.466}{43.120}
\emmoveto{111.574}{43.071}
\emlineto{111.718}{43.029}
\emmoveto{111.826}{42.979}
\emlineto{111.969}{42.936}
\emmoveto{112.077}{42.886}
\emlineto{112.220}{42.843}
\emmoveto{112.328}{42.793}
\emlineto{112.471}{42.749}
\emmoveto{112.578}{42.699}
\emlineto{112.720}{42.655}
\emmoveto{112.827}{42.604}
\emlineto{112.969}{42.559}
\emmoveto{113.075}{42.508}
\emlineto{113.217}{42.464}
\emmoveto{113.323}{42.412}
\emlineto{113.464}{42.367}
\emmoveto{113.570}{42.315}
\emlineto{113.711}{42.270}
\emmoveto{113.816}{42.218}
\emlineto{113.956}{42.172}
\emmoveto{114.061}{42.120}
\emlineto{114.201}{42.074}
\emmoveto{114.306}{42.021}
\emlineto{114.446}{41.974}
\emmoveto{114.550}{41.922}
\emlineto{114.689}{41.875}
\emmoveto{114.793}{41.822}
\emlineto{114.932}{41.774}
\emmoveto{115.036}{41.721}
\emlineto{115.174}{41.673}
\emmoveto{115.277}{41.620}
\emlineto{115.415}{41.571}
\emmoveto{115.518}{41.517}
\emlineto{115.656}{41.469}
\emmoveto{115.758}{41.415}
\emlineto{115.895}{41.366}
\emmoveto{115.998}{41.311}
\emlineto{116.134}{41.262}
\emmoveto{116.236}{41.208}
\emlineto{116.372}{41.158}
\emmoveto{116.474}{41.103}
\emlineto{116.610}{41.053}
\emmoveto{116.711}{40.998}
\emlineto{116.846}{40.947}
\emmoveto{116.947}{40.892}
\emlineto{117.082}{40.841}
\emmoveto{117.182}{40.785}
\emlineto{117.317}{40.734}
\emmoveto{117.417}{40.678}
\emlineto{117.551}{40.627}
\emmoveto{117.651}{40.570}
\emlineto{117.784}{40.519}
\emmoveto{117.884}{40.462}
\emlineto{118.016}{40.410}
\emmoveto{118.116}{40.353}
\emlineto{118.248}{40.300}
\emmoveto{118.347}{40.243}
\emlineto{118.478}{40.190}
\emmoveto{118.577}{40.133}
\emlineto{118.708}{40.080}
\emmoveto{118.807}{40.022}
\emlineto{118.937}{39.969}
\emmoveto{119.035}{39.911}
\emlineto{119.166}{39.857}
\emmoveto{119.263}{39.799}
\emlineto{119.393}{39.744}
\emmoveto{119.490}{39.686}
\emlineto{119.619}{39.631}
\emmoveto{119.716}{39.573}
\emlineto{119.845}{39.518}
\emmoveto{119.941}{39.459}
\emlineto{120.070}{39.403}
\emmoveto{120.166}{39.344}
\emlineto{120.294}{39.288}
\emmoveto{120.389}{39.229}
\emlineto{120.517}{39.173}
\emmoveto{120.612}{39.113}
\emlineto{120.739}{39.057}
\emmoveto{120.834}{38.997}
\emlineto{120.960}{38.940}
\emmoveto{121.054}{38.880}
\emlineto{121.180}{38.823}
\emmoveto{121.274}{38.762}
\emlineto{121.400}{38.705}
\emmoveto{121.493}{38.644}
\emlineto{121.618}{38.587}
\emmoveto{121.712}{38.526}
\emlineto{121.836}{38.468}
\emmoveto{121.929}{38.406}
\emlineto{122.052}{38.348}
\emmoveto{122.145}{38.286}
\emlineto{122.268}{38.228}
\emmoveto{122.360}{38.166}
\emlineto{122.483}{38.107}
\emmoveto{122.575}{38.045}
\emlineto{122.697}{37.986}
\emmoveto{122.788}{37.923}
\emlineto{122.910}{37.864}
\emmoveto{123.001}{37.801}
\emlineto{123.122}{37.741}
\emmoveto{123.213}{37.679}
\emlineto{123.333}{37.618}
\emmoveto{123.424}{37.555}
\emlineto{123.544}{37.495}
\emmoveto{123.633}{37.431}
\emlineto{123.753}{37.370}
\emmoveto{123.842}{37.307}
\emlineto{123.961}{37.246}
\emmoveto{124.050}{37.182}
\emlineto{124.168}{37.120}
\emmoveto{124.257}{37.056}
\emlineto{124.375}{36.994}
\emmoveto{124.463}{36.930}
\emlineto{124.580}{36.868}
\emmoveto{124.668}{36.804}
\emlineto{124.785}{36.741}
\emmoveto{124.872}{36.676}
\emlineto{125.017}{36.595}
\emmoveto{125.133}{36.512}
\emlineto{125.277}{36.430}
\emmoveto{125.392}{36.347}
\emlineto{125.536}{36.265}
\emmoveto{125.650}{36.181}
\emlineto{125.793}{36.098}
\emmoveto{125.906}{36.014}
\emlineto{126.048}{35.931}
\emmoveto{126.161}{35.846}
\emlineto{126.301}{35.762}
\emmoveto{126.413}{35.677}
\emlineto{126.553}{35.593}
\emmoveto{126.664}{35.507}
\emlineto{126.803}{35.423}
\emmoveto{126.914}{35.337}
\emlineto{127.052}{35.252}
\emmoveto{127.162}{35.166}
\emlineto{127.299}{35.080}
\emmoveto{127.408}{34.993}
\emlineto{127.544}{34.907}
\emmoveto{127.652}{34.820}
\emlineto{127.787}{34.734}
\emmoveto{127.895}{34.646}
\emlineto{128.029}{34.559}
\emmoveto{128.136}{34.472}
\emlineto{128.269}{34.384}
\emmoveto{128.375}{34.296}
\emlineto{128.507}{34.208}
\emmoveto{128.612}{34.120}
\emlineto{128.744}{34.031}
\emmoveto{128.848}{33.942}
\emlineto{128.978}{33.854}
\emmoveto{129.082}{33.764}
\emlineto{129.212}{33.675}
\emmoveto{129.315}{33.586}
\emlineto{129.443}{33.496}
\emmoveto{129.545}{33.406}
\emlineto{129.672}{33.316}
\emmoveto{129.774}{33.225}
\emlineto{129.900}{33.135}
\emshow{107.580}{45.700}{ v=0.43}
\emmoveto{12.000}{10.014}
\emlineto{12.001}{10.178}
\emmoveto{12.002}{10.296}
\emlineto{12.005}{10.460}
\emmoveto{12.007}{10.578}
\emlineto{12.012}{10.742}
\emmoveto{12.016}{10.861}
\emlineto{12.022}{11.025}
\emmoveto{12.028}{11.143}
\emlineto{12.036}{11.307}
\emmoveto{12.044}{11.425}
\emlineto{12.054}{11.589}
\emmoveto{12.063}{11.707}
\emlineto{12.074}{11.871}
\emmoveto{12.085}{11.989}
\emlineto{12.099}{12.153}
\emmoveto{12.111}{12.271}
\emlineto{12.126}{12.435}
\emmoveto{12.140}{12.553}
\emlineto{12.157}{12.717}
\emmoveto{12.172}{12.835}
\emlineto{12.191}{12.998}
\emmoveto{12.208}{13.116}
\emlineto{12.229}{13.280}
\emmoveto{12.247}{13.398}
\emlineto{12.270}{13.561}
\emmoveto{12.290}{13.679}
\emlineto{12.314}{13.842}
\emmoveto{12.336}{13.960}
\emlineto{12.362}{14.123}
\emmoveto{12.385}{14.241}
\emlineto{12.413}{14.404}
\emmoveto{12.437}{14.522}
\emlineto{12.468}{14.685}
\emmoveto{12.493}{14.802}
\emlineto{12.525}{14.965}
\emmoveto{12.553}{15.082}
\emlineto{12.587}{15.245}
\emmoveto{12.616}{15.362}
\emlineto{12.651}{15.525}
\emmoveto{12.682}{15.642}
\emlineto{12.719}{15.804}
\emmoveto{12.751}{15.921}
\emlineto{12.790}{16.083}
\emmoveto{12.824}{16.200}
\emlineto{12.865}{16.362}
\emmoveto{12.900}{16.479}
\emlineto{12.943}{16.641}
\emmoveto{12.979}{16.757}
\emlineto{13.024}{16.919}
\emmoveto{13.062}{17.035}
\emlineto{13.109}{17.197}
\emmoveto{13.148}{17.313}
\emlineto{13.197}{17.474}
\emmoveto{13.238}{17.590}
\emlineto{13.288}{17.752}
\emmoveto{13.331}{17.867}
\emlineto{13.383}{18.028}
\emmoveto{13.427}{18.144}
\emlineto{13.481}{18.305}
\emmoveto{13.526}{18.420}
\emlineto{13.582}{18.581}
\emmoveto{13.629}{18.696}
\emlineto{13.686}{18.856}
\emmoveto{13.735}{18.971}
\emlineto{13.794}{19.131}
\emmoveto{13.844}{19.246}
\emlineto{13.905}{19.406}
\emmoveto{13.957}{19.520}
\emlineto{14.020}{19.680}
\emmoveto{14.073}{19.794}
\emlineto{14.137}{19.953}
\emmoveto{14.192}{20.068}
\emlineto{14.258}{20.226}
\emmoveto{14.314}{20.340}
\emlineto{14.382}{20.499}
\emmoveto{14.440}{20.613}
\emlineto{14.510}{20.771}
\emmoveto{14.569}{20.885}
\emlineto{14.641}{21.043}
\emmoveto{14.689}{21.131}
\emlineto{14.750}{21.264}
\emmoveto{14.799}{21.353}
\emlineto{14.862}{21.486}
\emmoveto{14.912}{21.574}
\emlineto{14.975}{21.707}
\emmoveto{15.027}{21.795}
\emlineto{15.091}{21.927}
\emmoveto{15.144}{22.015}
\emlineto{15.209}{22.147}
\emmoveto{15.263}{22.235}
\emlineto{15.330}{22.367}
\emmoveto{15.384}{22.455}
\emlineto{15.452}{22.586}
\emmoveto{15.507}{22.674}
\emlineto{15.577}{22.805}
\emmoveto{15.633}{22.892}
\emlineto{15.703}{23.024}
\emmoveto{15.760}{23.111}
\emlineto{15.832}{23.242}
\emmoveto{15.890}{23.329}
\emlineto{15.963}{23.459}
\emmoveto{16.022}{23.546}
\emlineto{16.096}{23.677}
\emmoveto{16.156}{23.763}
\emlineto{16.231}{23.893}
\emmoveto{16.292}{23.979}
\emlineto{16.369}{24.109}
\emmoveto{16.430}{24.195}
\emlineto{16.508}{24.325}
\emmoveto{16.571}{24.411}
\emlineto{16.650}{24.540}
\emmoveto{16.713}{24.626}
\emlineto{16.793}{24.755}
\emmoveto{16.858}{24.840}
\emlineto{16.939}{24.969}
\emmoveto{17.004}{25.054}
\emlineto{17.087}{25.183}
\emmoveto{17.153}{25.268}
\emlineto{17.237}{25.396}
\emmoveto{17.304}{25.481}
\emlineto{17.389}{25.609}
\emmoveto{17.457}{25.693}
\emlineto{17.543}{25.821}
\emmoveto{17.612}{25.905}
\emlineto{17.699}{26.033}
\emmoveto{17.769}{26.117}
\emlineto{17.858}{26.244}
\emmoveto{17.929}{26.327}
\emlineto{18.018}{26.454}
\emmoveto{18.090}{26.538}
\emlineto{18.180}{26.664}
\emmoveto{18.253}{26.747}
\emlineto{18.345}{26.874}
\emmoveto{18.419}{26.957}
\emlineto{18.511}{27.083}
\emmoveto{18.586}{27.165}
\emlineto{18.680}{27.291}
\emmoveto{18.755}{27.373}
\emlineto{18.851}{27.498}
\emmoveto{18.927}{27.581}
\emlineto{19.023}{27.706}
\emmoveto{19.101}{27.787}
\emlineto{19.198}{27.912}
\emmoveto{19.276}{27.994}
\emlineto{19.375}{28.118}
\emmoveto{19.454}{28.199}
\emlineto{19.553}{28.323}
\emmoveto{19.633}{28.404}
\emlineto{19.734}{28.528}
\emmoveto{19.815}{28.609}
\emlineto{19.917}{28.732}
\emmoveto{19.999}{28.812}
\emlineto{20.101}{28.935}
\emmoveto{20.184}{29.015}
\emlineto{20.288}{29.138}
\emmoveto{20.372}{29.218}
\emlineto{20.477}{29.340}
\emmoveto{20.561}{29.420}
\emlineto{20.668}{29.542}
\emmoveto{20.753}{29.621}
\emlineto{20.860}{29.742}
\emmoveto{20.946}{29.821}
\emlineto{21.055}{29.942}
\emmoveto{21.142}{30.021}
\emlineto{21.251}{30.142}
\emmoveto{21.339}{30.220}
\emlineto{21.450}{30.341}
\emmoveto{21.539}{30.419}
\emlineto{21.650}{30.539}
\emmoveto{21.740}{30.616}
\emlineto{21.853}{30.736}
\emmoveto{21.943}{30.813}
\emlineto{22.057}{30.933}
\emmoveto{22.149}{31.010}
\emlineto{22.263}{31.129}
\emmoveto{22.356}{31.205}
\emlineto{22.472}{31.324}
\emmoveto{22.565}{31.400}
\emlineto{22.682}{31.518}
\emmoveto{22.776}{31.595}
\emlineto{22.894}{31.712}
\emmoveto{22.989}{31.788}
\emlineto{23.108}{31.905}
\emmoveto{23.203}{31.981}
\emlineto{23.323}{32.098}
\emmoveto{23.420}{32.173}
\emlineto{23.541}{32.289}
\emmoveto{23.638}{32.364}
\emlineto{23.761}{32.480}
\emmoveto{23.859}{32.554}
\emlineto{23.982}{32.670}
\emmoveto{24.081}{32.744}
\emlineto{24.205}{32.859}
\emmoveto{24.305}{32.933}
\emlineto{24.430}{33.048}
\emmoveto{24.531}{33.121}
\emlineto{24.657}{33.235}
\emmoveto{24.759}{33.309}
\emlineto{24.886}{33.422}
\emmoveto{24.989}{33.495}
\emlineto{25.117}{33.609}
\emmoveto{25.220}{33.681}
\emlineto{25.349}{33.794}
\emmoveto{25.453}{33.866}
\emlineto{25.584}{33.978}
\emmoveto{25.688}{34.050}
\emlineto{25.820}{34.162}
\emmoveto{25.925}{34.234}
\emlineto{26.058}{34.345}
\emmoveto{26.164}{34.416}
\emlineto{26.297}{34.527}
\emmoveto{26.404}{34.598}
\emlineto{26.539}{34.708}
\emmoveto{26.647}{34.779}
\emlineto{26.782}{34.889}
\emmoveto{26.891}{34.959}
\emlineto{27.027}{35.068}
\emmoveto{27.136}{35.138}
\emlineto{27.274}{35.247}
\emmoveto{27.384}{35.316}
\emlineto{27.522}{35.425}
\emmoveto{27.633}{35.494}
\emlineto{27.772}{35.602}
\emmoveto{27.884}{35.670}
\emlineto{28.024}{35.778}
\emmoveto{28.137}{35.846}
\emlineto{28.278}{35.953}
\emmoveto{28.363}{36.002}
\emlineto{28.477}{36.089}
\emmoveto{28.562}{36.137}
\emlineto{28.676}{36.224}
\emmoveto{28.762}{36.272}
\emlineto{28.877}{36.359}
\emmoveto{28.963}{36.407}
\emlineto{29.078}{36.493}
\emmoveto{29.165}{36.540}
\emlineto{29.281}{36.627}
\emmoveto{29.368}{36.674}
\emlineto{29.485}{36.760}
\emmoveto{29.572}{36.807}
\emlineto{29.689}{36.892}
\emmoveto{29.777}{36.939}
\emlineto{29.895}{37.024}
\emmoveto{29.983}{37.070}
\emlineto{30.102}{37.155}
\emmoveto{30.190}{37.202}
\emlineto{30.309}{37.286}
\emmoveto{30.398}{37.332}
\emlineto{30.518}{37.416}
\emmoveto{30.608}{37.462}
\emlineto{30.727}{37.546}
\emmoveto{30.818}{37.592}
\emlineto{30.938}{37.675}
\emmoveto{31.029}{37.720}
\emlineto{31.150}{37.804}
\emmoveto{31.241}{37.849}
\emlineto{31.362}{37.932}
\emmoveto{31.454}{37.977}
\emlineto{31.576}{38.059}
\emmoveto{31.668}{38.104}
\emlineto{31.790}{38.186}
\emmoveto{31.883}{38.230}
\emlineto{32.006}{38.312}
\emmoveto{32.099}{38.356}
\emlineto{32.222}{38.438}
\emmoveto{32.315}{38.482}
\emlineto{32.440}{38.563}
\emmoveto{32.533}{38.607}
\emlineto{32.658}{38.688}
\emmoveto{32.752}{38.731}
\emlineto{32.877}{38.812}
\emmoveto{32.972}{38.855}
\emlineto{33.098}{38.935}
\emmoveto{33.192}{38.978}
\emlineto{33.319}{39.058}
\emmoveto{33.414}{39.100}
\emlineto{33.541}{39.180}
\emmoveto{33.637}{39.222}
\emlineto{33.764}{39.301}
\emmoveto{33.860}{39.343}
\emlineto{33.988}{39.422}
\emmoveto{34.084}{39.464}
\emlineto{34.213}{39.543}
\emmoveto{34.310}{39.584}
\emlineto{34.439}{39.662}
\emmoveto{34.536}{39.704}
\emlineto{34.666}{39.782}
\emmoveto{34.763}{39.822}
\emlineto{34.893}{39.900}
\emmoveto{34.991}{39.941}
\emlineto{35.122}{40.018}
\emmoveto{35.220}{40.058}
\emlineto{35.351}{40.135}
\emmoveto{35.450}{40.175}
\emlineto{35.582}{40.252}
\emmoveto{35.681}{40.292}
\emlineto{35.813}{40.368}
\emmoveto{35.912}{40.408}
\emlineto{36.045}{40.483}
\emmoveto{36.145}{40.523}
\emlineto{36.278}{40.598}
\emmoveto{36.378}{40.637}
\emlineto{36.512}{40.712}
\emmoveto{36.612}{40.751}
\emlineto{36.746}{40.826}
\emmoveto{36.847}{40.864}
\emlineto{36.982}{40.939}
\emmoveto{37.083}{40.977}
\emlineto{37.218}{41.051}
\emmoveto{37.320}{41.089}
\emlineto{37.456}{41.163}
\emmoveto{37.558}{41.201}
\emlineto{37.694}{41.274}
\emmoveto{37.796}{41.311}
\emlineto{37.933}{41.384}
\emmoveto{38.036}{41.421}
\emlineto{38.173}{41.494}
\emmoveto{38.276}{41.531}
\emlineto{38.413}{41.603}
\emmoveto{38.517}{41.640}
\emlineto{38.655}{41.712}
\emmoveto{38.759}{41.748}
\emlineto{38.897}{41.819}
\emmoveto{39.001}{41.855}
\emlineto{39.140}{41.926}
\emmoveto{39.245}{41.962}
\emlineto{39.384}{42.033}
\emmoveto{39.489}{42.068}
\emlineto{39.629}{42.139}
\emmoveto{39.734}{42.174}
\emlineto{39.875}{42.244}
\emmoveto{39.980}{42.279}
\emlineto{40.121}{42.348}
\emmoveto{40.227}{42.383}
\emlineto{40.368}{42.452}
\emmoveto{40.474}{42.487}
\emlineto{40.616}{42.556}
\emmoveto{40.723}{42.590}
\emlineto{40.865}{42.658}
\emmoveto{40.972}{42.692}
\emlineto{41.114}{42.760}
\emmoveto{41.222}{42.793}
\emlineto{41.365}{42.861}
\emmoveto{41.472}{42.894}
\emlineto{41.616}{42.962}
\emmoveto{41.724}{42.994}
\emlineto{41.868}{43.061}
\emmoveto{41.976}{43.094}
\emlineto{42.120}{43.161}
\emmoveto{42.229}{43.193}
\emlineto{42.374}{43.259}
\emmoveto{42.482}{43.291}
\emlineto{42.628}{43.357}
\emmoveto{42.737}{43.389}
\emlineto{42.882}{43.454}
\emmoveto{42.992}{43.485}
\emlineto{43.138}{43.550}
\emmoveto{43.248}{43.581}
\emlineto{43.394}{43.646}
\emmoveto{43.504}{43.677}
\emlineto{43.651}{43.741}
\emmoveto{43.762}{43.772}
\emlineto{43.909}{43.835}
\emmoveto{44.020}{43.866}
\emlineto{44.168}{43.929}
\emmoveto{44.279}{43.959}
\emlineto{44.427}{44.022}
\emmoveto{44.538}{44.052}
\emlineto{44.687}{44.114}
\emmoveto{44.798}{44.144}
\emlineto{44.947}{44.206}
\emmoveto{45.059}{44.235}
\emlineto{45.209}{44.297}
\emmoveto{45.321}{44.326}
\emlineto{45.470}{44.387}
\emmoveto{45.583}{44.415}
\emlineto{45.733}{44.476}
\emmoveto{45.846}{44.505}
\emlineto{45.996}{44.565}
\emmoveto{46.110}{44.593}
\emlineto{46.261}{44.653}
\emmoveto{46.374}{44.681}
\emlineto{46.525}{44.741}
\emmoveto{46.639}{44.768}
\emlineto{46.791}{44.827}
\emmoveto{46.904}{44.854}
\emlineto{47.057}{44.913}
\emmoveto{47.171}{44.940}
\emlineto{47.323}{44.998}
\emmoveto{47.438}{45.025}
\emlineto{47.591}{45.083}
\emmoveto{47.705}{45.109}
\emlineto{47.858}{45.167}
\emmoveto{47.973}{45.192}
\emlineto{48.127}{45.250}
\emmoveto{48.242}{45.275}
\emlineto{48.396}{45.332}
\emmoveto{48.512}{45.357}
\emlineto{48.666}{45.414}
\emmoveto{48.782}{45.438}
\emlineto{48.937}{45.494}
\emmoveto{49.053}{45.519}
\emlineto{49.208}{45.575}
\emmoveto{49.324}{45.599}
\emlineto{49.479}{45.654}
\emmoveto{49.596}{45.678}
\emlineto{49.752}{45.733}
\emmoveto{49.868}{45.756}
\emlineto{50.024}{45.811}
\emmoveto{50.142}{45.834}
\emlineto{50.298}{45.888}
\emmoveto{50.415}{45.911}
\emlineto{50.572}{45.964}
\emmoveto{50.690}{45.987}
\emlineto{50.847}{46.040}
\emmoveto{50.965}{46.062}
\emlineto{51.122}{46.115}
\emmoveto{51.240}{46.137}
\emlineto{51.398}{46.189}
\emmoveto{51.516}{46.211}
\emlineto{51.674}{46.263}
\emmoveto{51.793}{46.284}
\emlineto{51.951}{46.336}
\emmoveto{52.070}{46.357}
\emlineto{52.228}{46.408}
\emmoveto{52.347}{46.428}
\emlineto{52.506}{46.479}
\emmoveto{52.626}{46.499}
\emlineto{52.785}{46.550}
\emmoveto{52.904}{46.570}
\emlineto{53.064}{46.619}
\emmoveto{53.184}{46.639}
\emlineto{53.344}{46.688}
\emmoveto{53.464}{46.708}
\emlineto{53.624}{46.757}
\emmoveto{53.744}{46.776}
\emlineto{53.904}{46.824}
\emmoveto{54.025}{46.843}
\emlineto{54.186}{46.891}
\emmoveto{54.306}{46.910}
\emlineto{54.467}{46.957}
\emmoveto{54.588}{46.975}
\emlineto{54.749}{47.022}
\emmoveto{54.871}{47.040}
\emlineto{55.032}{47.087}
\emmoveto{55.153}{47.104}
\emlineto{55.315}{47.151}
\emmoveto{55.437}{47.168}
\emlineto{55.599}{47.214}
\emmoveto{55.721}{47.230}
\emlineto{55.883}{47.276}
\emmoveto{56.005}{47.292}
\emlineto{56.168}{47.337}
\emmoveto{56.290}{47.354}
\emlineto{56.453}{47.398}
\emmoveto{56.575}{47.414}
\emlineto{56.738}{47.458}
\emmoveto{56.861}{47.474}
\emlineto{57.024}{47.517}
\emmoveto{57.147}{47.532}
\emlineto{57.311}{47.576}
\emmoveto{57.433}{47.590}
\emlineto{57.597}{47.633}
\emmoveto{57.721}{47.648}
\emlineto{57.885}{47.690}
\emmoveto{58.008}{47.704}
\emlineto{58.173}{47.746}
\emmoveto{58.296}{47.760}
\emlineto{58.461}{47.802}
\emmoveto{58.584}{47.815}
\emlineto{58.749}{47.856}
\emmoveto{58.873}{47.869}
\emlineto{59.038}{47.910}
\emmoveto{59.162}{47.923}
\emlineto{59.328}{47.963}
\emmoveto{59.452}{47.975}
\emlineto{59.617}{48.015}
\emmoveto{59.742}{48.027}
\emlineto{59.908}{48.067}
\emmoveto{60.032}{48.078}
\emlineto{60.198}{48.117}
\emmoveto{60.323}{48.129}
\emlineto{60.489}{48.167}
\emmoveto{60.614}{48.178}
\emlineto{60.781}{48.216}
\emmoveto{60.905}{48.227}
\emlineto{61.072}{48.265}
\emmoveto{61.197}{48.275}
\emlineto{61.364}{48.312}
\emmoveto{61.490}{48.322}
\emlineto{61.657}{48.359}
\emmoveto{61.782}{48.369}
\emlineto{61.950}{48.405}
\emmoveto{62.075}{48.414}
\emlineto{62.243}{48.450}
\emmoveto{62.368}{48.459}
\emlineto{62.536}{48.494}
\emmoveto{62.662}{48.503}
\emlineto{62.830}{48.538}
\emmoveto{62.956}{48.546}
\emlineto{63.124}{48.581}
\emmoveto{63.250}{48.589}
\emlineto{63.419}{48.623}
\emmoveto{63.545}{48.631}
\emlineto{63.713}{48.664}
\emmoveto{63.840}{48.671}
\emlineto{64.008}{48.704}
\emmoveto{64.135}{48.712}
\emlineto{64.304}{48.744}
\emmoveto{64.431}{48.751}
\emlineto{64.600}{48.783}
\emmoveto{64.726}{48.789}
\emlineto{64.896}{48.821}
\emmoveto{65.023}{48.827}
\emlineto{65.192}{48.858}
\emmoveto{65.319}{48.864}
\emlineto{65.488}{48.895}
\emmoveto{65.616}{48.900}
\emlineto{65.785}{48.931}
\emmoveto{65.912}{48.936}
\emlineto{66.082}{48.967}
\emmoveto{66.209}{48.972}
\emlineto{66.379}{49.003}
\emmoveto{66.506}{49.008}
\emlineto{66.675}{49.039}
\emmoveto{66.802}{49.044}
\emlineto{66.972}{49.075}
\emmoveto{67.099}{49.081}
\emlineto{67.269}{49.111}
\emmoveto{67.396}{49.117}
\emlineto{67.565}{49.147}
\emmoveto{67.693}{49.153}
\emlineto{67.862}{49.183}
\emmoveto{67.989}{49.189}
\emlineto{68.159}{49.219}
\emmoveto{68.286}{49.225}
\emlineto{68.455}{49.255}
\emmoveto{68.583}{49.261}
\emlineto{68.752}{49.292}
\emmoveto{68.879}{49.297}
\emlineto{69.049}{49.328}
\emmoveto{69.176}{49.333}
\emlineto{69.346}{49.364}
\emmoveto{69.473}{49.369}
\emlineto{69.642}{49.400}
\emmoveto{69.769}{49.405}
\emlineto{69.939}{49.436}
\emmoveto{70.066}{49.441}
\emlineto{70.236}{49.472}
\emmoveto{70.363}{49.477}
\emlineto{70.532}{49.508}
\emmoveto{70.660}{49.513}
\emlineto{70.829}{49.544}
\emmoveto{70.956}{49.550}
\emlineto{71.126}{49.580}
\emmoveto{71.253}{49.586}
\emlineto{71.422}{49.616}
\emmoveto{71.550}{49.622}
\emlineto{71.719}{49.652}
\emmoveto{71.846}{49.658}
\emlineto{72.016}{49.688}
\emmoveto{72.143}{49.694}
\emlineto{72.313}{49.724}
\emmoveto{72.440}{49.730}
\emlineto{72.609}{49.760}
\emmoveto{72.736}{49.766}
\emlineto{72.906}{49.797}
\emmoveto{73.033}{49.802}
\emlineto{73.203}{49.833}
\emmoveto{73.330}{49.838}
\emlineto{73.499}{49.869}
\emmoveto{73.627}{49.874}
\emlineto{73.796}{49.905}
\emmoveto{73.923}{49.910}
\emlineto{74.093}{49.941}
\emmoveto{74.220}{49.946}
\emlineto{74.390}{49.977}
\emmoveto{74.517}{49.982}
\emlineto{74.686}{50.013}
\emmoveto{74.813}{50.018}
\emlineto{74.983}{50.049}
\emmoveto{75.110}{50.055}
\emlineto{75.280}{50.085}
\emmoveto{75.407}{50.091}
\emlineto{75.576}{50.121}
\emmoveto{75.703}{50.127}
\emlineto{75.873}{50.157}
\emmoveto{76.000}{50.163}
\emlineto{76.170}{50.193}
\emmoveto{76.297}{50.199}
\emlineto{76.466}{50.229}
\emmoveto{76.594}{50.235}
\emlineto{76.763}{50.266}
\emmoveto{76.890}{50.271}
\emlineto{77.060}{50.302}
\emmoveto{77.187}{50.307}
\emlineto{77.357}{50.338}
\emmoveto{77.484}{50.343}
\emlineto{77.653}{50.374}
\emmoveto{77.780}{50.379}
\emlineto{77.950}{50.410}
\emmoveto{78.077}{50.415}
\emlineto{78.247}{50.446}
\emmoveto{78.374}{50.451}
\emlineto{78.543}{50.482}
\emmoveto{78.671}{50.487}
\emlineto{78.840}{50.518}
\emmoveto{78.967}{50.524}
\emlineto{79.137}{50.554}
\emmoveto{79.264}{50.560}
\emlineto{79.433}{50.590}
\emmoveto{79.561}{50.596}
\emlineto{79.730}{50.626}
\emmoveto{79.857}{50.632}
\emlineto{80.027}{50.662}
\emmoveto{80.154}{50.668}
\emlineto{80.324}{50.698}
\emmoveto{80.451}{50.704}
\emlineto{80.620}{50.735}
\emmoveto{80.747}{50.740}
\emlineto{80.917}{50.771}
\emmoveto{81.044}{50.776}
\emlineto{81.214}{50.807}
\emmoveto{81.341}{50.812}
\emlineto{81.510}{50.843}
\emmoveto{81.638}{50.848}
\emlineto{81.807}{50.879}
\emmoveto{81.934}{50.884}
\emlineto{82.104}{50.915}
\emmoveto{82.231}{50.920}
\emlineto{82.400}{50.951}
\emmoveto{82.528}{50.956}
\emlineto{82.697}{50.987}
\emmoveto{82.824}{50.992}
\emlineto{82.994}{51.023}
\emmoveto{83.121}{51.029}
\emlineto{83.291}{51.059}
\emmoveto{83.418}{51.065}
\emlineto{83.587}{51.095}
\emmoveto{83.714}{51.101}
\emlineto{83.884}{51.131}
\emmoveto{84.011}{51.137}
\emlineto{84.181}{51.167}
\emmoveto{84.308}{51.173}
\emlineto{84.477}{51.203}
\emmoveto{84.605}{51.209}
\emlineto{84.774}{51.240}
\emmoveto{84.901}{51.245}
\emlineto{85.071}{51.276}
\emmoveto{85.198}{51.281}
\emlineto{85.368}{51.312}
\emmoveto{85.495}{51.317}
\emlineto{85.664}{51.348}
\emmoveto{85.791}{51.353}
\emlineto{85.961}{51.384}
\emmoveto{86.088}{51.389}
\emlineto{86.258}{51.420}
\emmoveto{86.385}{51.425}
\emlineto{86.554}{51.456}
\emmoveto{86.681}{51.461}
\emlineto{86.851}{51.492}
\emmoveto{86.978}{51.498}
\emlineto{87.148}{51.528}
\emmoveto{87.275}{51.534}
\emlineto{87.444}{51.564}
\emmoveto{87.572}{51.570}
\emlineto{87.741}{51.600}
\emmoveto{87.868}{51.606}
\emlineto{88.038}{51.636}
\emmoveto{88.165}{51.642}
\emlineto{88.335}{51.672}
\emmoveto{88.462}{51.678}
\emlineto{88.631}{51.709}
\emmoveto{88.758}{51.714}
\emlineto{88.928}{51.745}
\emmoveto{89.055}{51.750}
\emlineto{89.225}{51.781}
\emmoveto{89.352}{51.786}
\emlineto{89.521}{51.817}
\emmoveto{89.648}{51.822}
\emlineto{89.818}{51.853}
\emmoveto{89.945}{51.858}
\emlineto{90.115}{51.889}
\emmoveto{90.242}{51.894}
\emlineto{90.411}{51.925}
\emmoveto{90.539}{51.930}
\emlineto{90.708}{51.961}
\emmoveto{90.835}{51.967}
\emlineto{91.005}{51.997}
\emmoveto{91.132}{52.003}
\emlineto{91.302}{52.033}
\emmoveto{91.429}{52.039}
\emlineto{91.598}{52.069}
\emmoveto{91.725}{52.075}
\emlineto{91.895}{52.105}
\emmoveto{92.022}{52.111}
\emlineto{92.192}{52.141}
\emmoveto{92.319}{52.147}
\emlineto{92.488}{52.178}
\emmoveto{92.616}{52.183}
\emlineto{92.785}{52.214}
\emmoveto{92.912}{52.219}
\emlineto{93.082}{52.250}
\emmoveto{93.209}{52.255}
\emlineto{93.378}{52.286}
\emmoveto{93.506}{52.291}
\emlineto{93.675}{52.322}
\emmoveto{93.802}{52.327}
\emlineto{93.972}{52.358}
\emmoveto{94.099}{52.363}
\emlineto{94.269}{52.394}
\emmoveto{94.396}{52.399}
\emlineto{94.565}{52.430}
\emmoveto{94.692}{52.435}
\emlineto{94.862}{52.466}
\emmoveto{94.989}{52.472}
\emlineto{95.159}{52.502}
\emmoveto{95.286}{52.508}
\emlineto{95.455}{52.538}
\emmoveto{95.583}{52.544}
\emlineto{95.752}{52.574}
\emmoveto{95.879}{52.580}
\emlineto{96.049}{52.610}
\emmoveto{96.176}{52.616}
\emlineto{96.345}{52.646}
\emmoveto{96.473}{52.652}
\emlineto{96.642}{52.683}
\emmoveto{96.769}{52.688}
\emlineto{96.939}{52.719}
\emmoveto{97.066}{52.724}
\emlineto{97.236}{52.755}
\emmoveto{97.363}{52.760}
\emlineto{97.532}{52.791}
\emmoveto{97.659}{52.796}
\emlineto{97.829}{52.827}
\emmoveto{97.956}{52.832}
\emlineto{98.126}{52.863}
\emmoveto{98.253}{52.868}
\emlineto{98.422}{52.899}
\emmoveto{98.550}{52.904}
\emlineto{98.719}{52.935}
\emmoveto{98.846}{52.941}
\emlineto{99.016}{52.971}
\emmoveto{99.143}{52.977}
\emlineto{99.313}{53.007}
\emmoveto{99.440}{53.013}
\emlineto{99.609}{53.043}
\emmoveto{99.736}{53.049}
\emlineto{99.906}{53.079}
\emmoveto{100.033}{53.085}
\emlineto{100.203}{53.115}
\emmoveto{100.330}{53.121}
\emlineto{100.499}{53.152}
\emmoveto{100.626}{53.157}
\emlineto{100.796}{53.188}
\emmoveto{100.923}{53.193}
\emlineto{101.093}{53.224}
\emmoveto{101.220}{53.229}
\emlineto{101.389}{53.260}
\emmoveto{101.517}{53.265}
\emlineto{101.686}{53.296}
\emmoveto{101.813}{53.301}
\emlineto{101.983}{53.332}
\emmoveto{102.110}{53.337}
\emlineto{102.280}{53.368}
\emmoveto{102.407}{53.373}
\emlineto{102.576}{53.404}
\emmoveto{102.703}{53.410}
\emlineto{102.873}{53.440}
\emmoveto{103.000}{53.446}
\emlineto{103.170}{53.476}
\emmoveto{103.297}{53.482}
\emlineto{103.466}{53.512}
\emmoveto{103.593}{53.518}
\emlineto{103.763}{53.548}
\emmoveto{103.890}{53.554}
\emlineto{104.060}{53.584}
\emmoveto{104.187}{53.590}
\emlineto{104.356}{53.620}
\emmoveto{104.484}{53.626}
\emlineto{104.653}{53.657}
\emmoveto{104.780}{53.662}
\emlineto{104.950}{53.693}
\emmoveto{105.077}{53.698}
\emlineto{105.247}{53.729}
\emmoveto{105.374}{53.734}
\emlineto{105.543}{53.765}
\emmoveto{105.670}{53.770}
\emlineto{105.840}{53.801}
\emmoveto{105.967}{53.806}
\emlineto{106.137}{53.837}
\emmoveto{106.264}{53.842}
\emlineto{106.433}{53.873}
\emmoveto{106.561}{53.878}
\emlineto{106.730}{53.909}
\emmoveto{106.857}{53.915}
\emlineto{107.027}{53.945}
\emmoveto{107.154}{53.951}
\emlineto{107.323}{53.981}
\emmoveto{107.451}{53.987}
\emlineto{107.620}{54.017}
\emmoveto{107.747}{54.023}
\emlineto{107.917}{54.053}
\emmoveto{108.044}{54.059}
\emlineto{108.214}{54.089}
\emmoveto{108.341}{54.095}
\emlineto{108.510}{54.126}
\emmoveto{108.637}{54.131}
\emlineto{108.807}{54.162}
\emmoveto{108.934}{54.167}
\emlineto{109.104}{54.198}
\emmoveto{109.231}{54.203}
\emlineto{109.400}{54.234}
\emmoveto{109.528}{54.239}
\emlineto{109.697}{54.270}
\emmoveto{109.824}{54.275}
\emlineto{109.994}{54.306}
\emmoveto{110.121}{54.311}
\emlineto{110.290}{54.342}
\emmoveto{110.418}{54.347}
\emlineto{110.587}{54.378}
\emmoveto{110.714}{54.384}
\emlineto{110.884}{54.414}
\emmoveto{111.011}{54.420}
\emlineto{111.181}{54.450}
\emmoveto{111.308}{54.456}
\emlineto{111.477}{54.486}
\emmoveto{111.604}{54.492}
\emlineto{111.774}{54.522}
\emmoveto{111.901}{54.528}
\emlineto{112.071}{54.558}
\emmoveto{112.198}{54.564}
\emlineto{112.367}{54.595}
\emmoveto{112.495}{54.600}
\emlineto{112.664}{54.631}
\emmoveto{112.791}{54.636}
\emlineto{112.961}{54.667}
\emmoveto{113.088}{54.672}
\emlineto{113.258}{54.703}
\emmoveto{113.385}{54.708}
\emlineto{113.554}{54.739}
\emmoveto{113.681}{54.744}
\emlineto{113.851}{54.775}
\emmoveto{113.978}{54.780}
\emlineto{114.148}{54.811}
\emmoveto{114.275}{54.816}
\emlineto{114.444}{54.847}
\emmoveto{114.571}{54.853}
\emlineto{114.741}{54.883}
\emmoveto{114.868}{54.889}
\emlineto{115.038}{54.919}
\emmoveto{115.165}{54.925}
\emlineto{115.334}{54.955}
\emmoveto{115.462}{54.961}
\emlineto{115.631}{54.991}
\emmoveto{115.758}{54.997}
\emlineto{115.928}{55.027}
\emmoveto{116.055}{55.033}
\emlineto{116.225}{55.063}
\emmoveto{116.352}{55.069}
\emlineto{116.521}{55.100}
\emmoveto{116.648}{55.105}
\emlineto{116.818}{55.136}
\emmoveto{116.945}{55.141}
\emlineto{117.115}{55.172}
\emmoveto{117.242}{55.177}
\emlineto{117.411}{55.208}
\emmoveto{117.539}{55.213}
\emlineto{117.708}{55.244}
\emmoveto{117.835}{55.249}
\emlineto{118.005}{55.280}
\emmoveto{118.132}{55.285}
\emlineto{118.301}{55.316}
\emmoveto{118.429}{55.321}
\emlineto{118.598}{55.352}
\emmoveto{118.725}{55.358}
\emlineto{118.895}{55.388}
\emmoveto{119.022}{55.394}
\emlineto{119.192}{55.424}
\emmoveto{119.319}{55.430}
\emlineto{119.488}{55.460}
\emmoveto{119.615}{55.466}
\emlineto{119.785}{55.496}
\emmoveto{119.912}{55.502}
\emlineto{120.082}{55.532}
\emmoveto{120.209}{55.538}
\emlineto{120.378}{55.569}
\emmoveto{120.506}{55.574}
\emlineto{120.675}{55.605}
\emmoveto{120.802}{55.610}
\emlineto{120.972}{55.641}
\emmoveto{121.099}{55.646}
\emlineto{121.268}{55.677}
\emmoveto{121.396}{55.682}
\emlineto{121.565}{55.713}
\emmoveto{121.692}{55.718}
\emlineto{121.862}{55.749}
\emmoveto{121.989}{55.754}
\emlineto{122.159}{55.785}
\emmoveto{122.286}{55.790}
\emlineto{122.455}{55.821}
\emmoveto{122.582}{55.827}
\emlineto{122.752}{55.857}
\emmoveto{122.879}{55.863}
\emlineto{123.049}{55.893}
\emmoveto{123.176}{55.899}
\emlineto{123.345}{55.929}
\emmoveto{123.473}{55.935}
\emlineto{123.642}{55.965}
\emmoveto{123.769}{55.971}
\emlineto{123.939}{56.001}
\emmoveto{124.066}{56.007}
\emlineto{124.235}{56.038}
\emmoveto{124.363}{56.043}
\emlineto{124.532}{56.074}
\emmoveto{124.659}{56.079}
\emlineto{124.829}{56.110}
\emmoveto{124.956}{56.115}
\emlineto{125.126}{56.146}
\emmoveto{125.253}{56.151}
\emlineto{125.422}{56.182}
\emmoveto{125.549}{56.187}
\emlineto{125.719}{56.218}
\emmoveto{125.846}{56.223}
\emlineto{126.016}{56.254}
\emmoveto{126.143}{56.259}
\emlineto{126.312}{56.290}
\emmoveto{126.440}{56.295}
\emlineto{126.609}{56.326}
\emmoveto{126.736}{56.332}
\emlineto{126.906}{56.362}
\emmoveto{127.033}{56.368}
\emlineto{127.203}{56.398}
\emmoveto{127.330}{56.404}
\emlineto{127.499}{56.434}
\emmoveto{127.626}{56.440}
\emlineto{127.796}{56.470}
\emmoveto{127.923}{56.476}
\emlineto{128.093}{56.506}
\emmoveto{128.220}{56.512}
\emlineto{128.389}{56.543}
\emmoveto{128.516}{56.548}
\emlineto{128.686}{56.579}
\emmoveto{128.813}{56.584}
\emlineto{128.983}{56.615}
\emmoveto{129.110}{56.620}
\emlineto{129.279}{56.651}
\emmoveto{129.407}{56.656}
\emlineto{129.576}{56.687}
\emmoveto{129.703}{56.692}
\emlineto{129.873}{56.723}
\emshow{60.380}{52.700}{Sommerfeld: v=0.44}
\emmoveto{12.000}{10.016}
\emlineto{12.002}{10.180}
\emmoveto{12.005}{10.298}
\emlineto{12.011}{10.462}
\emmoveto{12.017}{10.580}
\emlineto{12.027}{10.744}
\emmoveto{12.038}{10.862}
\emlineto{12.052}{11.026}
\emmoveto{12.066}{11.144}
\emlineto{12.084}{11.308}
\emmoveto{12.102}{11.426}
\emlineto{12.125}{11.589}
\emmoveto{12.146}{11.707}
\emlineto{12.173}{11.870}
\emmoveto{12.197}{11.988}
\emlineto{12.229}{12.150}
\emmoveto{12.257}{12.268}
\emlineto{12.292}{12.430}
\emmoveto{12.324}{12.547}
\emlineto{12.364}{12.709}
\emmoveto{12.399}{12.826}
\emlineto{12.443}{12.988}
\emmoveto{12.482}{13.104}
\emlineto{12.530}{13.266}
\emmoveto{12.572}{13.381}
\emlineto{12.625}{13.542}
\emmoveto{12.670}{13.658}
\emlineto{12.727}{13.818}
\emmoveto{12.776}{13.933}
\emlineto{12.837}{14.093}
\emmoveto{12.890}{14.207}
\emlineto{12.955}{14.366}
\emmoveto{13.011}{14.480}
\emlineto{13.080}{14.639}
\emmoveto{13.139}{14.752}
\emlineto{13.200}{14.885}
\emmoveto{13.250}{14.974}
\emlineto{13.314}{15.106}
\emmoveto{13.366}{15.194}
\emlineto{13.433}{15.326}
\emmoveto{13.487}{15.414}
\emlineto{13.556}{15.545}
\emmoveto{13.613}{15.632}
\emlineto{13.685}{15.763}
\emmoveto{13.744}{15.850}
\emlineto{13.819}{15.980}
\emmoveto{13.880}{16.067}
\emlineto{13.957}{16.196}
\emmoveto{14.020}{16.282}
\emlineto{14.101}{16.411}
\emmoveto{14.166}{16.496}
\emlineto{14.249}{16.625}
\emmoveto{14.316}{16.709}
\emlineto{14.402}{16.837}
\emmoveto{14.472}{16.921}
\emlineto{14.560}{17.048}
\emmoveto{14.632}{17.132}
\emlineto{14.723}{17.258}
\emmoveto{14.797}{17.341}
\emlineto{14.890}{17.467}
\emmoveto{14.966}{17.549}
\emlineto{15.063}{17.674}
\emmoveto{15.141}{17.756}
\emlineto{15.240}{17.880}
\emmoveto{15.320}{17.961}
\emlineto{15.421}{18.084}
\emmoveto{15.504}{18.165}
\emlineto{15.608}{18.287}
\emmoveto{15.692}{18.367}
\emlineto{15.798}{18.488}
\emmoveto{15.885}{18.567}
\emlineto{15.994}{18.688}
\emmoveto{16.082}{18.767}
\emlineto{16.194}{18.886}
\emmoveto{16.284}{18.964}
\emlineto{16.398}{19.083}
\emmoveto{16.491}{19.160}
\emlineto{16.607}{19.278}
\emmoveto{16.701}{19.354}
\emlineto{16.821}{19.471}
\emmoveto{16.917}{19.547}
\emlineto{17.038}{19.663}
\emmoveto{17.136}{19.737}
\emlineto{17.260}{19.853}
\emmoveto{17.360}{19.926}
\emlineto{17.487}{20.041}
\emmoveto{17.589}{20.114}
\emlineto{17.717}{20.227}
\emmoveto{17.821}{20.299}
\emlineto{17.952}{20.411}
\emmoveto{18.058}{20.482}
\emlineto{18.191}{20.594}
\emmoveto{18.299}{20.664}
\emlineto{18.434}{20.774}
\emmoveto{18.543}{20.844}
\emlineto{18.681}{20.953}
\emmoveto{18.792}{21.021}
\emlineto{18.933}{21.129}
\emmoveto{19.045}{21.197}
\emlineto{19.159}{21.284}
\emmoveto{19.245}{21.332}
\emlineto{19.360}{21.419}
\emmoveto{19.447}{21.466}
\emlineto{19.563}{21.552}
\emmoveto{19.651}{21.599}
\emlineto{19.769}{21.684}
\emmoveto{19.858}{21.730}
\emlineto{19.977}{21.815}
\emmoveto{20.067}{21.860}
\emlineto{20.187}{21.944}
\emmoveto{20.278}{21.989}
\emlineto{20.400}{22.072}
\emmoveto{20.491}{22.116}
\emlineto{20.614}{22.199}
\emmoveto{20.707}{22.243}
\emlineto{20.831}{22.324}
\emmoveto{20.925}{22.367}
\emlineto{21.051}{22.448}
\emmoveto{21.145}{22.491}
\emlineto{21.272}{22.571}
\emmoveto{21.368}{22.613}
\emlineto{21.496}{22.692}
\emmoveto{21.592}{22.733}
\emlineto{21.721}{22.812}
\emmoveto{21.819}{22.853}
\emlineto{21.949}{22.930}
\emmoveto{22.047}{22.970}
\emlineto{22.179}{23.047}
\emmoveto{22.278}{23.087}
\emlineto{22.411}{23.163}
\emmoveto{22.511}{23.202}
\emlineto{22.645}{23.277}
\emmoveto{22.746}{23.315}
\emlineto{22.881}{23.389}
\emmoveto{22.983}{23.427}
\emlineto{23.119}{23.501}
\emmoveto{23.221}{23.538}
\emlineto{23.359}{23.610}
\emmoveto{23.462}{23.647}
\emlineto{23.601}{23.718}
\emmoveto{23.705}{23.754}
\emlineto{23.844}{23.825}
\emmoveto{23.949}{23.860}
\emlineto{24.090}{23.930}
\emmoveto{24.196}{23.965}
\emlineto{24.337}{24.034}
\emmoveto{24.444}{24.068}
\emlineto{24.587}{24.136}
\emmoveto{24.694}{24.169}
\emlineto{24.838}{24.236}
\emmoveto{24.946}{24.269}
\emlineto{25.091}{24.335}
\emmoveto{25.199}{24.367}
\emlineto{25.345}{24.433}
\emmoveto{25.455}{24.464}
\emlineto{25.601}{24.528}
\emmoveto{25.712}{24.559}
\emlineto{25.859}{24.623}
\emmoveto{25.970}{24.653}
\emlineto{26.119}{24.715}
\emmoveto{26.231}{24.744}
\emlineto{26.380}{24.806}
\emmoveto{26.493}{24.835}
\emlineto{26.643}{24.895}
\emmoveto{26.756}{24.923}
\emlineto{26.907}{24.983}
\emmoveto{27.021}{25.010}
\emlineto{27.173}{25.069}
\emmoveto{27.288}{25.095}
\emlineto{27.441}{25.153}
\emmoveto{27.556}{25.179}
\emlineto{27.710}{25.236}
\emmoveto{27.825}{25.261}
\emlineto{27.980}{25.317}
\emmoveto{28.096}{25.341}
\emlineto{28.252}{25.396}
\emmoveto{28.369}{25.420}
\emlineto{28.525}{25.474}
\emmoveto{28.643}{25.497}
\emlineto{28.800}{25.550}
\emmoveto{28.918}{25.572}
\emlineto{29.075}{25.624}
\emmoveto{29.194}{25.645}
\emlineto{29.353}{25.696}
\emmoveto{29.472}{25.717}
\emlineto{29.631}{25.767}
\emmoveto{29.751}{25.787}
\emlineto{29.911}{25.836}
\emmoveto{30.031}{25.855}
\emlineto{30.192}{25.903}
\emmoveto{30.312}{25.922}
\emlineto{30.474}{25.969}
\emmoveto{30.595}{25.986}
\emlineto{30.757}{26.033}
\emmoveto{30.878}{26.049}
\emlineto{31.041}{26.095}
\emmoveto{31.163}{26.111}
\emlineto{31.326}{26.155}
\emmoveto{31.449}{26.170}
\emlineto{31.613}{26.213}
\emmoveto{31.736}{26.228}
\emlineto{31.900}{26.270}
\emmoveto{32.024}{26.284}
\emlineto{32.189}{26.325}
\emmoveto{32.313}{26.338}
\emlineto{32.478}{26.378}
\emmoveto{32.602}{26.390}
\emlineto{32.768}{26.429}
\emmoveto{32.893}{26.440}
\emlineto{33.060}{26.478}
\emmoveto{33.185}{26.489}
\emlineto{33.352}{26.526}
\emmoveto{33.477}{26.536}
\emlineto{33.645}{26.572}
\emmoveto{33.770}{26.581}
\emlineto{33.938}{26.616}
\emmoveto{34.064}{26.624}
\emlineto{34.233}{26.658}
\emmoveto{34.359}{26.665}
\emlineto{34.528}{26.698}
\emmoveto{34.655}{26.705}
\emlineto{34.824}{26.736}
\emmoveto{34.951}{26.742}
\emlineto{35.120}{26.773}
\emmoveto{35.248}{26.778}
\emlineto{35.417}{26.808}
\emmoveto{35.545}{26.812}
\emlineto{35.715}{26.841}
\emmoveto{35.843}{26.844}
\emlineto{36.013}{26.872}
\emmoveto{36.141}{26.874}
\emlineto{36.312}{26.901}
\emmoveto{36.441}{26.903}
\emlineto{36.612}{26.928}
\emmoveto{36.740}{26.929}
\emlineto{36.912}{26.954}
\emmoveto{37.040}{26.954}
\emlineto{37.212}{26.977}
\emmoveto{37.341}{26.977}
\emlineto{37.512}{26.999}
\emmoveto{37.641}{26.998}
\emlineto{37.813}{27.019}
\emmoveto{37.943}{27.017}
\emlineto{38.115}{27.037}
\emmoveto{38.244}{27.034}
\emlineto{38.416}{27.053}
\emmoveto{38.546}{27.049}
\emlineto{38.718}{27.067}
\emmoveto{38.848}{27.062}
\emlineto{39.021}{27.079}
\emmoveto{39.150}{27.074}
\emlineto{39.323}{27.090}
\emmoveto{39.453}{27.083}
\emlineto{39.625}{27.098}
\emmoveto{39.755}{27.091}
\emlineto{39.928}{27.105}
\emmoveto{40.058}{27.097}
\emlineto{40.231}{27.110}
\emmoveto{40.361}{27.101}
\emlineto{40.534}{27.112}
\emmoveto{40.663}{27.103}
\emlineto{40.837}{27.113}
\emmoveto{40.966}{27.103}
\emlineto{41.139}{27.113}
\emmoveto{41.269}{27.102}
\emlineto{41.442}{27.110}
\emmoveto{41.572}{27.098}
\emlineto{41.745}{27.105}
\emmoveto{41.875}{27.093}
\emlineto{42.048}{27.099}
\emmoveto{42.177}{27.085}
\emlineto{42.350}{27.090}
\emmoveto{42.480}{27.076}
\emlineto{42.653}{27.080}
\emmoveto{42.782}{27.065}
\emlineto{42.955}{27.068}
\emmoveto{43.084}{27.052}
\emlineto{43.257}{27.054}
\emmoveto{43.386}{27.037}
\emlineto{43.558}{27.038}
\emmoveto{43.688}{27.020}
\emlineto{43.860}{27.020}
\emmoveto{43.989}{27.002}
\emlineto{44.161}{27.000}
\emmoveto{44.290}{26.981}
\emlineto{44.462}{26.978}
\emmoveto{44.590}{26.959}
\emlineto{44.762}{26.955}
\emmoveto{44.890}{26.934}
\emlineto{45.062}{26.930}
\emmoveto{45.190}{26.908}
\emlineto{45.361}{26.902}
\emmoveto{45.489}{26.880}
\emlineto{45.660}{26.873}
\emmoveto{45.788}{26.850}
\emlineto{45.958}{26.842}
\emmoveto{46.086}{26.819}
\emlineto{46.256}{26.810}
\emmoveto{46.383}{26.785}
\emlineto{46.553}{26.775}
\emmoveto{46.680}{26.750}
\emlineto{46.850}{26.739}
\emmoveto{46.977}{26.712}
\emlineto{47.146}{26.700}
\emmoveto{47.272}{26.673}
\emlineto{47.441}{26.660}
\emmoveto{47.567}{26.632}
\emlineto{47.735}{26.618}
\emmoveto{47.861}{26.589}
\emlineto{48.029}{26.574}
\emmoveto{48.155}{26.545}
\emlineto{48.322}{26.529}
\emmoveto{48.447}{26.498}
\emlineto{48.614}{26.481}
\emmoveto{48.739}{26.450}
\emlineto{48.905}{26.432}
\emmoveto{49.030}{26.400}
\emlineto{49.196}{26.381}
\emmoveto{49.320}{26.348}
\emlineto{49.485}{26.328}
\emmoveto{49.609}{26.294}
\emlineto{49.774}{26.273}
\emmoveto{49.897}{26.239}
\emlineto{50.061}{26.216}
\emmoveto{50.184}{26.182}
\emlineto{50.348}{26.158}
\emmoveto{50.470}{26.123}
\emlineto{50.633}{26.098}
\emmoveto{50.755}{26.062}
\emlineto{50.917}{26.036}
\emmoveto{51.039}{25.999}
\emlineto{51.201}{25.973}
\emmoveto{51.322}{25.935}
\emlineto{51.483}{25.907}
\emmoveto{51.603}{25.869}
\emlineto{51.764}{25.840}
\emmoveto{51.884}{25.801}
\emlineto{52.043}{25.771}
\emmoveto{52.163}{25.731}
\emlineto{52.322}{25.701}
\emmoveto{52.441}{25.660}
\emlineto{52.599}{25.628}
\emmoveto{52.718}{25.587}
\emlineto{52.875}{25.554}
\emmoveto{52.993}{25.512}
\emlineto{53.150}{25.478}
\emmoveto{53.267}{25.435}
\emlineto{53.423}{25.401}
\emmoveto{53.540}{25.357}
\emlineto{53.695}{25.322}
\emmoveto{53.811}{25.277}
\emlineto{53.965}{25.241}
\emmoveto{54.081}{25.195}
\emlineto{54.234}{25.158}
\emmoveto{54.349}{25.112}
\emlineto{54.502}{25.074}
\emmoveto{54.616}{25.027}
\emlineto{54.768}{24.988}
\emmoveto{54.881}{24.941}
\emlineto{55.032}{24.900}
\emmoveto{55.145}{24.852}
\emlineto{55.295}{24.811}
\emmoveto{55.407}{24.762}
\emlineto{55.557}{24.720}
\emmoveto{55.668}{24.671}
\emlineto{55.816}{24.628}
\emmoveto{55.927}{24.578}
\emlineto{56.074}{24.534}
\emmoveto{56.184}{24.483}
\emlineto{56.331}{24.438}
\emmoveto{56.440}{24.387}
\emlineto{56.585}{24.341}
\emmoveto{56.694}{24.289}
\emlineto{56.838}{24.242}
\emmoveto{56.946}{24.189}
\emlineto{57.089}{24.142}
\emmoveto{57.197}{24.088}
\emlineto{57.339}{24.040}
\emmoveto{57.445}{23.986}
\emlineto{57.586}{23.936}
\emmoveto{57.692}{23.881}
\emlineto{57.832}{23.831}
\emmoveto{57.937}{23.776}
\emlineto{58.076}{23.725}
\emmoveto{58.180}{23.668}
\emlineto{58.318}{23.616}
\emmoveto{58.421}{23.560}
\emlineto{58.558}{23.507}
\emmoveto{58.660}{23.449}
\emlineto{58.796}{23.396}
\emmoveto{58.897}{23.338}
\emlineto{59.032}{23.283}
\emmoveto{59.133}{23.224}
\emlineto{59.266}{23.169}
\emmoveto{59.366}{23.110}
\emlineto{59.498}{23.054}
\emmoveto{59.597}{22.994}
\emlineto{59.728}{22.937}
\emmoveto{59.826}{22.876}
\emlineto{59.956}{22.818}
\emmoveto{60.053}{22.757}
\emlineto{60.182}{22.699}
\emmoveto{60.278}{22.637}
\emlineto{60.406}{22.577}
\emmoveto{60.501}{22.515}
\emlineto{60.627}{22.455}
\emmoveto{60.721}{22.392}
\emlineto{60.846}{22.331}
\emmoveto{60.940}{22.268}
\emlineto{61.064}{22.206}
\emmoveto{61.156}{22.142}
\emlineto{61.278}{22.079}
\emmoveto{61.370}{22.014}
\emlineto{61.491}{21.951}
\emmoveto{61.582}{21.886}
\emlineto{61.701}{21.822}
\emmoveto{61.791}{21.756}
\emlineto{61.910}{21.691}
\emmoveto{61.998}{21.625}
\emlineto{62.115}{21.559}
\emmoveto{62.203}{21.493}
\emlineto{62.319}{21.426}
\emmoveto{62.405}{21.359}
\emlineto{62.520}{21.292}
\emmoveto{62.605}{21.224}
\emlineto{62.718}{21.156}
\emmoveto{62.831}{21.068}
\emlineto{62.970}{20.980}
\emmoveto{63.081}{20.891}
\emlineto{63.218}{20.802}
\emmoveto{63.327}{20.712}
\emlineto{63.462}{20.622}
\emmoveto{63.569}{20.531}
\emlineto{63.701}{20.440}
\emmoveto{63.807}{20.348}
\emlineto{63.937}{20.256}
\emmoveto{64.040}{20.163}
\emlineto{64.168}{20.070}
\emmoveto{64.270}{19.976}
\emlineto{64.395}{19.882}
\emmoveto{64.495}{19.788}
\emlineto{64.618}{19.692}
\emmoveto{64.715}{19.598}
\emlineto{64.836}{19.501}
\emmoveto{64.932}{19.406}
\emlineto{65.050}{19.308}
\emmoveto{65.144}{19.212}
\emlineto{65.260}{19.113}
\emmoveto{65.352}{19.016}
\emlineto{65.465}{18.917}
\emmoveto{65.555}{18.819}
\emlineto{65.666}{18.719}
\emmoveto{65.753}{18.621}
\emlineto{65.862}{18.519}
\emmoveto{65.947}{18.420}
\emlineto{66.053}{18.318}
\emmoveto{66.137}{18.218}
\emlineto{66.240}{18.116}
\emmoveto{66.322}{18.015}
\emlineto{66.423}{17.912}
\emmoveto{66.502}{17.810}
\emlineto{66.600}{17.706}
\emmoveto{66.678}{17.604}
\emlineto{66.773}{17.499}
\emmoveto{66.849}{17.397}
\emlineto{66.942}{17.291}
\emmoveto{67.015}{17.188}
\emlineto{67.105}{17.081}
\emmoveto{67.176}{16.977}
\emlineto{67.264}{16.870}
\emmoveto{67.333}{16.766}
\emlineto{67.418}{16.658}
\emmoveto{67.485}{16.553}
\emlineto{67.567}{16.444}
\emmoveto{67.632}{16.339}
\emlineto{67.711}{16.230}
\emmoveto{67.774}{16.124}
\emlineto{67.850}{16.014}
\emmoveto{67.911}{15.908}
\emlineto{67.985}{15.797}
\emmoveto{68.043}{15.690}
\emlineto{68.114}{15.579}
\emmoveto{68.170}{15.472}
\emlineto{68.239}{15.360}
\emmoveto{68.292}{15.253}
\emlineto{68.358}{15.140}
\emmoveto{68.410}{15.032}
\emlineto{68.473}{14.920}
\emmoveto{68.522}{14.811}
\emlineto{68.594}{14.673}
\emmoveto{68.652}{14.540}
\emlineto{68.720}{14.401}
\emmoveto{68.775}{14.267}
\emlineto{68.839}{14.127}
\emmoveto{68.890}{13.993}
\emlineto{68.950}{13.853}
\emmoveto{68.998}{13.718}
\emlineto{69.053}{13.577}
\emmoveto{69.097}{13.442}
\emlineto{69.149}{13.301}
\emmoveto{69.190}{13.165}
\emlineto{69.237}{13.023}
\emmoveto{69.274}{12.887}
\emlineto{69.317}{12.745}
\emmoveto{69.351}{12.608}
\emlineto{69.389}{12.466}
\emmoveto{69.420}{12.329}
\emlineto{69.454}{12.186}
\emmoveto{69.481}{12.049}
\emlineto{69.511}{11.906}
\emmoveto{69.534}{11.768}
\emlineto{69.560}{11.625}
\emmoveto{69.580}{11.487}
\emlineto{69.601}{11.343}
\emmoveto{69.617}{11.205}
\emlineto{69.635}{11.062}
\emmoveto{69.647}{10.924}
\emlineto{69.660}{10.780}
\emmoveto{69.669}{10.642}
\emlineto{69.678}{10.498}
\emmoveto{69.684}{10.359}
\emlineto{69.688}{10.215}
\emmoveto{69.690}{10.077}
\emlineto{69.690}{10.010}
\emshow{60.380}{24.700}{Classic: v=0.44}
\emshow{1.000}{10.000}{2.61e0}
\emshow{1.000}{17.000}{1.80e3}
\emshow{1.000}{24.000}{3.60e3}
\emshow{1.000}{31.000}{5.40e3}
\emshow{1.000}{38.000}{7.20e3}
\emshow{1.000}{45.000}{9.00e3}
\emshow{1.000}{52.000}{1.08e4}
\emshow{1.000}{59.000}{1.26e4}
\emshow{1.000}{66.000}{1.44e4}
\emshow{1.000}{73.000}{1.62e4}
\emshow{1.000}{80.000}{1.80e4}
\emshow{12.000}{5.000}{0.00e0}
\emshow{23.800}{5.000}{1.80e3}
\emshow{35.600}{5.000}{3.60e3}
\emshow{47.400}{5.000}{5.40e3}
\emshow{59.200}{5.000}{7.20e3}
\emshow{71.000}{5.000}{9.00e3}
\emshow{82.800}{5.000}{1.08e4}
\emshow{94.600}{5.000}{1.26e4}
\emshow{106.400}{5.000}{1.44e4}
\emshow{118.200}{5.000}{1.62e4}
\emshow{130.000}{5.000}{1.80e4}
{\centerline {\bf Fig. A}}
\eject

\newcount\numpoint
\newcount\numpointo
\numpoint=1 \numpointo=1
\def\emmoveto#1#2{\offinterlineskip
\hbox to 0 true cm{\vbox to 0
true cm{\vskip - #2 true mm
\hskip #1 true mm \special{em:point
\the\numpoint}\vss}\hss}
\numpointo=\numpoint
\global\advance \numpoint by 1}
\def\emlineto#1#2{\offinterlineskip
\hbox to 0 true cm{\vbox to 0
true cm{\vskip - #2 true mm
\hskip #1 true mm \special{em:point
\the\numpoint}\vss}\hss}
\special{em:line
\the\numpointo,\the\numpoint}
\numpointo=\numpoint
\global\advance \numpoint by 1}
\def\emshow#1#2#3{\offinterlineskip
\hbox to 0 true cm{\vbox to 0
true cm{\vskip - #2 true mm
\hskip #1 true mm \vbox to 0
true cm{\vss\hbox{#3\hss
}}\vss}\hss}}
\special{em:linewidth 0.8pt}

\vrule width 0 mm height                0 mm depth 90.000 true mm

\special{em:linewidth 0.8pt}
\emmoveto{130.000}{10.000}
\emlineto{12.000}{10.000}
\emlineto{12.000}{80.000}
\emmoveto{71.000}{10.000}
\emlineto{71.000}{80.000}
\emmoveto{12.000}{45.000}
\emlineto{130.000}{45.000}
\emmoveto{130.000}{10.000}
\emlineto{130.000}{80.000}
\emlineto{12.000}{80.000}
\emlineto{12.000}{10.000}
\emlineto{130.000}{10.000}
\special{em:linewidth 0.4pt}
\emmoveto{12.000}{17.000}
\emlineto{130.000}{17.000}
\emmoveto{12.000}{24.000}
\emlineto{130.000}{24.000}
\emmoveto{12.000}{31.000}
\emlineto{130.000}{31.000}
\emmoveto{12.000}{38.000}
\emlineto{130.000}{38.000}
\emmoveto{12.000}{45.000}
\emlineto{130.000}{45.000}
\emmoveto{12.000}{52.000}
\emlineto{130.000}{52.000}
\emmoveto{12.000}{59.000}
\emlineto{130.000}{59.000}
\emmoveto{12.000}{66.000}
\emlineto{130.000}{66.000}
\emmoveto{12.000}{73.000}
\emlineto{130.000}{73.000}
\emmoveto{23.800}{10.000}
\emlineto{23.800}{80.000}
\emmoveto{35.600}{10.000}
\emlineto{35.600}{80.000}
\emmoveto{47.400}{10.000}
\emlineto{47.400}{80.000}
\emmoveto{59.200}{10.000}
\emlineto{59.200}{80.000}
\emmoveto{71.000}{10.000}
\emlineto{71.000}{80.000}
\emmoveto{82.800}{10.000}
\emlineto{82.800}{80.000}
\emmoveto{94.600}{10.000}
\emlineto{94.600}{80.000}
\emmoveto{106.400}{10.000}
\emlineto{106.400}{80.000}
\emmoveto{118.200}{10.000}
\emlineto{118.200}{80.000}
\special{em:linewidth 0.8pt}
\emmoveto{71.000}{45.026}
\emlineto{71.000}{45.246}
\emmoveto{71.000}{45.236}
\emlineto{71.000}{45.430}
\emmoveto{71.000}{45.420}
\emlineto{71.000}{45.664}
\emmoveto{71.000}{45.654}
\emlineto{71.010}{45.911}
\emmoveto{71.010}{45.901}
\emlineto{71.033}{46.159}
\emmoveto{71.033}{46.149}
\emlineto{71.068}{46.405}
\emmoveto{71.068}{46.395}
\emlineto{71.115}{46.651}
\emmoveto{71.115}{46.641}
\emlineto{71.174}{46.896}
\emmoveto{71.174}{46.886}
\emlineto{71.247}{47.140}
\emmoveto{71.247}{47.130}
\emlineto{71.331}{47.382}
\emmoveto{71.331}{47.372}
\emlineto{71.427}{47.623}
\emmoveto{71.427}{47.613}
\emlineto{71.536}{47.862}
\emmoveto{71.536}{47.852}
\emlineto{71.657}{48.099}
\emmoveto{71.657}{48.089}
\emlineto{71.790}{48.333}
\emmoveto{71.790}{48.323}
\emlineto{71.934}{48.566}
\emmoveto{71.934}{48.556}
\emlineto{72.090}{48.795}
\emmoveto{72.090}{48.785}
\emlineto{72.258}{49.022}
\emmoveto{72.258}{49.012}
\emlineto{72.437}{49.245}
\emmoveto{72.437}{49.235}
\emlineto{72.627}{49.465}
\emmoveto{72.627}{49.455}
\emlineto{72.828}{49.682}
\emmoveto{72.828}{49.672}
\emlineto{73.041}{49.895}
\emmoveto{73.041}{49.885}
\emlineto{73.263}{50.104}
\emmoveto{73.263}{50.094}
\emlineto{73.497}{50.310}
\emmoveto{73.497}{50.300}
\emlineto{73.740}{50.511}
\emmoveto{73.740}{50.501}
\emlineto{73.994}{50.707}
\emmoveto{73.994}{50.697}
\emlineto{74.257}{50.899}
\emmoveto{74.257}{50.889}
\emlineto{74.531}{51.086}
\emmoveto{74.531}{51.076}
\emlineto{74.813}{51.268}
\emmoveto{74.813}{51.258}
\emlineto{75.104}{51.445}
\emmoveto{75.104}{51.435}
\emlineto{75.405}{51.617}
\emmoveto{75.405}{51.607}
\emlineto{75.714}{51.783}
\emmoveto{75.714}{51.773}
\emlineto{76.031}{51.944}
\emmoveto{76.031}{51.934}
\emlineto{76.356}{52.099}
\emmoveto{76.356}{52.089}
\emlineto{76.689}{52.248}
\emmoveto{76.689}{52.238}
\emlineto{77.029}{52.392}
\emmoveto{77.029}{52.382}
\emlineto{77.376}{52.529}
\emmoveto{77.376}{52.519}
\emlineto{77.730}{52.660}
\emmoveto{77.730}{52.650}
\emlineto{78.091}{52.784}
\emmoveto{78.091}{52.774}
\emlineto{78.457}{52.902}
\emmoveto{78.457}{52.892}
\emlineto{78.830}{53.014}
\emmoveto{78.830}{53.004}
\emlineto{79.208}{53.118}
\emmoveto{79.208}{53.108}
\emlineto{79.591}{53.216}
\emmoveto{79.591}{53.206}
\emlineto{79.979}{53.308}
\emmoveto{79.979}{53.298}
\emlineto{80.371}{53.392}
\emmoveto{80.371}{53.382}
\emlineto{80.767}{53.469}
\emmoveto{80.767}{53.459}
\emlineto{81.167}{53.539}
\emmoveto{81.167}{53.529}
\emlineto{81.571}{53.602}
\emmoveto{81.571}{53.592}
\emlineto{81.977}{53.658}
\emmoveto{81.977}{53.648}
\emlineto{82.386}{53.706}
\emmoveto{82.386}{53.696}
\emlineto{82.798}{53.748}
\emmoveto{82.798}{53.738}
\emlineto{83.211}{53.781}
\emmoveto{83.211}{53.771}
\emlineto{83.626}{53.808}
\emmoveto{83.626}{53.798}
\emlineto{84.042}{53.827}
\emmoveto{84.042}{53.817}
\emlineto{84.458}{53.839}
\emmoveto{84.458}{53.829}
\emlineto{84.876}{53.843}
\emmoveto{84.876}{53.833}
\emlineto{85.293}{53.840}
\emmoveto{85.293}{53.830}
\emlineto{85.709}{53.829}
\emmoveto{85.709}{53.819}
\emlineto{86.126}{53.811}
\emmoveto{86.126}{53.801}
\emlineto{86.541}{53.786}
\emmoveto{86.541}{53.776}
\emlineto{86.954}{53.753}
\emmoveto{86.954}{53.743}
\emlineto{87.366}{53.713}
\emmoveto{87.366}{53.703}
\emlineto{87.775}{53.666}
\emmoveto{87.775}{53.656}
\emlineto{88.182}{53.611}
\emmoveto{88.182}{53.601}
\emlineto{88.586}{53.549}
\emmoveto{88.586}{53.539}
\emlineto{88.987}{53.480}
\emmoveto{88.987}{53.470}
\emlineto{89.384}{53.404}
\emmoveto{89.384}{53.394}
\emlineto{89.776}{53.321}
\emmoveto{89.776}{53.311}
\emlineto{90.165}{53.230}
\emmoveto{90.165}{53.220}
\emlineto{90.549}{53.133}
\emmoveto{90.549}{53.123}
\emlineto{90.927}{53.030}
\emmoveto{90.927}{53.020}
\emlineto{91.301}{52.919}
\emmoveto{91.301}{52.909}
\emlineto{91.668}{52.802}
\emmoveto{91.668}{52.792}
\emlineto{92.030}{52.678}
\emmoveto{92.030}{52.668}
\emlineto{92.385}{52.549}
\emmoveto{92.385}{52.539}
\emlineto{92.733}{52.412}
\emmoveto{92.733}{52.402}
\emlineto{93.074}{52.270}
\emmoveto{93.074}{52.260}
\emlineto{93.408}{52.122}
\emmoveto{93.408}{52.112}
\emlineto{93.735}{51.967}
\emmoveto{93.735}{51.957}
\emlineto{94.053}{51.807}
\emmoveto{94.053}{51.797}
\emlineto{94.363}{51.642}
\emmoveto{94.363}{51.632}
\emlineto{94.665}{51.471}
\emmoveto{94.665}{51.461}
\emlineto{94.957}{51.295}
\emmoveto{94.957}{51.285}
\emlineto{95.241}{51.113}
\emmoveto{95.241}{51.103}
\emlineto{95.516}{50.927}
\emmoveto{95.516}{50.917}
\emlineto{95.780}{50.736}
\emmoveto{95.780}{50.726}
\emlineto{96.036}{50.540}
\emmoveto{96.036}{50.530}
\emlineto{96.281}{50.340}
\emmoveto{96.281}{50.330}
\emlineto{96.516}{50.135}
\emmoveto{96.516}{50.125}
\emlineto{96.740}{49.926}
\emmoveto{96.740}{49.916}
\emlineto{96.954}{49.714}
\emmoveto{96.954}{49.704}
\emlineto{97.157}{49.498}
\emmoveto{97.157}{49.488}
\emlineto{97.349}{49.278}
\emmoveto{97.349}{49.268}
\emlineto{97.529}{49.055}
\emmoveto{97.529}{49.045}
\emlineto{97.699}{48.829}
\emmoveto{97.699}{48.819}
\emlineto{97.857}{48.600}
\emmoveto{97.857}{48.590}
\emlineto{98.003}{48.368}
\emmoveto{98.003}{48.358}
\emlineto{98.137}{48.134}
\emmoveto{98.137}{48.124}
\emlineto{98.260}{47.897}
\emmoveto{98.260}{47.887}
\emlineto{98.370}{47.658}
\emmoveto{98.370}{47.648}
\emlineto{98.468}{47.418}
\emmoveto{98.468}{47.408}
\emlineto{98.555}{47.176}
\emmoveto{98.555}{47.166}
\emlineto{98.629}{46.932}
\emmoveto{98.629}{46.922}
\emlineto{98.690}{46.687}
\emmoveto{98.690}{46.677}
\emlineto{98.739}{46.442}
\emmoveto{98.739}{46.432}
\emlineto{98.776}{46.195}
\emmoveto{98.776}{46.185}
\emlineto{98.800}{45.948}
\emmoveto{98.800}{45.938}
\emlineto{98.812}{45.701}
\emmoveto{98.812}{45.691}
\emlineto{98.813}{45.445}
\emmoveto{98.813}{45.435}
\emlineto{98.813}{45.141}
\emmoveto{98.813}{45.131}
\emlineto{98.810}{44.795}
\emmoveto{98.810}{44.785}
\emlineto{98.790}{44.445}
\emmoveto{98.790}{44.435}
\emlineto{98.752}{44.096}
\emmoveto{98.752}{44.086}
\emlineto{98.697}{43.748}
\emmoveto{98.697}{43.738}
\emlineto{98.625}{43.400}
\emmoveto{98.625}{43.390}
\emlineto{98.535}{43.055}
\emmoveto{98.535}{43.045}
\emlineto{98.427}{42.710}
\emmoveto{98.427}{42.700}
\emlineto{98.302}{42.368}
\emmoveto{98.302}{42.358}
\emlineto{98.160}{42.029}
\emmoveto{98.160}{42.019}
\emlineto{98.001}{41.692}
\emmoveto{98.001}{41.682}
\emlineto{97.824}{41.358}
\emmoveto{97.824}{41.348}
\emlineto{97.631}{41.027}
\emmoveto{97.631}{41.017}
\emlineto{97.422}{40.700}
\emmoveto{97.422}{40.690}
\emlineto{97.195}{40.377}
\emmoveto{97.195}{40.367}
\emlineto{96.953}{40.058}
\emmoveto{96.953}{40.048}
\emlineto{96.695}{39.743}
\emmoveto{96.695}{39.733}
\emlineto{96.420}{39.433}
\emmoveto{96.420}{39.423}
\emlineto{96.131}{39.128}
\emmoveto{96.131}{39.118}
\emlineto{95.826}{38.829}
\emmoveto{95.826}{38.819}
\emlineto{95.505}{38.535}
\emmoveto{95.505}{38.525}
\emlineto{95.171}{38.246}
\emmoveto{95.171}{38.236}
\emlineto{94.821}{37.964}
\emmoveto{94.821}{37.954}
\emlineto{94.458}{37.689}
\emmoveto{94.458}{37.679}
\emlineto{94.081}{37.419}
\emmoveto{94.081}{37.409}
\emlineto{93.691}{37.157}
\emmoveto{93.691}{37.147}
\emlineto{93.287}{36.902}
\emmoveto{93.287}{36.892}
\emlineto{92.871}{36.654}
\emmoveto{92.871}{36.644}
\emlineto{92.442}{36.413}
\emmoveto{92.442}{36.403}
\emlineto{92.001}{36.181}
\emmoveto{92.001}{36.171}
\emlineto{91.549}{35.956}
\emmoveto{91.549}{35.946}
\emlineto{91.086}{35.739}
\emmoveto{91.086}{35.729}
\emlineto{90.611}{35.531}
\emmoveto{90.611}{35.521}
\emlineto{90.127}{35.331}
\emmoveto{90.127}{35.321}
\emlineto{89.633}{35.140}
\emmoveto{89.633}{35.130}
\emlineto{89.129}{34.958}
\emmoveto{89.129}{34.948}
\emlineto{88.616}{34.785}
\emmoveto{88.616}{34.775}
\emlineto{88.095}{34.621}
\emmoveto{88.095}{34.611}
\emlineto{87.566}{34.466}
\emmoveto{87.566}{34.456}
\emlineto{87.029}{34.321}
\emmoveto{87.029}{34.311}
\emlineto{86.485}{34.186}
\emmoveto{86.485}{34.176}
\emlineto{85.934}{34.060}
\emmoveto{85.934}{34.050}
\emlineto{85.377}{33.944}
\emmoveto{85.377}{33.934}
\emlineto{84.815}{33.838}
\emmoveto{84.815}{33.828}
\emlineto{84.247}{33.743}
\emmoveto{84.247}{33.733}
\emlineto{83.675}{33.657}
\emmoveto{83.675}{33.647}
\emlineto{83.099}{33.581}
\emmoveto{83.099}{33.571}
\emlineto{82.520}{33.516}
\emmoveto{82.520}{33.506}
\emlineto{81.937}{33.461}
\emmoveto{81.937}{33.451}
\emlineto{81.352}{33.417}
\emmoveto{81.352}{33.407}
\emlineto{80.765}{33.383}
\emmoveto{80.765}{33.373}
\emlineto{80.176}{33.359}
\emmoveto{80.176}{33.349}
\emlineto{79.586}{33.346}
\emmoveto{79.586}{33.336}
\emlineto{78.997}{33.334}
\emlineto{78.997}{33.344}
\emmoveto{78.997}{33.334}
\emlineto{78.407}{33.352}
\emmoveto{78.407}{33.342}
\emlineto{77.818}{33.370}
\emmoveto{77.818}{33.360}
\emlineto{77.230}{33.399}
\emmoveto{77.230}{33.389}
\emlineto{76.643}{33.438}
\emmoveto{76.643}{33.428}
\emlineto{76.059}{33.488}
\emmoveto{76.059}{33.478}
\emlineto{75.478}{33.548}
\emmoveto{75.478}{33.538}
\emlineto{74.900}{33.618}
\emmoveto{74.900}{33.608}
\emlineto{74.326}{33.699}
\emmoveto{74.326}{33.689}
\emlineto{73.756}{33.790}
\emmoveto{73.756}{33.780}
\emlineto{73.191}{33.891}
\emmoveto{73.191}{33.881}
\emlineto{72.632}{34.002}
\emmoveto{72.632}{33.992}
\emlineto{72.078}{34.123}
\emmoveto{72.078}{34.113}
\emlineto{71.531}{34.253}
\emmoveto{71.531}{34.243}
\emlineto{70.990}{34.394}
\emmoveto{70.990}{34.384}
\emlineto{70.457}{34.544}
\emmoveto{70.457}{34.534}
\emlineto{69.932}{34.703}
\emmoveto{69.932}{34.693}
\emlineto{69.415}{34.872}
\emmoveto{69.415}{34.862}
\emlineto{68.907}{35.049}
\emmoveto{68.907}{35.039}
\emlineto{68.408}{35.236}
\emmoveto{68.408}{35.226}
\emlineto{67.918}{35.431}
\emmoveto{67.918}{35.421}
\emlineto{67.439}{35.636}
\emmoveto{67.439}{35.626}
\emlineto{66.970}{35.848}
\emmoveto{66.970}{35.838}
\emlineto{66.513}{36.069}
\emmoveto{66.513}{36.059}
\emlineto{66.066}{36.298}
\emmoveto{66.066}{36.288}
\emlineto{65.631}{36.534}
\emmoveto{65.631}{36.524}
\emlineto{65.209}{36.779}
\emmoveto{65.209}{36.769}
\emlineto{64.799}{37.030}
\emmoveto{64.799}{37.020}
\emlineto{64.402}{37.289}
\emmoveto{64.402}{37.279}
\emlineto{64.018}{37.555}
\emmoveto{64.018}{37.545}
\emlineto{63.648}{37.828}
\emmoveto{63.648}{37.818}
\emlineto{63.292}{38.107}
\emmoveto{63.292}{38.097}
\emlineto{62.950}{38.392}
\emmoveto{62.950}{38.382}
\emlineto{62.623}{38.683}
\emmoveto{62.623}{38.673}
\emlineto{62.310}{38.980}
\emmoveto{62.310}{38.970}
\emlineto{62.013}{39.282}
\emmoveto{62.013}{39.272}
\emlineto{61.731}{39.590}
\emmoveto{61.731}{39.580}
\emlineto{61.465}{39.902}
\emmoveto{61.465}{39.892}
\emlineto{61.214}{40.219}
\emmoveto{61.214}{40.209}
\emlineto{60.980}{40.540}
\emmoveto{60.980}{40.530}
\emlineto{60.762}{40.865}
\emmoveto{60.762}{40.855}
\emlineto{60.561}{41.194}
\emmoveto{60.561}{41.184}
\emlineto{60.377}{41.527}
\emmoveto{60.377}{41.517}
\emlineto{60.209}{41.862}
\emmoveto{60.209}{41.852}
\emlineto{60.058}{42.201}
\emmoveto{60.058}{42.191}
\emlineto{59.925}{42.542}
\emmoveto{59.925}{42.532}
\emlineto{59.809}{42.885}
\emmoveto{59.809}{42.875}
\emlineto{59.710}{43.230}
\emmoveto{59.710}{43.220}
\emlineto{59.629}{43.576}
\emmoveto{59.629}{43.566}
\emlineto{59.565}{43.924}
\emmoveto{59.565}{43.914}
\emlineto{59.519}{44.273}
\emmoveto{59.519}{44.263}
\emlineto{59.490}{44.623}
\emmoveto{59.490}{44.613}
\emlineto{59.480}{44.973}
\emmoveto{59.480}{44.963}
\emlineto{59.480}{45.344}
\emmoveto{59.480}{45.334}
\emlineto{59.481}{45.763}
\emmoveto{59.481}{45.753}
\emlineto{59.501}{46.191}
\emmoveto{59.501}{46.181}
\emlineto{59.542}{46.619}
\emmoveto{59.542}{46.609}
\emlineto{59.604}{47.046}
\emmoveto{59.604}{47.036}
\emlineto{59.689}{47.472}
\emmoveto{59.689}{47.462}
\emlineto{59.794}{47.896}
\emmoveto{59.794}{47.886}
\emlineto{59.921}{48.318}
\emmoveto{59.921}{48.308}
\emlineto{60.070}{48.738}
\emmoveto{60.070}{48.728}
\emlineto{60.239}{49.154}
\emmoveto{60.239}{49.144}
\emlineto{60.430}{49.568}
\emmoveto{60.430}{49.558}
\emlineto{60.641}{49.978}
\emmoveto{60.641}{49.968}
\emlineto{60.873}{50.384}
\emmoveto{60.873}{50.374}
\emlineto{61.126}{50.785}
\emmoveto{61.126}{50.775}
\emlineto{61.398}{51.182}
\emmoveto{61.398}{51.172}
\emlineto{61.691}{51.574}
\emmoveto{61.691}{51.564}
\emlineto{62.003}{51.961}
\emmoveto{62.003}{51.951}
\emlineto{62.335}{52.341}
\emmoveto{62.335}{52.331}
\emlineto{62.686}{52.716}
\emmoveto{62.686}{52.706}
\emlineto{63.055}{53.085}
\emmoveto{63.055}{53.075}
\emlineto{63.443}{53.446}
\emmoveto{63.443}{53.436}
\emlineto{63.849}{53.801}
\emmoveto{63.849}{53.791}
\emlineto{64.273}{54.148}
\emmoveto{64.273}{54.138}
\emlineto{64.714}{54.487}
\emmoveto{64.714}{54.477}
\emlineto{65.173}{54.819}
\emmoveto{65.173}{54.809}
\emlineto{65.647}{55.142}
\emmoveto{65.647}{55.132}
\emlineto{66.138}{55.456}
\emmoveto{66.138}{55.446}
\emlineto{66.645}{55.762}
\emmoveto{66.645}{55.752}
\emlineto{67.167}{56.059}
\emmoveto{67.167}{56.049}
\emlineto{67.703}{56.346}
\emmoveto{67.703}{56.336}
\emlineto{68.254}{56.623}
\emmoveto{68.254}{56.613}
\emlineto{68.819}{56.891}
\emmoveto{68.819}{56.881}
\emlineto{69.396}{57.148}
\emmoveto{69.396}{57.138}
\emlineto{69.987}{57.395}
\emmoveto{69.987}{57.385}
\emlineto{70.590}{57.631}
\emmoveto{70.590}{57.621}
\emlineto{71.204}{57.857}
\emmoveto{71.204}{57.847}
\emlineto{71.830}{58.071}
\emmoveto{71.830}{58.061}
\emlineto{72.466}{58.274}
\emmoveto{72.466}{58.264}
\emlineto{73.112}{58.466}
\emmoveto{73.112}{58.456}
\emlineto{73.768}{58.646}
\emmoveto{73.768}{58.636}
\emlineto{74.433}{58.815}
\emmoveto{74.433}{58.805}
\emlineto{75.105}{58.971}
\emmoveto{75.105}{58.961}
\emlineto{75.785}{59.116}
\emmoveto{75.785}{59.106}
\emlineto{76.473}{59.248}
\emmoveto{76.473}{59.238}
\emlineto{77.166}{59.368}
\emmoveto{77.166}{59.358}
\emlineto{77.866}{59.476}
\emmoveto{77.866}{59.466}
\emlineto{78.570}{59.571}
\emmoveto{78.570}{59.561}
\emlineto{79.279}{59.654}
\emmoveto{79.279}{59.644}
\emlineto{79.992}{59.724}
\emmoveto{79.992}{59.714}
\emlineto{80.708}{59.781}
\emmoveto{80.708}{59.771}
\emlineto{81.427}{59.826}
\emmoveto{81.427}{59.816}
\emlineto{82.148}{59.857}
\emmoveto{82.148}{59.847}
\emlineto{82.869}{59.876}
\emmoveto{82.869}{59.866}
\emlineto{83.592}{59.882}
\emmoveto{83.592}{59.872}
\emlineto{84.314}{59.875}
\emmoveto{84.314}{59.865}
\emlineto{85.036}{59.855}
\emmoveto{85.036}{59.845}
\emlineto{85.757}{59.823}
\emmoveto{85.757}{59.813}
\emlineto{86.475}{59.777}
\emmoveto{86.475}{59.767}
\emlineto{87.191}{59.719}
\emmoveto{87.191}{59.709}
\emlineto{87.904}{59.648}
\emmoveto{87.904}{59.638}
\emlineto{88.612}{59.565}
\emmoveto{88.612}{59.555}
\emlineto{89.317}{59.469}
\emmoveto{89.317}{59.459}
\emlineto{90.016}{59.360}
\emmoveto{90.016}{59.350}
\emlineto{90.709}{59.239}
\emmoveto{90.709}{59.229}
\emlineto{91.396}{59.106}
\emmoveto{91.396}{59.096}
\emlineto{92.075}{58.961}
\emmoveto{92.075}{58.951}
\emlineto{92.747}{58.803}
\emmoveto{92.747}{58.793}
\emlineto{93.411}{58.634}
\emmoveto{93.411}{58.624}
\emlineto{94.066}{58.453}
\emmoveto{94.066}{58.443}
\emlineto{94.712}{58.260}
\emmoveto{94.712}{58.250}
\emlineto{95.347}{58.056}
\emmoveto{95.347}{58.046}
\emlineto{95.972}{57.841}
\emmoveto{95.972}{57.831}
\emlineto{96.586}{57.615}
\emmoveto{96.586}{57.605}
\emlineto{97.188}{57.378}
\emmoveto{97.188}{57.368}
\emlineto{97.778}{57.130}
\emmoveto{97.778}{57.120}
\emlineto{98.354}{56.872}
\emmoveto{98.354}{56.862}
\emlineto{98.918}{56.604}
\emmoveto{98.918}{56.594}
\emlineto{99.468}{56.326}
\emmoveto{99.468}{56.316}
\emlineto{100.003}{56.038}
\emmoveto{100.003}{56.028}
\emlineto{100.524}{55.741}
\emmoveto{100.524}{55.731}
\emlineto{101.030}{55.434}
\emmoveto{101.030}{55.424}
\emlineto{101.519}{55.119}
\emmoveto{101.519}{55.109}
\emlineto{101.993}{54.795}
\emmoveto{101.993}{54.785}
\emlineto{102.450}{54.463}
\emmoveto{102.450}{54.453}
\emlineto{102.890}{54.123}
\emmoveto{102.890}{54.113}
\emlineto{103.313}{53.776}
\emmoveto{103.313}{53.766}
\emlineto{103.717}{53.421}
\emmoveto{103.717}{53.411}
\emlineto{104.104}{53.059}
\emmoveto{104.104}{53.049}
\emlineto{104.472}{52.690}
\emmoveto{104.472}{52.680}
\emlineto{104.822}{52.315}
\emmoveto{104.822}{52.305}
\emlineto{105.152}{51.933}
\emmoveto{105.152}{51.923}
\emlineto{105.463}{51.546}
\emmoveto{105.463}{51.536}
\emlineto{105.754}{51.154}
\emmoveto{105.754}{51.144}
\emlineto{106.025}{50.757}
\emmoveto{106.025}{50.747}
\emlineto{106.276}{50.355}
\emmoveto{106.276}{50.345}
\emlineto{106.507}{49.949}
\emmoveto{106.507}{49.939}
\emlineto{106.717}{49.538}
\emmoveto{106.717}{49.528}
\emlineto{106.906}{49.125}
\emmoveto{106.906}{49.115}
\emlineto{107.074}{48.708}
\emmoveto{107.074}{48.698}
\emlineto{107.221}{48.288}
\emmoveto{107.221}{48.278}
\emlineto{107.346}{47.866}
\emmoveto{107.346}{47.856}
\emlineto{107.450}{47.442}
\emmoveto{107.450}{47.432}
\emlineto{107.533}{47.016}
\emmoveto{107.533}{47.006}
\emlineto{107.594}{46.589}
\emmoveto{107.594}{46.579}
\emlineto{107.634}{46.161}
\emmoveto{107.634}{46.151}
\emlineto{107.652}{45.732}
\emmoveto{107.652}{45.722}
\emlineto{107.653}{45.292}
\emmoveto{107.653}{45.282}
\emlineto{107.651}{44.806}
\emmoveto{107.651}{44.796}
\emlineto{107.628}{44.311}
\emmoveto{107.628}{44.301}
\emlineto{107.580}{43.817}
\emmoveto{107.580}{43.807}
\emlineto{107.507}{43.324}
\emmoveto{107.507}{43.314}
\emlineto{107.409}{42.832}
\emmoveto{107.409}{42.822}
\emlineto{107.287}{42.343}
\emmoveto{107.287}{42.333}
\emlineto{107.140}{41.855}
\emmoveto{107.140}{41.845}
\emlineto{106.968}{41.371}
\emmoveto{106.968}{41.361}
\emlineto{106.772}{40.890}
\emmoveto{106.772}{40.880}
\emlineto{106.551}{40.413}
\emmoveto{106.551}{40.403}
\emlineto{106.307}{39.939}
\emmoveto{106.307}{39.929}
\emlineto{106.039}{39.471}
\emmoveto{106.039}{39.461}
\emlineto{105.747}{39.007}
\emmoveto{105.747}{38.997}
\emlineto{105.432}{38.549}
\emmoveto{105.432}{38.539}
\emlineto{105.093}{38.096}
\emmoveto{105.093}{38.086}
\emlineto{104.732}{37.650}
\emmoveto{104.732}{37.640}
\emlineto{104.349}{37.210}
\emmoveto{104.349}{37.200}
\emlineto{103.944}{36.778}
\emmoveto{103.944}{36.768}
\emlineto{103.516}{36.353}
\emmoveto{103.516}{36.343}
\emlineto{103.068}{35.935}
\emmoveto{103.068}{35.925}
\emlineto{102.599}{35.526}
\emmoveto{102.599}{35.516}
\emlineto{102.109}{35.125}
\emmoveto{102.109}{35.115}
\emlineto{101.599}{34.734}
\emmoveto{101.599}{34.724}
\emlineto{101.070}{34.351}
\emmoveto{101.070}{34.341}
\emlineto{100.521}{33.978}
\emmoveto{100.521}{33.968}
\emlineto{99.954}{33.615}
\emmoveto{99.954}{33.605}
\emlineto{99.369}{33.262}
\emmoveto{99.369}{33.252}
\emlineto{98.766}{32.920}
\emmoveto{98.766}{32.910}
\emlineto{98.146}{32.589}
\emmoveto{98.146}{32.579}
\emlineto{97.510}{32.268}
\emmoveto{97.510}{32.258}
\emlineto{96.858}{31.960}
\emmoveto{96.858}{31.950}
\emlineto{96.190}{31.663}
\emmoveto{96.190}{31.653}
\emlineto{95.508}{31.378}
\emmoveto{95.508}{31.368}
\emlineto{94.812}{31.105}
\emmoveto{94.812}{31.095}
\emlineto{94.102}{30.845}
\emmoveto{94.102}{30.835}
\emlineto{93.379}{30.598}
\emmoveto{93.379}{30.588}
\emlineto{92.644}{30.363}
\emmoveto{92.644}{30.353}
\emlineto{91.898}{30.142}
\emmoveto{91.898}{30.132}
\emlineto{91.141}{29.934}
\emmoveto{91.141}{29.924}
\emlineto{90.373}{29.740}
\emmoveto{90.373}{29.730}
\emlineto{89.597}{29.559}
\emmoveto{89.597}{29.549}
\emlineto{88.811}{29.393}
\emmoveto{88.811}{29.383}
\emlineto{88.017}{29.240}
\emmoveto{88.017}{29.230}
\emlineto{87.216}{29.102}
\emmoveto{87.216}{29.092}
\emlineto{86.408}{28.978}
\emmoveto{86.408}{28.968}
\emlineto{85.595}{28.868}
\emmoveto{85.595}{28.858}
\emlineto{84.776}{28.773}
\emmoveto{84.776}{28.763}
\emlineto{83.953}{28.692}
\emmoveto{83.953}{28.682}
\emlineto{83.126}{28.626}
\emmoveto{83.126}{28.616}
\emlineto{82.296}{28.575}
\emmoveto{82.296}{28.565}
\emlineto{81.464}{28.539}
\emmoveto{81.464}{28.529}
\emlineto{80.630}{28.517}
\emmoveto{80.630}{28.507}
\emlineto{79.796}{28.501}
\emlineto{79.796}{28.511}
\emmoveto{79.796}{28.501}
\emlineto{78.962}{28.519}
\emmoveto{78.962}{28.509}
\emlineto{78.128}{28.542}
\emmoveto{78.128}{28.532}
\emlineto{77.296}{28.580}
\emmoveto{77.296}{28.570}
\emlineto{76.467}{28.633}
\emmoveto{76.467}{28.623}
\emlineto{75.640}{28.700}
\emmoveto{75.640}{28.690}
\emlineto{74.817}{28.782}
\emmoveto{74.817}{28.772}
\emlineto{73.999}{28.879}
\emmoveto{73.999}{28.869}
\emlineto{73.186}{28.990}
\emmoveto{73.186}{28.980}
\emlineto{72.379}{29.116}
\emmoveto{72.379}{29.106}
\emlineto{71.579}{29.256}
\emmoveto{71.579}{29.246}
\emlineto{70.786}{29.410}
\emmoveto{70.786}{29.400}
\emlineto{70.001}{29.578}
\emmoveto{70.001}{29.568}
\emlineto{69.225}{29.760}
\emmoveto{69.225}{29.750}
\emlineto{68.459}{29.955}
\emmoveto{68.459}{29.945}
\emlineto{67.703}{30.165}
\emmoveto{67.703}{30.155}
\emlineto{66.958}{30.387}
\emmoveto{66.958}{30.377}
\emlineto{66.224}{30.623}
\emmoveto{66.224}{30.613}
\emlineto{65.503}{30.872}
\emmoveto{65.503}{30.862}
\emlineto{64.794}{31.133}
\emmoveto{64.794}{31.123}
\emlineto{64.099}{31.407}
\emmoveto{64.099}{31.397}
\emlineto{63.419}{31.694}
\emmoveto{63.419}{31.684}
\emlineto{62.753}{31.992}
\emmoveto{62.753}{31.982}
\emlineto{62.102}{32.302}
\emmoveto{62.102}{32.292}
\emlineto{61.467}{32.623}
\emmoveto{61.467}{32.613}
\emlineto{60.849}{32.956}
\emmoveto{60.849}{32.946}
\emlineto{60.248}{33.299}
\emmoveto{60.248}{33.289}
\emlineto{59.665}{33.653}
\emmoveto{59.665}{33.643}
\emlineto{59.100}{34.017}
\emmoveto{59.100}{34.007}
\emlineto{58.553}{34.391}
\emmoveto{58.553}{34.381}
\emlineto{58.026}{34.774}
\emmoveto{58.026}{34.764}
\emlineto{57.518}{35.167}
\emmoveto{57.518}{35.157}
\emlineto{57.031}{35.569}
\emmoveto{57.031}{35.559}
\emlineto{56.564}{35.979}
\emmoveto{56.564}{35.969}
\emlineto{56.117}{36.397}
\emmoveto{56.117}{36.387}
\emlineto{55.693}{36.823}
\emmoveto{55.693}{36.813}
\emlineto{55.289}{37.257}
\emmoveto{55.289}{37.247}
\emlineto{54.908}{37.697}
\emmoveto{54.908}{37.687}
\emlineto{54.550}{38.144}
\emmoveto{54.550}{38.134}
\emlineto{54.214}{38.597}
\emmoveto{54.214}{38.587}
\emlineto{53.901}{39.056}
\emmoveto{53.901}{39.046}
\emlineto{53.612}{39.520}
\emmoveto{53.612}{39.510}
\emlineto{53.346}{39.989}
\emmoveto{53.346}{39.979}
\emlineto{53.104}{40.463}
\emmoveto{53.104}{40.453}
\emlineto{52.886}{40.941}
\emmoveto{52.886}{40.931}
\emlineto{52.693}{41.422}
\emmoveto{52.693}{41.412}
\emlineto{52.523}{41.907}
\emmoveto{52.523}{41.897}
\emlineto{52.379}{42.394}
\emmoveto{52.379}{42.384}
\emlineto{52.259}{42.884}
\emmoveto{52.259}{42.874}
\emlineto{52.164}{43.376}
\emmoveto{52.164}{43.366}
\emlineto{52.094}{43.869}
\emmoveto{52.094}{43.859}
\emlineto{52.049}{44.363}
\emmoveto{52.049}{44.353}
\emlineto{52.028}{44.858}
\emmoveto{52.028}{44.848}
\emlineto{52.027}{45.366}
\emmoveto{52.027}{45.356}
\emlineto{52.032}{45.914}
\emmoveto{52.032}{45.904}
\emlineto{52.062}{46.467}
\emmoveto{52.062}{46.457}
\emlineto{52.120}{47.020}
\emmoveto{52.120}{47.010}
\emlineto{52.206}{47.571}
\emmoveto{52.206}{47.561}
\emlineto{52.320}{48.120}
\emmoveto{52.320}{48.110}
\emlineto{52.462}{48.667}
\emmoveto{52.462}{48.657}
\emlineto{52.631}{49.211}
\emmoveto{52.631}{49.201}
\emlineto{52.828}{49.752}
\emmoveto{52.828}{49.742}
\emlineto{53.052}{50.289}
\emmoveto{53.052}{50.279}
\emlineto{53.303}{50.822}
\emmoveto{53.303}{50.812}
\emlineto{53.580}{51.350}
\emmoveto{53.580}{51.340}
\emlineto{53.885}{51.874}
\emmoveto{53.885}{51.864}
\emlineto{54.215}{52.391}
\emmoveto{54.215}{52.381}
\emlineto{54.572}{52.902}
\emmoveto{54.572}{52.892}
\emlineto{54.955}{53.407}
\emmoveto{54.955}{53.397}
\emlineto{55.362}{53.905}
\emmoveto{55.362}{53.895}
\emlineto{55.795}{54.395}
\emmoveto{55.795}{54.385}
\emlineto{56.252}{54.877}
\emmoveto{56.252}{54.867}
\emlineto{56.734}{55.351}
\emmoveto{56.734}{55.341}
\emlineto{57.239}{55.816}
\emmoveto{57.239}{55.806}
\emlineto{57.768}{56.272}
\emmoveto{57.768}{56.262}
\emlineto{58.319}{56.719}
\emmoveto{58.319}{56.709}
\emlineto{58.893}{57.155}
\emmoveto{58.893}{57.145}
\emlineto{59.488}{57.581}
\emmoveto{59.488}{57.571}
\emlineto{60.105}{57.996}
\emmoveto{60.105}{57.986}
\emlineto{60.743}{58.400}
\emmoveto{60.743}{58.390}
\emlineto{61.400}{58.793}
\emmoveto{61.400}{58.783}
\emlineto{62.078}{59.173}
\emmoveto{62.078}{59.163}
\emlineto{62.774}{59.542}
\emmoveto{62.774}{59.532}
\emlineto{63.488}{59.897}
\emmoveto{63.488}{59.887}
\emlineto{64.220}{60.240}
\emmoveto{64.220}{60.230}
\emlineto{64.969}{60.570}
\emmoveto{64.969}{60.560}
\emlineto{65.735}{60.886}
\emmoveto{65.735}{60.876}
\emlineto{66.516}{61.189}
\emmoveto{66.516}{61.179}
\emlineto{67.312}{61.477}
\emmoveto{67.312}{61.467}
\emlineto{68.122}{61.752}
\emmoveto{68.122}{61.742}
\emlineto{68.946}{62.011}
\emmoveto{68.946}{62.001}
\emlineto{69.783}{62.256}
\emmoveto{69.783}{62.246}
\emlineto{70.631}{62.486}
\emmoveto{70.631}{62.476}
\emlineto{71.491}{62.701}
\emmoveto{71.491}{62.691}
\emlineto{72.361}{62.900}
\emmoveto{72.361}{62.890}
\emlineto{73.241}{63.084}
\emmoveto{73.241}{63.074}
\emlineto{74.130}{63.252}
\emmoveto{74.130}{63.242}
\emlineto{75.027}{63.404}
\emmoveto{75.027}{63.394}
\emlineto{75.931}{63.540}
\emmoveto{75.931}{63.530}
\emlineto{76.842}{63.660}
\emmoveto{76.842}{63.650}
\emlineto{77.758}{63.763}
\emmoveto{77.758}{63.753}
\emlineto{78.679}{63.851}
\emmoveto{78.679}{63.841}
\emlineto{79.604}{63.921}
\emmoveto{79.604}{63.911}
\emlineto{80.533}{63.976}
\emmoveto{80.533}{63.966}
\emlineto{81.463}{64.014}
\emmoveto{81.463}{64.004}
\emlineto{82.395}{64.035}
\emmoveto{82.395}{64.025}
\emlineto{83.328}{64.040}
\emmoveto{83.328}{64.030}
\emlineto{84.261}{64.028}
\emmoveto{84.261}{64.018}
\emlineto{85.192}{63.999}
\emmoveto{85.192}{63.989}
\emlineto{86.122}{63.954}
\emmoveto{86.122}{63.944}
\emlineto{87.049}{63.892}
\emmoveto{87.049}{63.882}
\emlineto{87.973}{63.814}
\emmoveto{87.973}{63.804}
\emlineto{88.892}{63.720}
\emmoveto{88.892}{63.710}
\emlineto{89.806}{63.609}
\emmoveto{89.806}{63.599}
\emlineto{90.714}{63.482}
\emmoveto{90.714}{63.472}
\emlineto{91.615}{63.339}
\emmoveto{91.615}{63.329}
\emlineto{92.508}{63.180}
\emmoveto{92.508}{63.170}
\emlineto{93.393}{63.005}
\emmoveto{93.393}{62.995}
\emlineto{94.269}{62.814}
\emmoveto{94.269}{62.804}
\emlineto{95.135}{62.608}
\emmoveto{95.135}{62.598}
\emlineto{95.990}{62.387}
\emmoveto{95.990}{62.377}
\emlineto{96.833}{62.150}
\emmoveto{96.833}{62.140}
\emlineto{97.664}{61.899}
\emmoveto{97.664}{61.889}
\emlineto{98.482}{61.633}
\emmoveto{98.482}{61.623}
\emlineto{99.286}{61.352}
\emmoveto{99.286}{61.342}
\emlineto{100.076}{61.058}
\emmoveto{100.076}{61.048}
\emlineto{100.850}{60.749}
\emmoveto{100.850}{60.739}
\emlineto{101.608}{60.427}
\emmoveto{101.608}{60.417}
\emlineto{102.350}{60.091}
\emmoveto{102.350}{60.081}
\emlineto{103.074}{59.742}
\emmoveto{103.074}{59.732}
\emlineto{103.781}{59.381}
\emmoveto{103.781}{59.371}
\emlineto{104.469}{59.007}
\emmoveto{104.469}{58.997}
\emlineto{105.137}{58.621}
\emmoveto{105.137}{58.611}
\emlineto{105.786}{58.224}
\emmoveto{105.786}{58.214}
\emlineto{106.415}{57.815}
\emmoveto{106.415}{57.805}
\emlineto{107.022}{57.395}
\emmoveto{107.022}{57.385}
\emlineto{107.608}{56.964}
\emmoveto{107.608}{56.954}
\emlineto{108.172}{56.523}
\emmoveto{108.172}{56.513}
\emlineto{108.713}{56.073}
\emmoveto{108.713}{56.063}
\emlineto{109.232}{55.613}
\emmoveto{109.232}{55.603}
\emlineto{109.727}{55.144}
\emmoveto{109.727}{55.134}
\emlineto{110.198}{54.666}
\emmoveto{110.198}{54.656}
\emlineto{110.644}{54.180}
\emmoveto{110.644}{54.170}
\emlineto{111.066}{53.687}
\emmoveto{111.066}{53.677}
\emlineto{111.463}{53.186}
\emmoveto{111.463}{53.176}
\emlineto{111.834}{52.678}
\emmoveto{111.834}{52.668}
\emlineto{112.179}{52.164}
\emmoveto{112.179}{52.154}
\emlineto{112.498}{51.644}
\emmoveto{112.498}{51.634}
\emlineto{112.791}{51.118}
\emmoveto{112.791}{51.108}
\emlineto{113.057}{50.588}
\emmoveto{113.057}{50.578}
\emlineto{113.296}{50.053}
\emmoveto{113.296}{50.043}
\emlineto{113.508}{49.514}
\emmoveto{113.508}{49.504}
\emlineto{113.693}{48.972}
\emmoveto{113.693}{48.962}
\emlineto{113.850}{48.426}
\emmoveto{113.850}{48.416}
\emlineto{113.979}{47.878}
\emmoveto{113.979}{47.868}
\emlineto{114.081}{47.328}
\emmoveto{114.081}{47.318}
\emlineto{114.154}{46.777}
\emmoveto{114.154}{46.767}
\emlineto{114.200}{46.224}
\emmoveto{114.200}{46.214}
\emlineto{114.218}{45.671}
\emmoveto{114.218}{45.661}
\emlineto{114.219}{45.098}
\emmoveto{114.219}{45.088}
\emlineto{114.207}{44.492}
\emmoveto{114.207}{44.482}
\emlineto{114.166}{43.887}
\emmoveto{114.166}{43.877}
\emlineto{114.094}{43.282}
\emmoveto{114.094}{43.272}
\emlineto{113.991}{42.679}
\emmoveto{113.991}{42.669}
\emlineto{113.858}{42.078}
\emmoveto{113.858}{42.068}
\emlineto{113.695}{41.479}
\emmoveto{113.695}{41.469}
\emlineto{113.501}{40.884}
\emmoveto{113.501}{40.874}
\emlineto{113.277}{40.293}
\emmoveto{113.277}{40.283}
\emlineto{113.024}{39.705}
\emmoveto{113.024}{39.695}
\emlineto{112.741}{39.123}
\emmoveto{112.741}{39.113}
\emlineto{112.429}{38.546}
\emmoveto{112.429}{38.536}
\emlineto{112.087}{37.974}
\emmoveto{112.087}{37.964}
\emlineto{111.717}{37.409}
\emmoveto{111.717}{37.399}
\emlineto{111.319}{36.851}
\emmoveto{111.319}{36.841}
\emlineto{110.892}{36.300}
\emmoveto{110.892}{36.290}
\emlineto{110.438}{35.757}
\emmoveto{110.438}{35.747}
\emlineto{109.956}{35.222}
\emmoveto{109.956}{35.212}
\emlineto{109.448}{34.697}
\emmoveto{109.448}{34.687}
\emlineto{108.913}{34.180}
\emmoveto{108.913}{34.170}
\emlineto{108.353}{33.673}
\emmoveto{108.353}{33.663}
\emlineto{107.767}{33.177}
\emmoveto{107.767}{33.167}
\emlineto{107.156}{32.691}
\emmoveto{107.156}{32.681}
\emlineto{106.521}{32.216}
\emmoveto{106.521}{32.206}
\emlineto{105.862}{31.752}
\emmoveto{105.862}{31.742}
\emlineto{105.180}{31.301}
\emmoveto{105.180}{31.291}
\emlineto{104.476}{30.862}
\emmoveto{104.476}{30.852}
\emlineto{103.749}{30.435}
\emmoveto{103.749}{30.425}
\emlineto{103.002}{30.022}
\emmoveto{103.002}{30.012}
\emlineto{102.234}{29.622}
\emmoveto{102.234}{29.612}
\emlineto{101.446}{29.236}
\emmoveto{101.446}{29.226}
\emlineto{100.639}{28.864}
\emmoveto{100.639}{28.854}
\emlineto{99.813}{28.507}
\emmoveto{99.813}{28.497}
\emlineto{98.970}{28.165}
\emmoveto{98.970}{28.155}
\emlineto{98.109}{27.838}
\emmoveto{98.109}{27.828}
\emlineto{97.233}{27.526}
\emmoveto{97.233}{27.516}
\emlineto{96.341}{27.230}
\emmoveto{96.341}{27.220}
\emlineto{95.435}{26.950}
\emmoveto{95.435}{26.940}
\emlineto{94.515}{26.686}
\emmoveto{94.515}{26.676}
\emlineto{93.582}{26.439}
\emmoveto{93.582}{26.429}
\emlineto{92.637}{26.208}
\emmoveto{92.637}{26.198}
\emlineto{91.681}{25.995}
\emmoveto{91.681}{25.985}
\emlineto{90.714}{25.798}
\emmoveto{90.714}{25.788}
\emlineto{89.738}{25.619}
\emmoveto{89.738}{25.609}
\emlineto{88.753}{25.457}
\emmoveto{88.753}{25.447}
\emlineto{87.760}{25.313}
\emmoveto{87.760}{25.303}
\emlineto{86.761}{25.186}
\emmoveto{86.761}{25.176}
\emlineto{85.756}{25.078}
\emmoveto{85.756}{25.068}
\emlineto{84.745}{24.987}
\emmoveto{84.745}{24.977}
\emlineto{83.731}{24.914}
\emmoveto{83.731}{24.904}
\emlineto{82.713}{24.860}
\emmoveto{82.713}{24.850}
\emlineto{81.693}{24.823}
\emmoveto{81.693}{24.813}
\emlineto{80.672}{24.805}
\emmoveto{80.672}{24.795}
\emlineto{79.650}{24.795}
\emlineto{79.650}{24.805}
\emmoveto{79.650}{24.795}
\emlineto{78.628}{24.823}
\emmoveto{78.628}{24.813}
\emlineto{77.608}{24.859}
\emmoveto{77.608}{24.849}
\emlineto{76.591}{24.914}
\emmoveto{76.591}{24.904}
\emlineto{75.576}{24.986}
\emmoveto{75.576}{24.976}
\emlineto{74.566}{25.077}
\emmoveto{74.566}{25.067}
\emlineto{73.560}{25.185}
\emmoveto{73.560}{25.175}
\emlineto{72.561}{25.311}
\emmoveto{72.561}{25.301}
\emlineto{71.568}{25.455}
\emmoveto{71.568}{25.445}
\emlineto{70.583}{25.617}
\emmoveto{70.583}{25.607}
\emlineto{69.607}{25.796}
\emmoveto{69.607}{25.786}
\emlineto{68.640}{25.992}
\emmoveto{68.640}{25.982}
\emlineto{67.684}{26.206}
\emmoveto{67.684}{26.196}
\emlineto{66.739}{26.436}
\emmoveto{66.739}{26.426}
\emlineto{65.805}{26.683}
\emmoveto{65.805}{26.673}
\emlineto{64.885}{26.947}
\emmoveto{64.885}{26.937}
\emlineto{63.979}{27.227}
\emmoveto{63.979}{27.217}
\emlineto{63.087}{27.522}
\emmoveto{63.087}{27.512}
\emlineto{62.210}{27.834}
\emmoveto{62.210}{27.824}
\emlineto{61.350}{28.161}
\emmoveto{61.350}{28.151}
\emlineto{60.506}{28.503}
\emmoveto{60.506}{28.493}
\emlineto{59.680}{28.860}
\emmoveto{59.680}{28.850}
\emlineto{58.873}{29.232}
\emmoveto{58.873}{29.222}
\emlineto{58.085}{29.618}
\emmoveto{58.085}{29.608}
\emlineto{57.317}{30.017}
\emmoveto{57.317}{30.007}
\emlineto{56.569}{30.430}
\emmoveto{56.569}{30.420}
\emlineto{55.842}{30.857}
\emmoveto{55.842}{30.847}
\emlineto{55.137}{31.296}
\emmoveto{55.137}{31.286}
\emlineto{54.455}{31.747}
\emmoveto{54.455}{31.737}
\emlineto{53.796}{32.210}
\emmoveto{53.796}{32.200}
\emlineto{53.161}{32.685}
\emmoveto{53.161}{32.675}
\emlineto{52.550}{33.171}
\emmoveto{52.550}{33.161}
\emlineto{51.964}{33.667}
\emmoveto{51.964}{33.657}
\emlineto{51.403}{34.174}
\emmoveto{51.403}{34.164}
\emlineto{50.868}{34.691}
\emmoveto{50.868}{34.681}
\emlineto{50.359}{35.216}
\emmoveto{50.359}{35.206}
\emlineto{49.877}{35.751}
\emmoveto{49.877}{35.741}
\emlineto{49.423}{36.294}
\emmoveto{49.423}{36.284}
\emlineto{48.996}{36.845}
\emmoveto{48.996}{36.835}
\emlineto{48.597}{37.403}
\emmoveto{48.597}{37.393}
\emlineto{48.227}{37.968}
\emmoveto{48.227}{37.958}
\emlineto{47.885}{38.539}
\emmoveto{47.885}{38.529}
\emlineto{47.572}{39.116}
\emmoveto{47.572}{39.106}
\emlineto{47.289}{39.699}
\emmoveto{47.289}{39.689}
\emlineto{47.035}{40.286}
\emmoveto{47.035}{40.276}
\emlineto{46.811}{40.877}
\emmoveto{46.811}{40.867}
\emlineto{46.617}{41.473}
\emmoveto{46.617}{41.463}
\emlineto{46.453}{42.071}
\emmoveto{46.453}{42.061}
\emlineto{46.320}{42.672}
\emmoveto{46.320}{42.662}
\emlineto{46.217}{43.275}
\emmoveto{46.217}{43.265}
\emlineto{46.145}{43.880}
\emmoveto{46.145}{43.870}
\emlineto{46.103}{44.485}
\emmoveto{46.103}{44.475}
\emlineto{46.091}{45.092}
\emmoveto{46.091}{45.082}
\emlineto{46.092}{45.730}
\emmoveto{46.092}{45.720}
\emlineto{46.115}{46.385}
\emmoveto{46.115}{46.375}
\emlineto{46.172}{47.039}
\emmoveto{46.172}{47.029}
\emlineto{46.262}{47.691}
\emmoveto{46.262}{47.681}
\emlineto{46.384}{48.342}
\emmoveto{46.384}{48.332}
\emlineto{46.540}{48.990}
\emmoveto{46.540}{48.980}
\emlineto{46.728}{49.636}
\emmoveto{46.728}{49.626}
\emlineto{46.948}{50.277}
\emmoveto{46.948}{50.267}
\emlineto{47.201}{50.915}
\emmoveto{47.201}{50.905}
\emlineto{47.487}{51.547}
\emmoveto{47.487}{51.537}
\emlineto{47.803}{52.174}
\emmoveto{47.803}{52.164}
\emlineto{48.152}{52.796}
\emmoveto{48.152}{52.786}
\emlineto{48.532}{53.410}
\emmoveto{48.532}{53.400}
\emlineto{48.943}{54.018}
\emmoveto{48.943}{54.008}
\emlineto{49.384}{54.618}
\emmoveto{49.384}{54.608}
\emlineto{49.855}{55.210}
\emmoveto{49.855}{55.200}
\emlineto{50.356}{55.794}
\emmoveto{50.356}{55.784}
\emlineto{50.887}{56.368}
\emmoveto{50.887}{56.358}
\emlineto{51.446}{56.932}
\emmoveto{51.446}{56.922}
\emlineto{52.033}{57.487}
\emmoveto{52.033}{57.477}
\emlineto{52.649}{58.030}
\emmoveto{52.649}{58.020}
\emlineto{53.291}{58.563}
\emmoveto{53.291}{58.553}
\emlineto{53.960}{59.083}
\emmoveto{53.960}{59.073}
\emlineto{54.655}{59.592}
\emmoveto{54.655}{59.582}
\emlineto{55.376}{60.088}
\emmoveto{55.376}{60.078}
\emlineto{56.121}{60.571}
\emmoveto{56.121}{60.561}
\emlineto{56.891}{61.040}
\emmoveto{56.891}{61.030}
\emlineto{57.683}{61.496}
\emmoveto{57.683}{61.486}
\emlineto{58.499}{61.937}
\emmoveto{58.499}{61.927}
\emlineto{59.336}{62.364}
\emmoveto{59.336}{62.354}
\emlineto{60.195}{62.775}
\emmoveto{60.195}{62.765}
\emlineto{61.074}{63.171}
\emmoveto{61.074}{63.161}
\emlineto{61.972}{63.552}
\emmoveto{61.972}{63.542}
\emlineto{62.890}{63.916}
\emmoveto{62.890}{63.906}
\emlineto{63.825}{64.263}
\emmoveto{63.825}{64.253}
\emlineto{64.778}{64.594}
\emmoveto{64.778}{64.584}
\emlineto{65.747}{64.907}
\emmoveto{65.747}{64.897}
\emlineto{66.731}{65.204}
\emmoveto{66.731}{65.194}
\emlineto{67.730}{65.482}
\emmoveto{67.730}{65.472}
\emlineto{68.742}{65.743}
\emmoveto{68.742}{65.733}
\emlineto{69.767}{65.986}
\emmoveto{69.767}{65.976}
\emlineto{70.805}{66.210}
\emmoveto{70.805}{66.200}
\emlineto{71.852}{66.415}
\emmoveto{71.852}{66.405}
\emlineto{72.910}{66.602}
\emmoveto{72.910}{66.592}
\emlineto{73.977}{66.770}
\emmoveto{73.977}{66.760}
\emlineto{75.052}{66.919}
\emmoveto{75.052}{66.909}
\emlineto{76.134}{67.049}
\emmoveto{76.134}{67.039}
\emlineto{77.222}{67.160}
\emmoveto{77.222}{67.150}
\emlineto{78.315}{67.251}
\emmoveto{78.315}{67.241}
\emlineto{79.412}{67.322}
\emmoveto{79.412}{67.312}
\emlineto{80.512}{67.374}
\emmoveto{80.512}{67.364}
\emlineto{81.614}{67.407}
\emmoveto{81.614}{67.397}
\emlineto{82.718}{67.419}
\emmoveto{82.718}{67.409}
\emlineto{83.822}{67.413}
\emmoveto{83.822}{67.403}
\emlineto{84.924}{67.386}
\emmoveto{84.924}{67.376}
\emlineto{86.026}{67.340}
\emmoveto{86.026}{67.330}
\emlineto{87.124}{67.274}
\emmoveto{87.124}{67.264}
\emlineto{88.218}{67.189}
\emmoveto{88.218}{67.179}
\emlineto{89.308}{67.084}
\emmoveto{89.308}{67.074}
\emlineto{90.391}{66.960}
\emmoveto{90.391}{66.950}
\emlineto{91.468}{66.817}
\emmoveto{91.468}{66.807}
\emlineto{92.538}{66.655}
\emmoveto{92.538}{66.645}
\emlineto{93.598}{66.474}
\emmoveto{93.598}{66.464}
\emlineto{94.649}{66.274}
\emmoveto{94.649}{66.264}
\emlineto{95.690}{66.055}
\emmoveto{95.690}{66.045}
\emlineto{96.719}{65.818}
\emmoveto{96.719}{65.808}
\emlineto{97.735}{65.563}
\emmoveto{97.735}{65.553}
\emlineto{98.738}{65.289}
\emmoveto{98.738}{65.279}
\emlineto{99.727}{64.999}
\emmoveto{99.727}{64.989}
\emlineto{100.700}{64.690}
\emmoveto{100.700}{64.680}
\emlineto{101.658}{64.364}
\emmoveto{101.658}{64.354}
\emlineto{102.599}{64.022}
\emmoveto{102.599}{64.012}
\emlineto{103.522}{63.663}
\emmoveto{103.522}{63.653}
\emlineto{104.426}{63.288}
\emmoveto{104.426}{63.278}
\emlineto{105.311}{62.896}
\emmoveto{105.311}{62.886}
\emlineto{106.176}{62.489}
\emmoveto{106.176}{62.479}
\emlineto{107.020}{62.067}
\emmoveto{107.020}{62.057}
\emlineto{107.842}{61.630}
\emmoveto{107.842}{61.620}
\emlineto{108.642}{61.179}
\emmoveto{108.642}{61.169}
\emlineto{109.418}{60.714}
\emmoveto{109.418}{60.704}
\emlineto{110.171}{60.235}
\emmoveto{110.171}{60.225}
\emlineto{110.899}{59.743}
\emmoveto{110.899}{59.733}
\emlineto{111.601}{59.238}
\emmoveto{111.601}{59.228}
\emlineto{112.278}{58.721}
\emmoveto{112.278}{58.711}
\emlineto{112.929}{58.192}
\emmoveto{112.929}{58.182}
\emlineto{113.553}{57.652}
\emmoveto{113.553}{57.642}
\emlineto{114.148}{57.100}
\emmoveto{114.148}{57.090}
\emlineto{114.716}{56.539}
\emmoveto{114.716}{56.529}
\emlineto{115.255}{55.968}
\emmoveto{115.255}{55.958}
\emlineto{115.765}{55.387}
\emmoveto{115.765}{55.377}
\emlineto{116.246}{54.797}
\emmoveto{116.246}{54.787}
\emlineto{116.696}{54.200}
\emmoveto{116.696}{54.190}
\emlineto{117.116}{53.594}
\emmoveto{117.116}{53.584}
\emlineto{117.505}{52.981}
\emmoveto{117.505}{52.971}
\emlineto{117.863}{52.362}
\emmoveto{117.863}{52.352}
\emlineto{118.190}{51.737}
\emmoveto{118.190}{51.727}
\emlineto{118.484}{51.106}
\emmoveto{118.484}{51.096}
\emlineto{118.747}{50.470}
\emmoveto{118.747}{50.460}
\emlineto{118.978}{49.829}
\emmoveto{118.978}{49.819}
\emlineto{119.176}{49.185}
\emmoveto{119.176}{49.175}
\emlineto{119.341}{48.538}
\emmoveto{119.341}{48.528}
\emlineto{119.473}{47.888}
\emmoveto{119.473}{47.878}
\emlineto{119.573}{47.236}
\emmoveto{119.573}{47.226}
\emlineto{119.639}{46.582}
\emmoveto{119.639}{46.572}
\emlineto{119.673}{45.928}
\emmoveto{119.673}{45.918}
\emlineto{119.677}{45.266}
\emmoveto{119.677}{45.256}
\emlineto{119.670}{44.570}
\emmoveto{119.670}{44.560}
\emlineto{119.630}{43.871}
\emmoveto{119.630}{43.861}
\emlineto{119.555}{43.172}
\emmoveto{119.555}{43.162}
\emlineto{119.444}{42.475}
\emmoveto{119.444}{42.465}
\emlineto{119.299}{41.781}
\emmoveto{119.299}{41.771}
\emlineto{119.118}{41.089}
\emmoveto{119.118}{41.079}
\emlineto{118.902}{40.401}
\emmoveto{118.902}{40.391}
\emlineto{118.652}{39.717}
\emmoveto{118.652}{39.707}
\emlineto{118.367}{39.037}
\emmoveto{118.367}{39.027}
\emlineto{118.048}{38.363}
\emmoveto{118.048}{38.353}
\emlineto{117.695}{37.696}
\emmoveto{117.695}{37.686}
\emlineto{117.308}{37.034}
\emmoveto{117.308}{37.024}
\emlineto{116.888}{36.380}
\emmoveto{116.888}{36.370}
\emlineto{116.435}{35.734}
\emmoveto{116.435}{35.724}
\emlineto{115.950}{35.096}
\emmoveto{115.950}{35.086}
\emlineto{115.433}{34.467}
\emmoveto{115.433}{34.457}
\emlineto{114.884}{33.847}
\emmoveto{114.884}{33.837}
\emlineto{114.304}{33.237}
\emmoveto{114.304}{33.227}
\emlineto{113.693}{32.638}
\emmoveto{113.693}{32.628}
\emlineto{113.053}{32.050}
\emmoveto{113.053}{32.040}
\emlineto{112.383}{31.474}
\emmoveto{112.383}{31.464}
\emlineto{111.684}{30.910}
\emmoveto{111.684}{30.900}
\emlineto{110.957}{30.359}
\emmoveto{110.957}{30.349}
\emlineto{110.202}{29.821}
\emmoveto{110.202}{29.811}
\emlineto{109.421}{29.296}
\emmoveto{109.421}{29.286}
\emlineto{108.613}{28.786}
\emmoveto{108.613}{28.776}
\emlineto{107.780}{28.290}
\emmoveto{107.780}{28.280}
\emlineto{106.922}{27.810}
\emmoveto{106.922}{27.800}
\emlineto{106.041}{27.345}
\emmoveto{106.041}{27.335}
\emlineto{105.136}{26.895}
\emmoveto{105.136}{26.885}
\emlineto{104.209}{26.462}
\emmoveto{104.209}{26.452}
\emlineto{103.260}{26.046}
\emmoveto{103.260}{26.036}
\emlineto{102.291}{25.647}
\emmoveto{102.291}{25.637}
\emlineto{101.302}{25.265}
\emmoveto{101.302}{25.255}
\emlineto{100.294}{24.901}
\emmoveto{100.294}{24.891}
\emlineto{99.268}{24.555}
\emmoveto{99.268}{24.545}
\emlineto{98.226}{24.227}
\emmoveto{98.226}{24.217}
\emlineto{97.167}{23.918}
\emmoveto{97.167}{23.908}
\emlineto{96.093}{23.629}
\emmoveto{96.093}{23.619}
\emlineto{95.004}{23.358}
\emmoveto{95.004}{23.348}
\emlineto{93.903}{23.107}
\emmoveto{93.903}{23.097}
\emlineto{92.789}{22.875}
\emmoveto{92.789}{22.865}
\emlineto{91.665}{22.664}
\emmoveto{91.665}{22.654}
\emlineto{90.530}{22.473}
\emmoveto{90.530}{22.463}
\emlineto{89.385}{22.302}
\emmoveto{89.385}{22.292}
\emlineto{88.233}{22.151}
\emmoveto{88.233}{22.141}
\emlineto{87.074}{22.021}
\emmoveto{87.074}{22.011}
\emlineto{85.908}{21.911}
\emmoveto{85.908}{21.901}
\emlineto{84.738}{21.823}
\emmoveto{84.738}{21.813}
\emlineto{83.563}{21.755}
\emmoveto{83.563}{21.745}
\emlineto{82.386}{21.708}
\emmoveto{82.386}{21.698}
\emlineto{81.207}{21.682}
\emmoveto{81.207}{21.672}
\emlineto{80.027}{21.667}
\emlineto{80.027}{21.677}
\emmoveto{80.027}{21.667}
\emlineto{78.847}{21.694}
\emmoveto{78.847}{21.684}
\emlineto{77.669}{21.731}
\emmoveto{77.669}{21.721}
\emlineto{76.493}{21.789}
\emmoveto{76.493}{21.779}
\emlineto{75.321}{21.868}
\emmoveto{75.321}{21.858}
\emlineto{74.153}{21.968}
\emmoveto{74.153}{21.958}
\emlineto{72.990}{22.088}
\emmoveto{72.990}{22.078}
\emlineto{71.835}{22.229}
\emmoveto{71.835}{22.219}
\emlineto{70.687}{22.391}
\emmoveto{70.687}{22.381}
\emlineto{69.547}{22.573}
\emmoveto{69.547}{22.563}
\emlineto{68.418}{22.775}
\emmoveto{68.418}{22.765}
\emlineto{67.299}{22.997}
\emmoveto{67.299}{22.987}
\emlineto{66.192}{23.239}
\emmoveto{66.192}{23.229}
\emlineto{65.097}{23.501}
\emmoveto{65.097}{23.491}
\emlineto{64.016}{23.782}
\emmoveto{64.016}{23.772}
\emlineto{62.950}{24.082}
\emmoveto{62.950}{24.072}
\emlineto{61.900}{24.401}
\emmoveto{61.900}{24.391}
\emlineto{60.866}{24.738}
\emmoveto{60.866}{24.728}
\emlineto{59.850}{25.094}
\emmoveto{59.850}{25.084}
\emlineto{58.852}{25.468}
\emmoveto{58.852}{25.458}
\emlineto{57.874}{25.859}
\emmoveto{57.874}{25.849}
\emlineto{56.916}{26.267}
\emmoveto{56.916}{26.257}
\emlineto{55.978}{26.693}
\emmoveto{55.978}{26.683}
\emlineto{55.063}{27.135}
\emmoveto{55.063}{27.125}
\emlineto{54.171}{27.592}
\emmoveto{54.171}{27.582}
\emlineto{53.302}{28.066}
\emmoveto{53.302}{28.056}
\emlineto{52.457}{28.555}
\emmoveto{52.457}{28.545}
\emlineto{51.638}{29.058}
\emmoveto{51.638}{29.048}
\emlineto{50.844}{29.576}
\emmoveto{50.844}{29.566}
\emlineto{50.077}{30.108}
\emmoveto{50.077}{30.098}
\emlineto{49.337}{30.653}
\emmoveto{49.337}{30.643}
\emlineto{48.625}{31.212}
\emmoveto{48.625}{31.202}
\emlineto{47.942}{31.782}
\emmoveto{47.942}{31.772}
\emlineto{47.287}{32.365}
\emmoveto{47.287}{32.355}
\emlineto{46.663}{32.959}
\emmoveto{46.663}{32.949}
\emlineto{46.069}{33.563}
\emmoveto{46.069}{33.553}
\emlineto{45.505}{34.178}
\emmoveto{45.505}{34.168}
\emlineto{44.973}{34.803}
\emmoveto{44.973}{34.793}
\emlineto{44.473}{35.437}
\emmoveto{44.473}{35.427}
\emlineto{44.005}{36.080}
\emmoveto{44.005}{36.070}
\emlineto{43.570}{36.730}
\emmoveto{43.570}{36.720}
\emlineto{43.168}{37.389}
\emmoveto{43.168}{37.379}
\emlineto{42.799}{38.053}
\emmoveto{42.799}{38.043}
\emlineto{42.464}{38.725}
\emmoveto{42.464}{38.715}
\emlineto{42.164}{39.401}
\emmoveto{42.164}{39.391}
\emlineto{41.897}{40.083}
\emmoveto{41.897}{40.073}
\emlineto{41.665}{40.770}
\emmoveto{41.665}{40.760}
\emlineto{41.468}{41.460}
\emmoveto{41.468}{41.450}
\emlineto{41.306}{42.153}
\emmoveto{41.306}{42.143}
\emlineto{41.180}{42.849}
\emmoveto{41.180}{42.839}
\emlineto{41.088}{43.547}
\emmoveto{41.088}{43.537}
\emlineto{41.032}{44.246}
\emmoveto{41.032}{44.236}
\emlineto{41.011}{44.946}
\emmoveto{41.011}{44.936}
\emlineto{41.011}{45.667}
\emmoveto{41.011}{45.657}
\emlineto{41.033}{46.410}
\emmoveto{41.033}{46.400}
\emlineto{41.093}{47.151}
\emmoveto{41.093}{47.141}
\emlineto{41.191}{47.891}
\emmoveto{41.191}{47.881}
\emlineto{41.326}{48.630}
\emmoveto{41.326}{48.620}
\emlineto{41.498}{49.365}
\emmoveto{41.498}{49.355}
\emlineto{41.707}{50.097}
\emmoveto{41.707}{50.087}
\emlineto{41.953}{50.825}
\emmoveto{41.953}{50.815}
\emlineto{42.236}{51.548}
\emmoveto{42.236}{51.538}
\emlineto{42.555}{52.266}
\emmoveto{42.555}{52.256}
\emlineto{42.911}{52.978}
\emmoveto{42.911}{52.968}
\emlineto{43.302}{53.683}
\emmoveto{43.302}{53.673}
\emlineto{43.729}{54.381}
\emmoveto{43.729}{54.371}
\emlineto{44.191}{55.071}
\emmoveto{44.191}{55.061}
\emlineto{44.687}{55.752}
\emmoveto{44.687}{55.742}
\emlineto{45.218}{56.425}
\emmoveto{45.218}{56.415}
\emlineto{45.783}{57.087}
\emmoveto{45.783}{57.077}
\emlineto{46.381}{57.740}
\emmoveto{46.381}{57.730}
\emlineto{47.011}{58.381}
\emmoveto{47.011}{58.371}
\emlineto{47.674}{59.011}
\emmoveto{47.674}{59.001}
\emlineto{48.368}{59.629}
\emmoveto{48.368}{59.619}
\emlineto{49.093}{60.234}
\emmoveto{49.093}{60.224}
\emlineto{49.849}{60.826}
\emmoveto{49.849}{60.816}
\emlineto{50.634}{61.404}
\emmoveto{50.634}{61.394}
\emlineto{51.447}{61.968}
\emmoveto{51.447}{61.958}
\emlineto{52.290}{62.517}
\emmoveto{52.290}{62.507}
\emlineto{53.159}{63.051}
\emmoveto{53.159}{63.041}
\emlineto{54.055}{63.570}
\emmoveto{54.055}{63.560}
\emlineto{54.977}{64.072}
\emmoveto{54.977}{64.062}
\emlineto{55.924}{64.557}
\emmoveto{55.924}{64.547}
\emlineto{56.895}{65.026}
\emmoveto{56.895}{65.016}
\emlineto{57.889}{65.477}
\emmoveto{57.889}{65.467}
\emlineto{58.905}{65.910}
\emmoveto{58.905}{65.900}
\emlineto{59.943}{66.324}
\emmoveto{59.943}{66.314}
\emlineto{61.002}{66.721}
\emmoveto{61.002}{66.711}
\emlineto{62.080}{67.098}
\emmoveto{62.080}{67.088}
\emlineto{63.177}{67.456}
\emmoveto{63.177}{67.446}
\emlineto{64.291}{67.794}
\emmoveto{64.291}{67.784}
\emlineto{65.422}{68.112}
\emmoveto{65.422}{68.102}
\emlineto{66.568}{68.410}
\emmoveto{66.568}{68.400}
\emlineto{67.729}{68.687}
\emmoveto{67.729}{68.677}
\emlineto{68.903}{68.943}
\emmoveto{68.903}{68.933}
\emlineto{70.090}{69.179}
\emmoveto{70.090}{69.169}
\emlineto{71.289}{69.393}
\emmoveto{71.289}{69.383}
\emlineto{72.497}{69.586}
\emmoveto{72.497}{69.576}
\emlineto{73.715}{69.757}
\emmoveto{73.715}{69.747}
\emlineto{74.941}{69.907}
\emmoveto{74.941}{69.897}
\emlineto{76.174}{70.035}
\emmoveto{76.174}{70.025}
\emlineto{77.413}{70.140}
\emmoveto{77.413}{70.130}
\emlineto{78.656}{70.224}
\emmoveto{78.656}{70.214}
\emlineto{79.903}{70.285}
\emmoveto{79.903}{70.275}
\emlineto{81.153}{70.324}
\emmoveto{81.153}{70.314}
\emlineto{82.404}{70.341}
\emmoveto{82.404}{70.331}
\emlineto{83.656}{70.336}
\emmoveto{83.656}{70.326}
\emlineto{84.906}{70.308}
\emmoveto{84.906}{70.298}
\emlineto{86.155}{70.259}
\emmoveto{86.155}{70.249}
\emlineto{87.401}{70.186}
\emmoveto{87.401}{70.176}
\emlineto{88.642}{70.092}
\emmoveto{88.642}{70.082}
\emlineto{89.878}{69.976}
\emmoveto{89.878}{69.966}
\emlineto{91.108}{69.838}
\emmoveto{91.108}{69.828}
\emlineto{92.330}{69.678}
\emmoveto{92.330}{69.668}
\emlineto{93.544}{69.496}
\emmoveto{93.544}{69.486}
\emlineto{94.747}{69.293}
\emmoveto{94.747}{69.283}
\emlineto{95.940}{69.069}
\emmoveto{95.940}{69.059}
\emlineto{97.121}{68.823}
\emmoveto{97.121}{68.813}
\emlineto{98.289}{68.556}
\emmoveto{98.289}{68.546}
\emlineto{99.444}{68.269}
\emmoveto{99.444}{68.259}
\emlineto{100.583}{67.962}
\emmoveto{100.583}{67.952}
\emlineto{101.706}{67.634}
\emmoveto{101.706}{67.624}
\emlineto{102.812}{67.286}
\emmoveto{102.812}{67.276}
\emlineto{103.899}{66.919}
\emmoveto{103.899}{66.909}
\emlineto{104.968}{66.533}
\emmoveto{104.968}{66.523}
\emlineto{106.017}{66.128}
\emmoveto{106.017}{66.118}
\emlineto{107.045}{65.704}
\emmoveto{107.045}{65.694}
\emlineto{108.051}{65.262}
\emmoveto{108.051}{65.252}
\emlineto{109.034}{64.803}
\emmoveto{109.034}{64.793}
\emlineto{109.993}{64.326}
\emmoveto{109.993}{64.316}
\emlineto{110.928}{63.833}
\emmoveto{110.928}{63.823}
\emlineto{111.838}{63.323}
\emmoveto{111.838}{63.313}
\emlineto{112.721}{62.797}
\emmoveto{112.721}{62.787}
\emlineto{113.578}{62.255}
\emmoveto{113.578}{62.245}
\emlineto{114.406}{61.699}
\emmoveto{114.406}{61.689}
\emlineto{115.206}{61.128}
\emmoveto{115.206}{61.118}
\emlineto{115.977}{60.543}
\emmoveto{115.977}{60.533}
\emlineto{116.718}{59.945}
\emmoveto{116.718}{59.935}
\emlineto{117.429}{59.334}
\emmoveto{117.429}{59.324}
\emlineto{118.108}{58.710}
\emmoveto{118.108}{58.700}
\emlineto{118.755}{58.075}
\emmoveto{118.755}{58.065}
\emlineto{119.370}{57.428}
\emmoveto{119.370}{57.418}
\emlineto{119.952}{56.771}
\emmoveto{119.952}{56.761}
\emlineto{120.500}{56.103}
\emmoveto{120.500}{56.093}
\emlineto{121.015}{55.427}
\emmoveto{121.015}{55.417}
\emlineto{121.495}{54.741}
\emmoveto{121.495}{54.731}
\emlineto{121.940}{54.047}
\emmoveto{121.940}{54.037}
\emlineto{122.350}{53.346}
\emmoveto{122.350}{53.336}
\emlineto{122.724}{52.637}
\emmoveto{122.724}{52.627}
\emlineto{123.062}{51.922}
\emmoveto{123.062}{51.912}
\emlineto{123.364}{51.202}
\emmoveto{123.364}{51.192}
\emlineto{123.629}{50.476}
\emmoveto{123.629}{50.466}
\emlineto{123.858}{49.746}
\emmoveto{123.858}{49.736}
\emlineto{124.049}{49.012}
\emmoveto{124.049}{49.002}
\emlineto{124.203}{48.276}
\emmoveto{124.203}{48.266}
\emlineto{124.320}{47.537}
\emmoveto{124.320}{47.527}
\emlineto{124.400}{46.796}
\emmoveto{124.400}{46.786}
\emlineto{124.442}{46.054}
\emmoveto{124.442}{46.044}
\emlineto{124.449}{45.307}
\emmoveto{124.449}{45.297}
\emlineto{124.442}{44.529}
\emmoveto{124.442}{44.519}
\emlineto{124.397}{43.747}
\emmoveto{124.397}{43.737}
\emlineto{124.314}{42.965}
\emmoveto{124.314}{42.955}
\emlineto{124.191}{42.186}
\emmoveto{124.191}{42.176}
\emlineto{124.028}{41.410}
\emmoveto{124.028}{41.400}
\emlineto{123.827}{40.636}
\emmoveto{123.827}{40.626}
\emlineto{123.586}{39.867}
\emmoveto{123.586}{39.857}
\emlineto{123.307}{39.102}
\emmoveto{123.307}{39.092}
\emlineto{122.989}{38.342}
\emmoveto{122.989}{38.332}
\emlineto{122.632}{37.589}
\emmoveto{122.632}{37.579}
\emlineto{122.238}{36.842}
\emmoveto{122.238}{36.832}
\emlineto{121.806}{36.103}
\emmoveto{121.806}{36.093}
\emlineto{121.337}{35.371}
\emmoveto{121.337}{35.361}
\emlineto{120.831}{34.648}
\emmoveto{120.831}{34.638}
\emlineto{120.289}{33.935}
\emmoveto{120.289}{33.925}
\emlineto{119.711}{33.231}
\emmoveto{119.711}{33.221}
\emlineto{119.098}{32.538}
\emmoveto{119.098}{32.528}
\emlineto{118.450}{31.857}
\emmoveto{118.450}{31.847}
\emlineto{117.768}{31.187}
\emmoveto{117.768}{31.177}
\emlineto{117.052}{30.530}
\emmoveto{117.052}{30.520}
\emlineto{116.304}{29.885}
\emmoveto{116.304}{29.875}
\emlineto{115.523}{29.254}
\emmoveto{115.523}{29.244}
\emlineto{114.710}{28.638}
\emmoveto{114.710}{28.628}
\emlineto{113.867}{28.036}
\emmoveto{113.867}{28.026}
\emlineto{112.994}{27.449}
\emmoveto{112.994}{27.439}
\emlineto{112.091}{26.879}
\emmoveto{112.091}{26.869}
\emlineto{111.160}{26.324}
\emmoveto{111.160}{26.314}
\emlineto{110.202}{25.786}
\emmoveto{110.202}{25.776}
\emlineto{109.216}{25.266}
\emmoveto{109.216}{25.256}
\emlineto{108.205}{24.764}
\emmoveto{108.205}{24.754}
\emlineto{107.169}{24.279}
\emmoveto{107.169}{24.269}
\emlineto{106.109}{23.814}
\emmoveto{106.109}{23.804}
\emlineto{105.025}{23.367}
\emmoveto{105.025}{23.357}
\emlineto{103.920}{22.940}
\emmoveto{103.920}{22.930}
\emlineto{102.793}{22.533}
\emmoveto{102.793}{22.523}
\emlineto{101.647}{22.145}
\emmoveto{101.647}{22.135}
\emlineto{100.481}{21.779}
\emmoveto{100.481}{21.769}
\emlineto{99.297}{21.433}
\emmoveto{99.297}{21.423}
\emlineto{98.097}{21.109}
\emmoveto{98.097}{21.099}
\emlineto{96.880}{20.806}
\emmoveto{96.880}{20.796}
\emlineto{95.649}{20.525}
\emshow{24.980}{17.700}{}
\emshow{1.000}{10.000}{-6.00e-1}
\emshow{1.000}{17.000}{-4.80e-1}
\emshow{1.000}{24.000}{-3.60e-1}
\emshow{1.000}{31.000}{-2.40e-1}
\emshow{1.000}{38.000}{-1.20e-1}
\emshow{1.000}{45.000}{0.00e0}
\emshow{1.000}{52.000}{1.20e-1}
\emshow{1.000}{59.000}{2.40e-1}
\emshow{1.000}{66.000}{3.60e-1}
\emshow{1.000}{73.000}{4.80e-1}
\emshow{1.000}{80.000}{6.00e-1}
\emshow{12.000}{5.000}{-6.00e-1}
\emshow{23.800}{5.000}{-4.80e-1}
\emshow{35.600}{5.000}{-3.60e-1}
\emshow{47.400}{5.000}{-2.40e-1}
\emshow{59.200}{5.000}{-1.20e-1}
\emshow{71.000}{5.000}{0.00e0}
\emshow{82.800}{5.000}{1.20e-1}
\emshow{94.600}{5.000}{2.40e-1}
\emshow{106.400}{5.000}{3.60e-1}
\emshow{118.200}{5.000}{4.80e-1}
\emshow{130.000}{5.000}{6.00e-1}
{\centerline {\bf Fig. B.1.}}
\eject
\newcount\numpoint
\newcount\numpointo
\numpoint=1 \numpointo=1
\def\emmoveto#1#2{\offinterlineskip
\hbox to 0 true cm{\vbox to 0
true cm{\vskip - #2 true mm
\hskip #1 true mm \special{em:point
\the\numpoint}\vss}\hss}
\numpointo=\numpoint
\global\advance \numpoint by 1}
\def\emlineto#1#2{\offinterlineskip
\hbox to 0 true cm{\vbox to 0
true cm{\vskip - #2 true mm
\hskip #1 true mm \special{em:point
\the\numpoint}\vss}\hss}
\special{em:line
\the\numpointo,\the\numpoint}
\numpointo=\numpoint
\global\advance \numpoint by 1}
\def\emshow#1#2#3{\offinterlineskip
\hbox to 0 true cm{\vbox to 0
true cm{\vskip - #2 true mm
\hskip #1 true mm \vbox to 0
true cm{\vss\hbox{#3\hss
}}\vss}\hss}}
\special{em:linewidth 0.8pt}

\vrule width 0 mm height                0 mm depth 90.000 true mm

\special{em:linewidth 0.8pt}
\emmoveto{130.000}{10.000}
\emlineto{12.000}{10.000}
\emlineto{12.000}{80.000}
\emmoveto{71.000}{10.000}
\emlineto{71.000}{80.000}
\emmoveto{12.000}{45.000}
\emlineto{130.000}{45.000}
\emmoveto{130.000}{10.000}
\emlineto{130.000}{80.000}
\emlineto{12.000}{80.000}
\emlineto{12.000}{10.000}
\emlineto{130.000}{10.000}
\special{em:linewidth 0.4pt}
\emmoveto{12.000}{17.000}
\emlineto{130.000}{17.000}
\emmoveto{12.000}{24.000}
\emlineto{130.000}{24.000}
\emmoveto{12.000}{31.000}
\emlineto{130.000}{31.000}
\emmoveto{12.000}{38.000}
\emlineto{130.000}{38.000}
\emmoveto{12.000}{45.000}
\emlineto{130.000}{45.000}
\emmoveto{12.000}{52.000}
\emlineto{130.000}{52.000}
\emmoveto{12.000}{59.000}
\emlineto{130.000}{59.000}
\emmoveto{12.000}{66.000}
\emlineto{130.000}{66.000}
\emmoveto{12.000}{73.000}
\emlineto{130.000}{73.000}
\emmoveto{23.800}{10.000}
\emlineto{23.800}{80.000}
\emmoveto{35.600}{10.000}
\emlineto{35.600}{80.000}
\emmoveto{47.400}{10.000}
\emlineto{47.400}{80.000}
\emmoveto{59.200}{10.000}
\emlineto{59.200}{80.000}
\emmoveto{71.000}{10.000}
\emlineto{71.000}{80.000}
\emmoveto{82.800}{10.000}
\emlineto{82.800}{80.000}
\emmoveto{94.600}{10.000}
\emlineto{94.600}{80.000}
\emmoveto{106.400}{10.000}
\emlineto{106.400}{80.000}
\emmoveto{118.200}{10.000}
\emlineto{118.200}{80.000}
\special{em:linewidth 0.8pt}
\emmoveto{71.000}{45.016}
\emlineto{71.000}{45.273}
\emmoveto{71.000}{45.263}
\emlineto{71.000}{45.420}
\emmoveto{71.000}{45.410}
\emlineto{71.000}{45.601}
\emmoveto{71.000}{45.591}
\emlineto{71.003}{45.796}
\emmoveto{71.003}{45.786}
\emlineto{71.013}{45.992}
\emmoveto{71.013}{45.982}
\emlineto{71.029}{46.187}
\emmoveto{71.029}{46.177}
\emlineto{71.050}{46.382}
\emmoveto{71.050}{46.372}
\emlineto{71.078}{46.577}
\emmoveto{71.078}{46.567}
\emlineto{71.113}{46.772}
\emmoveto{71.113}{46.762}
\emlineto{71.153}{46.966}
\emmoveto{71.153}{46.956}
\emlineto{71.199}{47.160}
\emmoveto{71.199}{47.150}
\emlineto{71.252}{47.353}
\emmoveto{71.252}{47.343}
\emlineto{71.311}{47.546}
\emmoveto{71.311}{47.536}
\emlineto{71.375}{47.737}
\emmoveto{71.375}{47.727}
\emlineto{71.446}{47.929}
\emmoveto{71.446}{47.919}
\emlineto{71.523}{48.119}
\emmoveto{71.523}{48.109}
\emlineto{71.606}{48.308}
\emmoveto{71.606}{48.298}
\emlineto{71.695}{48.497}
\emmoveto{71.695}{48.487}
\emlineto{71.789}{48.684}
\emmoveto{71.789}{48.674}
\emlineto{71.890}{48.870}
\emmoveto{71.890}{48.860}
\emlineto{71.996}{49.056}
\emmoveto{71.996}{49.046}
\emlineto{72.109}{49.239}
\emmoveto{72.109}{49.229}
\emlineto{72.227}{49.422}
\emmoveto{72.227}{49.412}
\emlineto{72.351}{49.603}
\emmoveto{72.351}{49.593}
\emlineto{72.480}{49.783}
\emmoveto{72.480}{49.773}
\emlineto{72.616}{49.962}
\emmoveto{72.616}{49.952}
\emlineto{72.757}{50.139}
\emmoveto{72.757}{50.129}
\emlineto{72.903}{50.314}
\emmoveto{72.903}{50.304}
\emlineto{73.055}{50.488}
\emmoveto{73.055}{50.478}
\emlineto{73.212}{50.660}
\emmoveto{73.212}{50.650}
\emlineto{73.375}{50.830}
\emmoveto{73.375}{50.820}
\emlineto{73.543}{50.998}
\emmoveto{73.543}{50.988}
\emlineto{73.717}{51.164}
\emmoveto{73.717}{51.154}
\emlineto{73.896}{51.329}
\emmoveto{73.896}{51.319}
\emlineto{74.079}{51.491}
\emmoveto{74.079}{51.481}
\emlineto{74.268}{51.652}
\emmoveto{74.268}{51.642}
\emlineto{74.462}{51.810}
\emmoveto{74.462}{51.800}
\emlineto{74.661}{51.966}
\emmoveto{74.661}{51.956}
\emlineto{74.865}{52.120}
\emmoveto{74.865}{52.110}
\emlineto{75.074}{52.271}
\emmoveto{75.074}{52.261}
\emlineto{75.287}{52.420}
\emmoveto{75.287}{52.410}
\emlineto{75.505}{52.567}
\emmoveto{75.505}{52.557}
\emlineto{75.728}{52.712}
\emmoveto{75.728}{52.702}
\emlineto{75.955}{52.853}
\emmoveto{75.955}{52.843}
\emlineto{76.187}{52.993}
\emmoveto{76.187}{52.983}
\emlineto{76.423}{53.129}
\emmoveto{76.423}{53.119}
\emlineto{76.663}{53.263}
\emmoveto{76.663}{53.253}
\emlineto{76.908}{53.395}
\emmoveto{76.908}{53.385}
\emlineto{77.156}{53.523}
\emmoveto{77.156}{53.513}
\emlineto{77.409}{53.649}
\emmoveto{77.409}{53.639}
\emlineto{77.666}{53.772}
\emmoveto{77.666}{53.762}
\emlineto{77.926}{53.892}
\emmoveto{77.926}{53.882}
\emlineto{78.190}{54.009}
\emmoveto{78.190}{53.999}
\emlineto{78.458}{54.123}
\emmoveto{78.458}{54.113}
\emlineto{78.729}{54.234}
\emmoveto{78.729}{54.224}
\emlineto{79.004}{54.343}
\emmoveto{79.004}{54.333}
\emlineto{79.282}{54.448}
\emmoveto{79.282}{54.438}
\emlineto{79.564}{54.550}
\emmoveto{79.564}{54.540}
\emlineto{79.849}{54.648}
\emmoveto{79.849}{54.638}
\emlineto{80.136}{54.744}
\emmoveto{80.136}{54.734}
\emlineto{80.427}{54.837}
\emmoveto{80.427}{54.827}
\emlineto{80.721}{54.926}
\emmoveto{80.721}{54.916}
\emlineto{81.017}{55.012}
\emmoveto{81.017}{55.002}
\emlineto{81.316}{55.094}
\emmoveto{81.316}{55.084}
\emlineto{81.617}{55.173}
\emmoveto{81.617}{55.163}
\emlineto{81.921}{55.249}
\emmoveto{81.921}{55.239}
\emlineto{82.228}{55.322}
\emmoveto{82.228}{55.312}
\emlineto{82.536}{55.391}
\emmoveto{82.536}{55.381}
\emlineto{82.847}{55.456}
\emmoveto{82.847}{55.446}
\emlineto{83.160}{55.519}
\emmoveto{83.160}{55.509}
\emlineto{83.474}{55.577}
\emmoveto{83.474}{55.567}
\emlineto{83.791}{55.632}
\emmoveto{83.791}{55.622}
\emlineto{84.109}{55.684}
\emmoveto{84.109}{55.674}
\emlineto{84.429}{55.732}
\emmoveto{84.429}{55.722}
\emlineto{84.750}{55.777}
\emmoveto{84.750}{55.767}
\emlineto{85.072}{55.818}
\emmoveto{85.072}{55.808}
\emlineto{85.396}{55.855}
\emmoveto{85.396}{55.845}
\emlineto{85.721}{55.889}
\emmoveto{85.721}{55.879}
\emlineto{86.047}{55.919}
\emmoveto{86.047}{55.909}
\emlineto{86.374}{55.945}
\emmoveto{86.374}{55.935}
\emlineto{86.701}{55.968}
\emmoveto{86.701}{55.958}
\emlineto{87.030}{55.987}
\emmoveto{87.030}{55.977}
\emlineto{87.358}{56.003}
\emmoveto{87.358}{55.993}
\emlineto{87.688}{56.015}
\emmoveto{87.688}{56.005}
\emlineto{88.017}{56.023}
\emmoveto{88.017}{56.013}
\emlineto{88.347}{56.027}
\emmoveto{88.347}{56.017}
\emlineto{88.677}{56.028}
\emmoveto{88.677}{56.018}
\emlineto{89.006}{56.025}
\emmoveto{89.006}{56.015}
\emlineto{89.336}{56.019}
\emmoveto{89.336}{56.009}
\emlineto{89.665}{56.009}
\emmoveto{89.665}{55.999}
\emlineto{89.994}{55.995}
\emmoveto{89.994}{55.985}
\emlineto{90.323}{55.978}
\emmoveto{90.323}{55.968}
\emlineto{90.651}{55.956}
\emmoveto{90.651}{55.946}
\emlineto{90.978}{55.932}
\emmoveto{90.978}{55.922}
\emlineto{91.304}{55.903}
\emmoveto{91.304}{55.893}
\emlineto{91.630}{55.871}
\emmoveto{91.630}{55.861}
\emlineto{91.954}{55.836}
\emmoveto{91.954}{55.826}
\emlineto{92.277}{55.797}
\emmoveto{92.277}{55.787}
\emlineto{92.599}{55.754}
\emmoveto{92.599}{55.744}
\emlineto{92.919}{55.707}
\emmoveto{92.919}{55.697}
\emlineto{93.238}{55.657}
\emmoveto{93.238}{55.647}
\emlineto{93.556}{55.604}
\emmoveto{93.556}{55.594}
\emlineto{93.871}{55.547}
\emmoveto{93.871}{55.537}
\emlineto{94.185}{55.487}
\emmoveto{94.185}{55.477}
\emlineto{94.496}{55.423}
\emmoveto{94.496}{55.413}
\emlineto{94.806}{55.355}
\emmoveto{94.806}{55.345}
\emlineto{95.113}{55.284}
\emmoveto{95.113}{55.274}
\emlineto{95.419}{55.210}
\emmoveto{95.419}{55.200}
\emlineto{95.721}{55.132}
\emmoveto{95.721}{55.122}
\emlineto{96.022}{55.051}
\emmoveto{96.022}{55.041}
\emlineto{96.319}{54.967}
\emmoveto{96.319}{54.957}
\emlineto{96.614}{54.880}
\emmoveto{96.614}{54.870}
\emlineto{96.906}{54.789}
\emmoveto{96.906}{54.779}
\emlineto{97.195}{54.695}
\emmoveto{97.195}{54.685}
\emlineto{97.481}{54.597}
\emmoveto{97.481}{54.587}
\emlineto{97.764}{54.497}
\emmoveto{97.764}{54.487}
\emlineto{98.044}{54.393}
\emmoveto{98.044}{54.383}
\emlineto{98.321}{54.286}
\emmoveto{98.321}{54.276}
\emlineto{98.594}{54.177}
\emmoveto{98.594}{54.167}
\emlineto{98.863}{54.064}
\emmoveto{98.863}{54.054}
\emlineto{99.129}{53.948}
\emmoveto{99.129}{53.938}
\emlineto{99.391}{53.830}
\emmoveto{99.391}{53.820}
\emlineto{99.650}{53.708}
\emmoveto{99.650}{53.698}
\emlineto{99.904}{53.584}
\emmoveto{99.904}{53.574}
\emlineto{100.155}{53.456}
\emmoveto{100.155}{53.446}
\emlineto{100.401}{53.326}
\emmoveto{100.401}{53.316}
\emlineto{100.644}{53.194}
\emmoveto{100.644}{53.184}
\emlineto{100.882}{53.058}
\emmoveto{100.882}{53.048}
\emlineto{101.115}{52.920}
\emmoveto{101.115}{52.910}
\emlineto{101.345}{52.780}
\emmoveto{101.345}{52.770}
\emlineto{101.570}{52.636}
\emmoveto{101.570}{52.626}
\emlineto{101.790}{52.491}
\emmoveto{101.790}{52.481}
\emlineto{102.006}{52.343}
\emmoveto{102.006}{52.333}
\emlineto{102.217}{52.192}
\emmoveto{102.217}{52.182}
\emlineto{102.423}{52.040}
\emmoveto{102.423}{52.030}
\emlineto{102.624}{51.885}
\emmoveto{102.624}{51.875}
\emlineto{102.821}{51.727}
\emmoveto{102.821}{51.717}
\emlineto{103.012}{51.568}
\emmoveto{103.012}{51.558}
\emlineto{103.198}{51.407}
\emmoveto{103.198}{51.397}
\emlineto{103.379}{51.243}
\emmoveto{103.379}{51.233}
\emlineto{103.555}{51.078}
\emmoveto{103.555}{51.068}
\emlineto{103.726}{50.910}
\emmoveto{103.726}{50.900}
\emlineto{103.891}{50.741}
\emmoveto{103.891}{50.731}
\emlineto{104.051}{50.570}
\emmoveto{104.051}{50.560}
\emlineto{104.206}{50.397}
\emmoveto{104.206}{50.387}
\emlineto{104.355}{50.223}
\emmoveto{104.355}{50.213}
\emlineto{104.499}{50.046}
\emmoveto{104.499}{50.036}
\emlineto{104.637}{49.869}
\emmoveto{104.637}{49.859}
\emlineto{104.769}{49.690}
\emmoveto{104.769}{49.680}
\emlineto{104.896}{49.509}
\emmoveto{104.896}{49.499}
\emlineto{105.016}{49.327}
\emmoveto{105.016}{49.317}
\emlineto{105.132}{49.144}
\emmoveto{105.132}{49.134}
\emlineto{105.241}{48.959}
\emmoveto{105.241}{48.949}
\emlineto{105.344}{48.773}
\emmoveto{105.344}{48.763}
\emlineto{105.442}{48.586}
\emmoveto{105.442}{48.576}
\emlineto{105.533}{48.398}
\emmoveto{105.533}{48.388}
\emlineto{105.619}{48.209}
\emmoveto{105.619}{48.199}
\emlineto{105.699}{48.019}
\emmoveto{105.699}{48.009}
\emlineto{105.772}{47.829}
\emmoveto{105.772}{47.819}
\emlineto{105.840}{47.637}
\emmoveto{105.840}{47.627}
\emlineto{105.902}{47.445}
\emmoveto{105.902}{47.435}
\emlineto{105.957}{47.252}
\emmoveto{105.957}{47.242}
\emlineto{106.006}{47.059}
\emmoveto{106.006}{47.049}
\emlineto{106.050}{46.865}
\emmoveto{106.050}{46.855}
\emlineto{106.087}{46.670}
\emmoveto{106.087}{46.660}
\emlineto{106.118}{46.476}
\emmoveto{106.118}{46.466}
\emlineto{106.143}{46.280}
\emmoveto{106.143}{46.270}
\emlineto{106.161}{46.085}
\emmoveto{106.161}{46.075}
\emlineto{106.174}{45.890}
\emmoveto{106.174}{45.880}
\emlineto{106.180}{45.694}
\emmoveto{106.180}{45.684}
\emlineto{106.181}{45.494}
\emmoveto{106.181}{45.484}
\emlineto{106.181}{45.266}
\emmoveto{106.181}{45.256}
\emlineto{106.181}{45.006}
\emmoveto{106.181}{44.996}
\emlineto{106.176}{44.729}
\emmoveto{106.176}{44.719}
\emlineto{106.163}{44.452}
\emmoveto{106.163}{44.442}
\emlineto{106.141}{44.176}
\emmoveto{106.141}{44.166}
\emlineto{106.110}{43.900}
\emmoveto{106.110}{43.890}
\emlineto{106.071}{43.624}
\emmoveto{106.071}{43.614}
\emlineto{106.023}{43.349}
\emmoveto{106.023}{43.339}
\emlineto{105.966}{43.074}
\emmoveto{105.966}{43.064}
\emlineto{105.900}{42.800}
\emmoveto{105.900}{42.790}
\emlineto{105.826}{42.527}
\emmoveto{105.826}{42.517}
\emlineto{105.743}{42.255}
\emmoveto{105.743}{42.245}
\emlineto{105.652}{41.984}
\emmoveto{105.652}{41.974}
\emlineto{105.552}{41.713}
\emmoveto{105.552}{41.703}
\emlineto{105.444}{41.444}
\emmoveto{105.444}{41.434}
\emlineto{105.327}{41.176}
\emmoveto{105.327}{41.166}
\emlineto{105.201}{40.910}
\emmoveto{105.201}{40.900}
\emlineto{105.068}{40.645}
\emmoveto{105.068}{40.635}
\emlineto{104.925}{40.381}
\emmoveto{104.925}{40.371}
\emlineto{104.775}{40.119}
\emmoveto{104.775}{40.109}
\emlineto{104.616}{39.859}
\emmoveto{104.616}{39.849}
\emlineto{104.449}{39.601}
\emmoveto{104.449}{39.591}
\emlineto{104.274}{39.344}
\emmoveto{104.274}{39.334}
\emlineto{104.091}{39.090}
\emmoveto{104.091}{39.080}
\emlineto{103.900}{38.837}
\emmoveto{103.900}{38.827}
\emlineto{103.701}{38.587}
\emmoveto{103.701}{38.577}
\emlineto{103.494}{38.339}
\emmoveto{103.494}{38.329}
\emlineto{103.279}{38.094}
\emmoveto{103.279}{38.084}
\emlineto{103.057}{37.850}
\emmoveto{103.057}{37.840}
\emlineto{102.827}{37.610}
\emmoveto{102.827}{37.600}
\emlineto{102.589}{37.372}
\emmoveto{102.589}{37.362}
\emlineto{102.344}{37.136}
\emmoveto{102.344}{37.126}
\emlineto{102.092}{36.904}
\emmoveto{102.092}{36.894}
\emlineto{101.832}{36.674}
\emmoveto{101.832}{36.664}
\emlineto{101.565}{36.447}
\emmoveto{101.565}{36.437}
\emlineto{101.290}{36.223}
\emmoveto{101.290}{36.213}
\emlineto{101.009}{36.002}
\emmoveto{101.009}{35.992}
\emlineto{100.721}{35.785}
\emmoveto{100.721}{35.775}
\emlineto{100.426}{35.571}
\emmoveto{100.426}{35.561}
\emlineto{100.124}{35.359}
\emmoveto{100.124}{35.349}
\emlineto{99.816}{35.152}
\emmoveto{99.816}{35.142}
\emlineto{99.501}{34.948}
\emmoveto{99.501}{34.938}
\emlineto{99.180}{34.747}
\emmoveto{99.180}{34.737}
\emlineto{98.852}{34.550}
\emmoveto{98.852}{34.540}
\emlineto{98.519}{34.357}
\emmoveto{98.519}{34.347}
\emlineto{98.179}{34.167}
\emmoveto{98.179}{34.157}
\emlineto{97.833}{33.981}
\emmoveto{97.833}{33.971}
\emlineto{97.482}{33.799}
\emmoveto{97.482}{33.789}
\emlineto{97.125}{33.621}
\emmoveto{97.125}{33.611}
\emlineto{96.762}{33.448}
\emmoveto{96.762}{33.438}
\emlineto{96.394}{33.278}
\emmoveto{96.394}{33.268}
\emlineto{96.020}{33.112}
\emmoveto{96.020}{33.102}
\emlineto{95.642}{32.950}
\emmoveto{95.642}{32.940}
\emlineto{95.258}{32.793}
\emmoveto{95.258}{32.783}
\emlineto{94.870}{32.640}
\emmoveto{94.870}{32.630}
\emlineto{94.476}{32.491}
\emmoveto{94.476}{32.481}
\emlineto{94.078}{32.347}
\emmoveto{94.078}{32.337}
\emlineto{93.676}{32.207}
\emmoveto{93.676}{32.197}
\emlineto{93.269}{32.072}
\emmoveto{93.269}{32.062}
\emlineto{92.858}{31.941}
\emmoveto{92.858}{31.931}
\emlineto{92.443}{31.815}
\emmoveto{92.443}{31.805}
\emlineto{92.024}{31.693}
\emmoveto{92.024}{31.683}
\emlineto{91.601}{31.576}
\emmoveto{91.601}{31.566}
\emlineto{91.175}{31.464}
\emmoveto{91.175}{31.454}
\emlineto{90.745}{31.357}
\emmoveto{90.745}{31.347}
\emlineto{90.312}{31.254}
\emmoveto{90.312}{31.244}
\emlineto{89.875}{31.156}
\emmoveto{89.875}{31.146}
\emlineto{89.436}{31.063}
\emmoveto{89.436}{31.053}
\emlineto{88.994}{30.975}
\emmoveto{88.994}{30.965}
\emlineto{88.549}{30.892}
\emmoveto{88.549}{30.882}
\emlineto{88.101}{30.814}
\emmoveto{88.101}{30.804}
\emlineto{87.651}{30.741}
\emmoveto{87.651}{30.731}
\emlineto{87.199}{30.673}
\emmoveto{87.199}{30.663}
\emlineto{86.745}{30.610}
\emmoveto{86.745}{30.600}
\emlineto{86.289}{30.552}
\emmoveto{86.289}{30.542}
\emlineto{85.831}{30.499}
\emmoveto{85.831}{30.489}
\emlineto{85.372}{30.451}
\emmoveto{85.372}{30.441}
\emlineto{84.911}{30.409}
\emmoveto{84.911}{30.399}
\emlineto{84.449}{30.371}
\emmoveto{84.449}{30.361}
\emlineto{83.986}{30.339}
\emmoveto{83.986}{30.329}
\emlineto{83.522}{30.311}
\emmoveto{83.522}{30.301}
\emlineto{83.057}{30.289}
\emmoveto{83.057}{30.279}
\emlineto{82.591}{30.272}
\emmoveto{82.591}{30.262}
\emlineto{82.125}{30.261}
\emmoveto{82.125}{30.251}
\emlineto{81.659}{30.254}
\emmoveto{81.659}{30.244}
\emlineto{81.192}{30.243}
\emlineto{81.192}{30.253}
\emmoveto{81.192}{30.243}
\emlineto{80.726}{30.257}
\emmoveto{80.726}{30.247}
\emlineto{80.260}{30.266}
\emmoveto{80.260}{30.256}
\emlineto{79.794}{30.280}
\emmoveto{79.794}{30.270}
\emlineto{79.329}{30.299}
\emmoveto{79.329}{30.289}
\emlineto{78.864}{30.324}
\emmoveto{78.864}{30.314}
\emlineto{78.400}{30.354}
\emmoveto{78.400}{30.344}
\emlineto{77.938}{30.388}
\emmoveto{77.938}{30.378}
\emlineto{77.476}{30.428}
\emmoveto{77.476}{30.418}
\emlineto{77.016}{30.474}
\emmoveto{77.016}{30.464}
\emlineto{76.557}{30.524}
\emmoveto{76.557}{30.514}
\emlineto{76.100}{30.579}
\emmoveto{76.100}{30.569}
\emlineto{75.645}{30.640}
\emmoveto{75.645}{30.630}
\emlineto{75.192}{30.705}
\emmoveto{75.192}{30.695}
\emlineto{74.741}{30.776}
\emmoveto{74.741}{30.766}
\emlineto{74.292}{30.851}
\emmoveto{74.292}{30.841}
\emlineto{73.846}{30.932}
\emmoveto{73.846}{30.922}
\emlineto{73.402}{31.017}
\emmoveto{73.402}{31.007}
\emlineto{72.961}{31.107}
\emmoveto{72.961}{31.097}
\emlineto{72.523}{31.203}
\emmoveto{72.523}{31.193}
\emlineto{72.089}{31.303}
\emmoveto{72.089}{31.293}
\emlineto{71.657}{31.408}
\emmoveto{71.657}{31.398}
\emlineto{71.229}{31.517}
\emmoveto{71.229}{31.507}
\emlineto{70.804}{31.632}
\emmoveto{70.804}{31.622}
\emlineto{70.383}{31.751}
\emmoveto{70.383}{31.741}
\emlineto{69.966}{31.875}
\emmoveto{69.966}{31.865}
\emlineto{69.553}{32.003}
\emmoveto{69.553}{31.993}
\emlineto{69.144}{32.136}
\emmoveto{69.144}{32.126}
\emlineto{68.739}{32.274}
\emmoveto{68.739}{32.264}
\emlineto{68.339}{32.416}
\emmoveto{68.339}{32.406}
\emlineto{67.943}{32.562}
\emmoveto{67.943}{32.552}
\emlineto{67.552}{32.713}
\emmoveto{67.552}{32.703}
\emlineto{67.166}{32.868}
\emmoveto{67.166}{32.858}
\emlineto{66.784}{33.027}
\emmoveto{66.784}{33.017}
\emlineto{66.408}{33.191}
\emmoveto{66.408}{33.181}
\emlineto{66.037}{33.359}
\emmoveto{66.037}{33.349}
\emlineto{65.672}{33.530}
\emmoveto{65.672}{33.520}
\emlineto{65.312}{33.706}
\emmoveto{65.312}{33.696}
\emlineto{64.957}{33.886}
\emmoveto{64.957}{33.876}
\emlineto{64.609}{34.070}
\emmoveto{64.609}{34.060}
\emlineto{64.266}{34.258}
\emmoveto{64.266}{34.248}
\emlineto{63.929}{34.449}
\emmoveto{63.929}{34.439}
\emlineto{63.598}{34.644}
\emmoveto{63.598}{34.634}
\emlineto{63.274}{34.843}
\emmoveto{63.274}{34.833}
\emlineto{62.955}{35.045}
\emmoveto{62.955}{35.035}
\emlineto{62.644}{35.251}
\emmoveto{62.644}{35.241}
\emlineto{62.339}{35.460}
\emmoveto{62.339}{35.450}
\emlineto{62.040}{35.673}
\emmoveto{62.040}{35.663}
\emlineto{61.748}{35.889}
\emmoveto{61.748}{35.879}
\emlineto{61.464}{36.108}
\emmoveto{61.464}{36.098}
\emlineto{61.186}{36.330}
\emmoveto{61.186}{36.320}
\emlineto{60.915}{36.555}
\emmoveto{60.915}{36.545}
\emlineto{60.651}{36.784}
\emmoveto{60.651}{36.774}
\emlineto{60.395}{37.015}
\emmoveto{60.395}{37.005}
\emlineto{60.146}{37.249}
\emmoveto{60.146}{37.239}
\emlineto{59.904}{37.486}
\emmoveto{59.904}{37.476}
\emlineto{59.670}{37.725}
\emmoveto{59.670}{37.715}
\emlineto{59.444}{37.967}
\emmoveto{59.444}{37.957}
\emlineto{59.225}{38.211}
\emmoveto{59.225}{38.201}
\emlineto{59.014}{38.458}
\emmoveto{59.014}{38.448}
\emlineto{58.811}{38.707}
\emmoveto{58.811}{38.697}
\emlineto{58.616}{38.958}
\emmoveto{58.616}{38.948}
\emlineto{58.428}{39.212}
\emmoveto{58.428}{39.202}
\emlineto{58.249}{39.467}
\emmoveto{58.249}{39.457}
\emlineto{58.078}{39.724}
\emmoveto{58.078}{39.714}
\emlineto{57.915}{39.984}
\emmoveto{57.915}{39.974}
\emlineto{57.760}{40.245}
\emmoveto{57.760}{40.235}
\emlineto{57.614}{40.507}
\emmoveto{57.614}{40.497}
\emlineto{57.476}{40.772}
\emmoveto{57.476}{40.762}
\emlineto{57.346}{41.038}
\emmoveto{57.346}{41.028}
\emlineto{57.225}{41.305}
\emmoveto{57.225}{41.295}
\emlineto{57.112}{41.573}
\emmoveto{57.112}{41.563}
\emlineto{57.007}{41.843}
\emmoveto{57.007}{41.833}
\emlineto{56.911}{42.114}
\emmoveto{56.911}{42.104}
\emlineto{56.824}{42.385}
\emmoveto{56.824}{42.375}
\emlineto{56.746}{42.658}
\emmoveto{56.746}{42.648}
\emlineto{56.676}{42.932}
\emmoveto{56.676}{42.922}
\emlineto{56.614}{43.206}
\emmoveto{56.614}{43.196}
\emlineto{56.562}{43.481}
\emmoveto{56.562}{43.471}
\emlineto{56.518}{43.756}
\emmoveto{56.518}{43.746}
\emlineto{56.482}{44.032}
\emmoveto{56.482}{44.022}
\emlineto{56.456}{44.309}
\emmoveto{56.456}{44.299}
\emlineto{56.438}{44.585}
\emmoveto{56.438}{44.575}
\emlineto{56.429}{44.862}
\emmoveto{56.429}{44.852}
\emlineto{56.428}{45.142}
\emmoveto{56.428}{45.132}
\emlineto{56.428}{45.450}
\emmoveto{56.428}{45.440}
\emlineto{56.429}{45.786}
\emmoveto{56.429}{45.776}
\emlineto{56.441}{46.125}
\emmoveto{56.441}{46.115}
\emlineto{56.463}{46.464}
\emmoveto{56.463}{46.454}
\emlineto{56.496}{46.802}
\emmoveto{56.496}{46.792}
\emlineto{56.539}{47.140}
\emmoveto{56.539}{47.130}
\emlineto{56.594}{47.477}
\emmoveto{56.594}{47.467}
\emlineto{56.658}{47.814}
\emmoveto{56.658}{47.804}
\emlineto{56.734}{48.150}
\emmoveto{56.734}{48.140}
\emlineto{56.820}{48.485}
\emmoveto{56.820}{48.475}
\emlineto{56.917}{48.819}
\emmoveto{56.917}{48.809}
\emlineto{57.024}{49.152}
\emmoveto{57.024}{49.142}
\emlineto{57.142}{49.483}
\emmoveto{57.142}{49.473}
\emlineto{57.270}{49.813}
\emmoveto{57.270}{49.803}
\emlineto{57.408}{50.142}
\emmoveto{57.408}{50.132}
\emlineto{57.557}{50.469}
\emmoveto{57.557}{50.459}
\emlineto{57.717}{50.795}
\emmoveto{57.717}{50.785}
\emlineto{57.886}{51.118}
\emmoveto{57.886}{51.108}
\emlineto{58.066}{51.440}
\emmoveto{58.066}{51.430}
\emlineto{58.256}{51.760}
\emmoveto{58.256}{51.750}
\emlineto{58.456}{52.077}
\emmoveto{58.456}{52.067}
\emlineto{58.666}{52.392}
\emmoveto{58.666}{52.382}
\emlineto{58.885}{52.705}
\emmoveto{58.885}{52.695}
\emlineto{59.115}{53.015}
\emmoveto{59.115}{53.005}
\emlineto{59.354}{53.323}
\emmoveto{59.354}{53.313}
\emlineto{59.604}{53.628}
\emmoveto{59.604}{53.618}
\emlineto{59.862}{53.930}
\emmoveto{59.862}{53.920}
\emlineto{60.131}{54.229}
\emmoveto{60.131}{54.219}
\emlineto{60.408}{54.526}
\emmoveto{60.408}{54.516}
\emlineto{60.695}{54.819}
\emmoveto{60.695}{54.809}
\emlineto{60.991}{55.108}
\emmoveto{60.991}{55.098}
\emlineto{61.297}{55.395}
\emmoveto{61.297}{55.385}
\emlineto{61.611}{55.678}
\emmoveto{61.611}{55.668}
\emlineto{61.934}{55.957}
\emmoveto{61.934}{55.947}
\emlineto{62.266}{56.233}
\emmoveto{62.266}{56.223}
\emlineto{62.607}{56.505}
\emmoveto{62.607}{56.495}
\emlineto{62.956}{56.773}
\emmoveto{62.956}{56.763}
\emlineto{63.313}{57.038}
\emmoveto{63.313}{57.028}
\emlineto{63.679}{57.298}
\emmoveto{63.679}{57.288}
\emlineto{64.053}{57.554}
\emmoveto{64.053}{57.544}
\emlineto{64.435}{57.806}
\emmoveto{64.435}{57.796}
\emlineto{64.825}{58.054}
\emmoveto{64.825}{58.044}
\emlineto{65.223}{58.297}
\emmoveto{65.223}{58.287}
\emlineto{65.628}{58.536}
\emmoveto{65.628}{58.526}
\emlineto{66.041}{58.770}
\emmoveto{66.041}{58.760}
\emlineto{66.461}{59.000}
\emmoveto{66.461}{58.990}
\emlineto{66.888}{59.224}
\emmoveto{66.888}{59.214}
\emlineto{67.323}{59.445}
\emmoveto{67.323}{59.435}
\emlineto{67.764}{59.660}
\emmoveto{67.764}{59.650}
\emlineto{68.212}{59.870}
\emmoveto{68.212}{59.860}
\emlineto{68.666}{60.075}
\emmoveto{68.666}{60.065}
\emlineto{69.127}{60.276}
\emmoveto{69.127}{60.266}
\emlineto{69.594}{60.471}
\emmoveto{69.594}{60.461}
\emlineto{70.068}{60.660}
\emmoveto{70.068}{60.650}
\emlineto{70.547}{60.845}
\emmoveto{70.547}{60.835}
\emlineto{71.032}{61.024}
\emmoveto{71.032}{61.014}
\emlineto{71.522}{61.198}
\emmoveto{71.522}{61.188}
\emlineto{72.018}{61.366}
\emmoveto{72.018}{61.356}
\emlineto{72.519}{61.529}
\emmoveto{72.519}{61.519}
\emlineto{73.025}{61.686}
\emmoveto{73.025}{61.676}
\emlineto{73.536}{61.837}
\emmoveto{73.536}{61.827}
\emlineto{74.052}{61.983}
\emmoveto{74.052}{61.973}
\emlineto{74.572}{62.123}
\emmoveto{74.572}{62.113}
\emlineto{75.097}{62.257}
\emmoveto{75.097}{62.247}
\emlineto{75.626}{62.385}
\emmoveto{75.626}{62.375}
\emlineto{76.158}{62.508}
\emmoveto{76.158}{62.498}
\emlineto{76.695}{62.624}
\emmoveto{76.695}{62.614}
\emlineto{77.235}{62.735}
\emmoveto{77.235}{62.725}
\emlineto{77.778}{62.839}
\emmoveto{77.778}{62.829}
\emlineto{78.325}{62.938}
\emmoveto{78.325}{62.928}
\emlineto{78.875}{63.030}
\emmoveto{78.875}{63.020}
\emlineto{79.427}{63.116}
\emmoveto{79.427}{63.106}
\emlineto{79.982}{63.196}
\emmoveto{79.982}{63.186}
\emlineto{80.540}{63.270}
\emmoveto{80.540}{63.260}
\emlineto{81.100}{63.337}
\emmoveto{81.100}{63.327}
\emlineto{81.662}{63.399}
\emmoveto{81.662}{63.389}
\emlineto{82.225}{63.454}
\emmoveto{82.225}{63.444}
\emlineto{82.791}{63.502}
\emmoveto{82.791}{63.492}
\emlineto{83.357}{63.545}
\emmoveto{83.357}{63.535}
\emlineto{83.925}{63.581}
\emmoveto{83.925}{63.571}
\emlineto{84.494}{63.611}
\emmoveto{84.494}{63.601}
\emlineto{85.064}{63.635}
\emmoveto{85.064}{63.625}
\emlineto{85.635}{63.652}
\emmoveto{85.635}{63.642}
\emlineto{86.206}{63.663}
\emmoveto{86.206}{63.653}
\emlineto{86.777}{63.667}
\emmoveto{86.777}{63.657}
\emlineto{87.348}{63.665}
\emmoveto{87.348}{63.655}
\emlineto{87.919}{63.657}
\emmoveto{87.919}{63.647}
\emlineto{88.490}{63.642}
\emmoveto{88.490}{63.632}
\emlineto{89.060}{63.622}
\emmoveto{89.060}{63.612}
\emlineto{89.630}{63.594}
\emmoveto{89.630}{63.584}
\emlineto{90.198}{63.561}
\emmoveto{90.198}{63.551}
\emlineto{90.765}{63.521}
\emmoveto{90.765}{63.511}
\emlineto{91.331}{63.475}
\emmoveto{91.331}{63.465}
\emlineto{91.896}{63.422}
\emmoveto{91.896}{63.412}
\emlineto{92.458}{63.363}
\emmoveto{92.458}{63.353}
\emlineto{93.019}{63.298}
\emmoveto{93.019}{63.288}
\emlineto{93.577}{63.227}
\emmoveto{93.577}{63.217}
\emlineto{94.133}{63.150}
\emmoveto{94.133}{63.140}
\emlineto{94.687}{63.066}
\emmoveto{94.687}{63.056}
\emlineto{95.238}{62.976}
\emmoveto{95.238}{62.966}
\emlineto{95.786}{62.881}
\emmoveto{95.786}{62.871}
\emlineto{96.331}{62.779}
\emmoveto{96.331}{62.769}
\emlineto{96.872}{62.671}
\emmoveto{96.872}{62.661}
\emlineto{97.410}{62.557}
\emmoveto{97.410}{62.547}
\emlineto{97.944}{62.437}
\emmoveto{97.944}{62.427}
\emlineto{98.475}{62.311}
\emmoveto{98.475}{62.301}
\emlineto{99.001}{62.179}
\emmoveto{99.001}{62.169}
\emlineto{99.523}{62.042}
\emmoveto{99.523}{62.032}
\emlineto{100.041}{61.898}
\emmoveto{100.041}{61.888}
\emlineto{100.554}{61.749}
\emmoveto{100.554}{61.739}
\emlineto{101.062}{61.595}
\emmoveto{101.062}{61.585}
\emlineto{101.565}{61.434}
\emmoveto{101.565}{61.424}
\emlineto{102.063}{61.268}
\emmoveto{102.063}{61.258}
\emlineto{102.556}{61.097}
\emmoveto{102.556}{61.087}
\emlineto{103.043}{60.920}
\emmoveto{103.043}{60.910}
\emlineto{103.525}{60.737}
\emmoveto{103.525}{60.727}
\emlineto{104.001}{60.550}
\emmoveto{104.001}{60.540}
\emlineto{104.470}{60.357}
\emmoveto{104.470}{60.347}
\emlineto{104.934}{60.159}
\emmoveto{104.934}{60.149}
\emlineto{105.391}{59.956}
\emmoveto{105.391}{59.946}
\emlineto{105.842}{59.748}
\emmoveto{105.842}{59.738}
\emlineto{106.286}{59.534}
\emmoveto{106.286}{59.524}
\emlineto{106.723}{59.316}
\emmoveto{106.723}{59.306}
\emlineto{107.153}{59.093}
\emmoveto{107.153}{59.083}
\emlineto{107.576}{58.866}
\emmoveto{107.576}{58.856}
\emlineto{107.992}{58.633}
\emmoveto{107.992}{58.623}
\emlineto{108.400}{58.396}
\emmoveto{108.400}{58.386}
\emlineto{108.801}{58.155}
\emmoveto{108.801}{58.145}
\emlineto{109.195}{57.909}
\emmoveto{109.195}{57.899}
\emlineto{109.580}{57.659}
\emmoveto{109.580}{57.649}
\emlineto{109.957}{57.404}
\emmoveto{109.957}{57.394}
\emlineto{110.326}{57.146}
\emmoveto{110.326}{57.136}
\emlineto{110.687}{56.883}
\emmoveto{110.687}{56.873}
\emlineto{111.040}{56.617}
\emmoveto{111.040}{56.607}
\emlineto{111.384}{56.346}
\emmoveto{111.384}{56.336}
\emlineto{111.720}{56.072}
\emmoveto{111.720}{56.062}
\emlineto{112.046}{55.794}
\emmoveto{112.046}{55.784}
\emlineto{112.365}{55.512}
\emmoveto{112.365}{55.502}
\emlineto{112.674}{55.227}
\emmoveto{112.674}{55.217}
\emlineto{112.973}{54.939}
\emmoveto{112.973}{54.929}
\emlineto{113.264}{54.647}
\emmoveto{113.264}{54.637}
\emlineto{113.546}{54.352}
\emmoveto{113.546}{54.342}
\emlineto{113.818}{54.054}
\emmoveto{113.818}{54.044}
\emlineto{114.081}{53.753}
\emmoveto{114.081}{53.743}
\emlineto{114.334}{53.450}
\emmoveto{114.334}{53.440}
\emlineto{114.577}{53.143}
\emmoveto{114.577}{53.133}
\emlineto{114.811}{52.834}
\emmoveto{114.811}{52.824}
\emlineto{115.035}{52.522}
\emmoveto{115.035}{52.512}
\emlineto{115.249}{52.208}
\emmoveto{115.249}{52.198}
\emlineto{115.453}{51.891}
\emmoveto{115.453}{51.881}
\emlineto{115.647}{51.573}
\emmoveto{115.647}{51.563}
\emlineto{115.831}{51.252}
\emmoveto{115.831}{51.242}
\emlineto{116.005}{50.929}
\emmoveto{116.005}{50.919}
\emlineto{116.168}{50.604}
\emmoveto{116.168}{50.594}
\emlineto{116.321}{50.278}
\emmoveto{116.321}{50.268}
\emlineto{116.464}{49.950}
\emmoveto{116.464}{49.940}
\emlineto{116.597}{49.620}
\emmoveto{116.597}{49.610}
\emlineto{116.719}{49.289}
\emmoveto{116.719}{49.279}
\emlineto{116.830}{48.957}
\emmoveto{116.830}{48.947}
\emlineto{116.931}{48.623}
\emmoveto{116.931}{48.613}
\emlineto{117.022}{48.289}
\emmoveto{117.022}{48.279}
\emlineto{117.102}{47.953}
\emmoveto{117.102}{47.943}
\emlineto{117.171}{47.617}
\emmoveto{117.171}{47.607}
\emlineto{117.230}{47.279}
\emmoveto{117.230}{47.269}
\emlineto{117.278}{46.942}
\emmoveto{117.278}{46.932}
\emlineto{117.315}{46.604}
\emmoveto{117.315}{46.594}
\emlineto{117.342}{46.265}
\emmoveto{117.342}{46.255}
\emlineto{117.357}{45.926}
\emmoveto{117.357}{45.916}
\emlineto{117.363}{45.588}
\emmoveto{117.363}{45.578}
\emlineto{117.363}{45.232}
\emmoveto{117.363}{45.222}
\emlineto{117.362}{44.846}
\emmoveto{117.362}{44.836}
\emlineto{117.350}{44.455}
\emmoveto{117.350}{44.445}
\emlineto{117.326}{44.064}
\emmoveto{117.326}{44.054}
\emlineto{117.290}{43.673}
\emmoveto{117.290}{43.663}
\emlineto{117.242}{43.283}
\emmoveto{117.242}{43.273}
\emlineto{117.181}{42.893}
\emmoveto{117.181}{42.883}
\emlineto{117.108}{42.504}
\emmoveto{117.108}{42.494}
\emlineto{117.023}{42.116}
\emmoveto{117.023}{42.106}
\emlineto{116.925}{41.729}
\emmoveto{116.925}{41.719}
\emlineto{116.815}{41.343}
\emmoveto{116.815}{41.333}
\emlineto{116.693}{40.959}
\emmoveto{116.693}{40.949}
\emlineto{116.559}{40.576}
\emmoveto{116.559}{40.566}
\emlineto{116.413}{40.194}
\emmoveto{116.413}{40.184}
\emlineto{116.255}{39.814}
\emmoveto{116.255}{39.804}
\emlineto{116.085}{39.436}
\emmoveto{116.085}{39.426}
\emlineto{115.902}{39.060}
\emmoveto{115.902}{39.050}
\emlineto{115.708}{38.686}
\emmoveto{115.708}{38.676}
\emlineto{115.503}{38.314}
\emmoveto{115.503}{38.304}
\emlineto{115.285}{37.945}
\emmoveto{115.285}{37.935}
\emlineto{115.056}{37.578}
\emmoveto{115.056}{37.568}
\emlineto{114.815}{37.213}
\emmoveto{114.815}{37.203}
\emlineto{114.563}{36.852}
\emmoveto{114.563}{36.842}
\emlineto{114.300}{36.493}
\emmoveto{114.300}{36.483}
\emlineto{114.025}{36.137}
\emmoveto{114.025}{36.127}
\emlineto{113.739}{35.785}
\emmoveto{113.739}{35.775}
\emlineto{113.442}{35.435}
\emmoveto{113.442}{35.425}
\emlineto{113.134}{35.089}
\emmoveto{113.134}{35.079}
\emlineto{112.815}{34.747}
\emmoveto{112.815}{34.737}
\emlineto{112.485}{34.408}
\emmoveto{112.485}{34.398}
\emlineto{112.145}{34.073}
\emmoveto{112.145}{34.063}
\emlineto{111.794}{33.741}
\emmoveto{111.794}{33.731}
\emlineto{111.432}{33.414}
\emmoveto{111.432}{33.404}
\emlineto{111.061}{33.091}
\emmoveto{111.061}{33.081}
\emlineto{110.679}{32.772}
\emmoveto{110.679}{32.762}
\emlineto{110.287}{32.457}
\emmoveto{110.287}{32.447}
\emlineto{109.885}{32.147}
\emmoveto{109.885}{32.137}
\emlineto{109.474}{31.841}
\emmoveto{109.474}{31.831}
\emlineto{109.053}{31.539}
\emmoveto{109.053}{31.529}
\emlineto{108.622}{31.243}
\emmoveto{108.622}{31.233}
\emlineto{108.183}{30.951}
\emmoveto{108.183}{30.941}
\emlineto{107.734}{30.665}
\emmoveto{107.734}{30.655}
\emlineto{107.276}{30.383}
\emmoveto{107.276}{30.373}
\emlineto{106.809}{30.106}
\emmoveto{106.809}{30.096}
\emlineto{106.334}{29.835}
\emmoveto{106.334}{29.825}
\emlineto{105.850}{29.569}
\emmoveto{105.850}{29.559}
\emlineto{105.358}{29.309}
\emmoveto{105.358}{29.299}
\emlineto{104.857}{29.054}
\emmoveto{104.857}{29.044}
\emlineto{104.349}{28.804}
\emmoveto{104.349}{28.794}
\emlineto{103.833}{28.560}
\emmoveto{103.833}{28.550}
\emlineto{103.309}{28.323}
\emmoveto{103.309}{28.313}
\emlineto{102.778}{28.090}
\emmoveto{102.778}{28.080}
\emlineto{102.240}{27.864}
\emmoveto{102.240}{27.854}
\emlineto{101.694}{27.644}
\emmoveto{101.694}{27.634}
\emlineto{101.142}{27.430}
\emmoveto{101.142}{27.420}
\emlineto{100.583}{27.223}
\emmoveto{100.583}{27.213}
\emlineto{100.018}{27.021}
\emmoveto{100.018}{27.011}
\emlineto{99.446}{26.826}
\emmoveto{99.446}{26.816}
\emlineto{98.868}{26.637}
\emmoveto{98.868}{26.627}
\emlineto{98.285}{26.455}
\emmoveto{98.285}{26.445}
\emlineto{97.696}{26.279}
\emmoveto{97.696}{26.269}
\emlineto{97.101}{26.110}
\emmoveto{97.101}{26.100}
\emlineto{96.501}{25.947}
\emmoveto{96.501}{25.937}
\emlineto{95.896}{25.791}
\emmoveto{95.896}{25.781}
\emlineto{95.286}{25.642}
\emmoveto{95.286}{25.632}
\emlineto{94.671}{25.499}
\emmoveto{94.671}{25.489}
\emlineto{94.053}{25.364}
\emmoveto{94.053}{25.354}
\emlineto{93.430}{25.235}
\emmoveto{93.430}{25.225}
\emlineto{92.803}{25.114}
\emmoveto{92.803}{25.104}
\emlineto{92.172}{24.999}
\emmoveto{92.172}{24.989}
\emlineto{91.538}{24.892}
\emmoveto{91.538}{24.882}
\emlineto{90.900}{24.791}
\emmoveto{90.900}{24.781}
\emlineto{90.260}{24.698}
\emmoveto{90.260}{24.688}
\emlineto{89.616}{24.611}
\emmoveto{89.616}{24.601}
\emlineto{88.970}{24.532}
\emmoveto{88.970}{24.522}
\emlineto{88.322}{24.460}
\emmoveto{88.322}{24.450}
\emlineto{87.671}{24.396}
\emmoveto{87.671}{24.386}
\emlineto{87.019}{24.338}
\emmoveto{87.019}{24.328}
\emlineto{86.365}{24.288}
\emmoveto{86.365}{24.278}
\emlineto{85.709}{24.245}
\emmoveto{85.709}{24.235}
\emlineto{85.052}{24.210}
\emmoveto{85.052}{24.200}
\emlineto{84.394}{24.181}
\emmoveto{84.394}{24.171}
\emlineto{83.736}{24.160}
\emmoveto{83.736}{24.150}
\emlineto{83.076}{24.147}
\emmoveto{83.076}{24.137}
\emlineto{82.417}{24.130}
\emlineto{82.417}{24.140}
\emmoveto{82.417}{24.130}
\emlineto{81.757}{24.142}
\emmoveto{81.757}{24.132}
\emlineto{81.098}{24.150}
\emmoveto{81.098}{24.140}
\emlineto{80.439}{24.166}
\emmoveto{80.439}{24.156}
\emlineto{79.780}{24.189}
\emmoveto{79.780}{24.179}
\emlineto{79.123}{24.219}
\emmoveto{79.123}{24.209}
\emlineto{78.466}{24.257}
\emmoveto{78.466}{24.247}
\emlineto{77.811}{24.302}
\emmoveto{77.811}{24.292}
\emlineto{77.157}{24.354}
\emmoveto{77.157}{24.344}
\emlineto{76.505}{24.414}
\emmoveto{76.505}{24.404}
\emlineto{75.855}{24.480}
\emmoveto{75.855}{24.470}
\emlineto{75.207}{24.554}
\emmoveto{75.207}{24.544}
\emlineto{74.562}{24.635}
\emmoveto{74.562}{24.625}
\emlineto{73.919}{24.724}
\emmoveto{73.919}{24.714}
\emlineto{73.280}{24.819}
\emmoveto{73.280}{24.809}
\emlineto{72.643}{24.922}
\emmoveto{72.643}{24.912}
\emlineto{72.010}{25.031}
\emmoveto{72.010}{25.021}
\emlineto{71.380}{25.148}
\emmoveto{71.380}{25.138}
\emlineto{70.754}{25.272}
\emmoveto{70.754}{25.262}
\emlineto{70.133}{25.402}
\emmoveto{70.133}{25.392}
\emlineto{69.515}{25.540}
\emmoveto{69.515}{25.530}
\emlineto{68.902}{25.684}
\emmoveto{68.902}{25.674}
\emlineto{68.293}{25.835}
\emmoveto{68.293}{25.825}
\emlineto{67.690}{25.993}
\emmoveto{67.690}{25.983}
\emlineto{67.091}{26.157}
\emmoveto{67.091}{26.147}
\emlineto{66.498}{26.329}
\emmoveto{66.498}{26.319}
\emlineto{65.910}{26.506}
\emmoveto{65.910}{26.496}
\emlineto{65.329}{26.691}
\emmoveto{65.329}{26.681}
\emlineto{64.752}{26.881}
\emmoveto{64.752}{26.871}
\emlineto{64.183}{27.078}
\emmoveto{64.183}{27.068}
\emlineto{63.619}{27.282}
\emmoveto{63.619}{27.272}
\emlineto{63.062}{27.491}
\emmoveto{63.062}{27.481}
\emlineto{62.512}{27.707}
\emmoveto{62.512}{27.697}
\emlineto{61.968}{27.929}
\emmoveto{61.968}{27.919}
\emlineto{61.432}{28.157}
\emmoveto{61.432}{28.147}
\emlineto{60.903}{28.390}
\emmoveto{60.903}{28.380}
\emlineto{60.381}{28.630}
\emmoveto{60.381}{28.620}
\emlineto{59.867}{28.875}
\emmoveto{59.867}{28.865}
\emlineto{59.361}{29.126}
\emmoveto{59.361}{29.116}
\emlineto{58.864}{29.383}
\emmoveto{58.864}{29.373}
\emlineto{58.374}{29.645}
\emmoveto{58.374}{29.635}
\emlineto{57.892}{29.912}
\emmoveto{57.892}{29.902}
\emlineto{57.419}{30.185}
\emmoveto{57.419}{30.175}
\emlineto{56.955}{30.463}
\emmoveto{56.955}{30.453}
\emlineto{56.500}{30.746}
\emmoveto{56.500}{30.736}
\emlineto{56.054}{31.035}
\emmoveto{56.054}{31.025}
\emlineto{55.616}{31.328}
\emmoveto{55.616}{31.318}
\emlineto{55.189}{31.626}
\emmoveto{55.189}{31.616}
\emlineto{54.770}{31.928}
\emmoveto{54.770}{31.918}
\emlineto{54.362}{32.235}
\emmoveto{54.362}{32.225}
\emlineto{53.963}{32.547}
\emmoveto{53.963}{32.537}
\emlineto{53.574}{32.863}
\emmoveto{53.574}{32.853}
\emlineto{53.195}{33.183}
\emmoveto{53.195}{33.173}
\emlineto{52.826}{33.508}
\emmoveto{52.826}{33.498}
\emlineto{52.468}{33.836}
\emmoveto{52.468}{33.826}
\emlineto{52.120}{34.169}
\emmoveto{52.120}{34.159}
\emlineto{51.783}{34.505}
\emmoveto{51.783}{34.495}
\emlineto{51.456}{34.845}
\emmoveto{51.456}{34.835}
\emlineto{51.140}{35.188}
\emmoveto{51.140}{35.178}
\emlineto{50.835}{35.535}
\emmoveto{50.835}{35.525}
\emlineto{50.541}{35.886}
\emmoveto{50.541}{35.876}
\emlineto{50.259}{36.239}
\emmoveto{50.259}{36.229}
\emlineto{49.987}{36.596}
\emmoveto{49.987}{36.586}
\emlineto{49.727}{36.955}
\emmoveto{49.727}{36.945}
\emlineto{49.478}{37.318}
\emmoveto{49.478}{37.308}
\emlineto{49.240}{37.683}
\emmoveto{49.240}{37.673}
\emlineto{49.015}{38.051}
\emmoveto{49.015}{38.041}
\emlineto{48.801}{38.421}
\emmoveto{48.801}{38.411}
\emlineto{48.598}{38.793}
\emmoveto{48.598}{38.783}
\emlineto{48.408}{39.168}
\emmoveto{48.408}{39.158}
\emlineto{48.229}{39.544}
\emmoveto{48.229}{39.534}
\emlineto{48.062}{39.923}
\emmoveto{48.062}{39.913}
\emlineto{47.907}{40.303}
\emmoveto{47.907}{40.293}
\emlineto{47.765}{40.685}
\emmoveto{47.765}{40.675}
\emlineto{47.634}{41.069}
\emmoveto{47.634}{41.059}
\emlineto{47.515}{41.454}
\emmoveto{47.515}{41.444}
\emlineto{47.409}{41.840}
\emmoveto{47.409}{41.830}
\emlineto{47.315}{42.227}
\emmoveto{47.315}{42.217}
\emlineto{47.233}{42.616}
\emmoveto{47.233}{42.606}
\emlineto{47.164}{43.005}
\emmoveto{47.164}{42.995}
\emlineto{47.106}{43.395}
\emmoveto{47.106}{43.385}
\emlineto{47.062}{43.785}
\emmoveto{47.062}{43.775}
\emlineto{47.029}{44.176}
\emmoveto{47.029}{44.166}
\emlineto{47.009}{44.567}
\emmoveto{47.009}{44.557}
\emlineto{47.001}{44.958}
\emmoveto{47.001}{44.948}
\emlineto{47.001}{45.362}
\emmoveto{47.001}{45.352}
\emlineto{47.002}{45.795}
\emmoveto{47.002}{45.785}
\emlineto{47.016}{46.232}
\emmoveto{47.016}{46.222}
\emlineto{47.043}{46.669}
\emmoveto{47.043}{46.659}
\emlineto{47.084}{47.106}
\emmoveto{47.084}{47.096}
\emlineto{47.138}{47.542}
\emmoveto{47.138}{47.532}
\emlineto{47.207}{47.978}
\emmoveto{47.207}{47.968}
\emlineto{47.289}{48.413}
\emmoveto{47.289}{48.403}
\emlineto{47.385}{48.847}
\emmoveto{47.385}{48.837}
\emlineto{47.495}{49.279}
\emmoveto{47.495}{49.269}
\emlineto{47.618}{49.711}
\emmoveto{47.618}{49.701}
\emlineto{47.755}{50.140}
\emmoveto{47.755}{50.130}
\emlineto{47.906}{50.569}
\emmoveto{47.906}{50.559}
\emlineto{48.070}{50.995}
\emmoveto{48.070}{50.985}
\emlineto{48.247}{51.420}
\emmoveto{48.247}{51.410}
\emlineto{48.438}{51.842}
\emmoveto{48.438}{51.832}
\emlineto{48.642}{52.263}
\emmoveto{48.642}{52.253}
\emlineto{48.859}{52.681}
\emmoveto{48.859}{52.671}
\emlineto{49.090}{53.096}
\emmoveto{49.090}{53.086}
\emlineto{49.334}{53.509}
\emmoveto{49.334}{53.499}
\emlineto{49.590}{53.920}
\emmoveto{49.590}{53.910}
\emlineto{49.860}{54.327}
\emmoveto{49.860}{54.317}
\emlineto{50.142}{54.731}
\emmoveto{50.142}{54.721}
\emlineto{50.438}{55.132}
\emmoveto{50.438}{55.122}
\emlineto{50.745}{55.529}
\emmoveto{50.745}{55.519}
\emlineto{51.066}{55.923}
\emmoveto{51.066}{55.913}
\emlineto{51.398}{56.314}
\emmoveto{51.398}{56.304}
\emlineto{51.743}{56.701}
\emmoveto{51.743}{56.691}
\emlineto{52.100}{57.083}
\emmoveto{52.100}{57.073}
\emlineto{52.470}{57.462}
\emmoveto{52.470}{57.452}
\emlineto{52.851}{57.837}
\emmoveto{52.851}{57.827}
\emlineto{53.244}{58.207}
\emmoveto{53.244}{58.197}
\emlineto{53.648}{58.573}
\emmoveto{53.648}{58.563}
\emlineto{54.064}{58.934}
\emmoveto{54.064}{58.924}
\emlineto{54.491}{59.291}
\emmoveto{54.491}{59.281}
\emlineto{54.930}{59.642}
\emmoveto{54.930}{59.632}
\emlineto{55.379}{59.989}
\emmoveto{55.379}{59.979}
\emlineto{55.840}{60.331}
\emmoveto{55.840}{60.321}
\emlineto{56.311}{60.667}
\emmoveto{56.311}{60.657}
\emlineto{56.793}{60.999}
\emmoveto{56.793}{60.989}
\emlineto{57.285}{61.325}
\emmoveto{57.285}{61.315}
\emlineto{57.787}{61.645}
\emmoveto{57.787}{61.635}
\emlineto{58.299}{61.960}
\emmoveto{58.299}{61.950}
\emlineto{58.822}{62.269}
\emmoveto{58.822}{62.259}
\emlineto{59.353}{62.572}
\emmoveto{59.353}{62.562}
\emlineto{59.895}{62.869}
\emmoveto{59.895}{62.859}
\emlineto{60.445}{63.160}
\emmoveto{60.445}{63.150}
\emlineto{61.005}{63.445}
\emmoveto{61.005}{63.435}
\emlineto{61.574}{63.723}
\emmoveto{61.574}{63.713}
\emlineto{62.151}{63.995}
\emmoveto{62.151}{63.985}
\emlineto{62.737}{64.261}
\emmoveto{62.737}{64.251}
\emlineto{63.331}{64.520}
\emmoveto{63.331}{64.510}
\emlineto{63.933}{64.773}
\emmoveto{63.933}{64.763}
\emlineto{64.543}{65.019}
\emmoveto{64.543}{65.009}
\emlineto{65.161}{65.258}
\emmoveto{65.161}{65.248}
\emlineto{65.786}{65.490}
\emmoveto{65.786}{65.480}
\emlineto{66.419}{65.715}
\emmoveto{66.419}{65.705}
\emlineto{67.058}{65.933}
\emmoveto{67.058}{65.923}
\emlineto{67.705}{66.143}
\emmoveto{67.705}{66.133}
\emlineto{68.357}{66.347}
\emmoveto{68.357}{66.337}
\emlineto{69.016}{66.543}
\emmoveto{69.016}{66.533}
\emlineto{69.682}{66.732}
\emmoveto{69.682}{66.722}
\emlineto{70.353}{66.914}
\emmoveto{70.353}{66.904}
\emlineto{71.029}{67.088}
\emmoveto{71.029}{67.078}
\emlineto{71.711}{67.254}
\emmoveto{71.711}{67.244}
\emlineto{72.398}{67.413}
\emmoveto{72.398}{67.403}
\emlineto{73.090}{67.564}
\emmoveto{73.090}{67.554}
\emlineto{73.787}{67.708}
\emmoveto{73.787}{67.698}
\emlineto{74.488}{67.843}
\emmoveto{74.488}{67.833}
\emlineto{75.194}{67.971}
\emmoveto{75.194}{67.961}
\emlineto{75.903}{68.091}
\emmoveto{75.903}{68.081}
\emlineto{76.616}{68.203}
\emmoveto{76.616}{68.193}
\emlineto{77.332}{68.307}
\emmoveto{77.332}{68.297}
\emlineto{78.052}{68.403}
\emmoveto{78.052}{68.393}
\emlineto{78.774}{68.492}
\emmoveto{78.774}{68.482}
\emlineto{79.499}{68.572}
\emmoveto{79.499}{68.562}
\emlineto{80.226}{68.644}
\emmoveto{80.226}{68.634}
\emlineto{80.956}{68.708}
\emmoveto{80.956}{68.698}
\emlineto{81.687}{68.763}
\emmoveto{81.687}{68.753}
\emlineto{82.421}{68.811}
\emmoveto{82.421}{68.801}
\emlineto{83.155}{68.850}
\emmoveto{83.155}{68.840}
\emlineto{83.891}{68.882}
\emmoveto{83.891}{68.872}
\emlineto{84.627}{68.905}
\emmoveto{84.627}{68.895}
\emlineto{85.364}{68.920}
\emmoveto{85.364}{68.910}
\emlineto{86.102}{68.926}
\emmoveto{86.102}{68.916}
\emlineto{86.839}{68.925}
\emmoveto{86.839}{68.915}
\emlineto{87.576}{68.915}
\emmoveto{87.576}{68.905}
\emlineto{88.313}{68.897}
\emmoveto{88.313}{68.887}
\emlineto{89.049}{68.871}
\emmoveto{89.049}{68.861}
\emlineto{89.785}{68.837}
\emmoveto{89.785}{68.827}
\emlineto{90.519}{68.794}
\emmoveto{90.519}{68.784}
\emlineto{91.251}{68.744}
\emmoveto{91.251}{68.734}
\emlineto{91.982}{68.685}
\emmoveto{91.982}{68.675}
\emlineto{92.711}{68.618}
\emmoveto{92.711}{68.608}
\emlineto{93.437}{68.543}
\emmoveto{93.437}{68.533}
\emlineto{94.161}{68.460}
\emmoveto{94.161}{68.450}
\emlineto{94.883}{68.369}
\emmoveto{94.883}{68.359}
\emlineto{95.601}{68.270}
\emmoveto{95.601}{68.260}
\emlineto{96.316}{68.163}
\emmoveto{96.316}{68.153}
\emlineto{97.028}{68.048}
\emmoveto{97.028}{68.038}
\emlineto{97.736}{67.925}
\emmoveto{97.736}{67.915}
\emlineto{98.439}{67.794}
\emmoveto{98.439}{67.784}
\emlineto{99.139}{67.656}
\emmoveto{99.139}{67.646}
\emlineto{99.834}{67.509}
\emmoveto{99.834}{67.499}
\emlineto{100.524}{67.355}
\emmoveto{100.524}{67.345}
\emlineto{101.209}{67.194}
\emmoveto{101.209}{67.184}
\emlineto{101.890}{67.024}
\emmoveto{101.890}{67.014}
\emlineto{102.564}{66.848}
\emmoveto{102.564}{66.838}
\emlineto{103.233}{66.663}
\emmoveto{103.233}{66.653}
\emlineto{103.896}{66.472}
\emmoveto{103.896}{66.462}
\emlineto{104.553}{66.273}
\emmoveto{104.553}{66.263}
\emlineto{105.203}{66.067}
\emmoveto{105.203}{66.057}
\emlineto{105.847}{65.853}
\emmoveto{105.847}{65.843}
\emlineto{106.484}{65.632}
\emmoveto{106.484}{65.622}
\emlineto{107.114}{65.405}
\emmoveto{107.114}{65.395}
\emlineto{107.736}{65.170}
\emmoveto{107.736}{65.160}
\emlineto{108.351}{64.929}
\emmoveto{108.351}{64.919}
\emlineto{108.958}{64.681}
\emmoveto{108.958}{64.671}
\emlineto{109.558}{64.425}
\emmoveto{109.558}{64.415}
\emlineto{110.149}{64.164}
\emmoveto{110.149}{64.154}
\emlineto{110.731}{63.896}
\emmoveto{110.731}{63.886}
\emlineto{111.306}{63.621}
\emmoveto{111.306}{63.611}
\emlineto{111.871}{63.340}
\emmoveto{111.871}{63.330}
\emlineto{112.427}{63.053}
\emmoveto{112.427}{63.043}
\emlineto{112.975}{62.760}
\emmoveto{112.975}{62.750}
\emlineto{113.513}{62.461}
\emmoveto{113.513}{62.451}
\emlineto{114.041}{62.155}
\emmoveto{114.041}{62.145}
\emlineto{114.559}{61.844}
\emmoveto{114.559}{61.834}
\emlineto{115.068}{61.527}
\emmoveto{115.068}{61.517}
\emlineto{115.567}{61.205}
\emmoveto{115.567}{61.195}
\emlineto{116.055}{60.877}
\emmoveto{116.055}{60.867}
\emlineto{116.533}{60.544}
\emmoveto{116.533}{60.534}
\emlineto{117.000}{60.205}
\emmoveto{117.000}{60.195}
\emlineto{117.456}{59.862}
\emmoveto{117.456}{59.852}
\emlineto{117.902}{59.513}
\emmoveto{117.902}{59.503}
\emlineto{118.336}{59.160}
\emmoveto{118.336}{59.150}
\emlineto{118.759}{58.801}
\emmoveto{118.759}{58.791}
\emlineto{119.171}{58.438}
\emmoveto{119.171}{58.428}
\emlineto{119.571}{58.071}
\emmoveto{119.571}{58.061}
\emlineto{119.960}{57.699}
\emmoveto{119.960}{57.689}
\emlineto{120.337}{57.323}
\emmoveto{120.337}{57.313}
\emlineto{120.701}{56.943}
\emmoveto{120.701}{56.933}
\emlineto{121.054}{56.558}
\emmoveto{121.054}{56.548}
\emlineto{121.394}{56.170}
\emmoveto{121.394}{56.160}
\emlineto{121.722}{55.779}
\emmoveto{121.722}{55.769}
\emlineto{122.038}{55.383}
\emmoveto{122.038}{55.373}
\emlineto{122.341}{54.984}
\emmoveto{122.341}{54.974}
\emlineto{122.632}{54.582}
\emmoveto{122.632}{54.572}
\emlineto{122.909}{54.177}
\emmoveto{122.909}{54.167}
\emlineto{123.174}{53.769}
\emmoveto{123.174}{53.759}
\emlineto{123.426}{53.357}
\emmoveto{123.426}{53.347}
\emlineto{123.665}{52.943}
\emmoveto{123.665}{52.933}
\emlineto{123.891}{52.527}
\emmoveto{123.891}{52.517}
\emlineto{124.103}{52.108}
\emmoveto{124.103}{52.098}
\emlineto{124.303}{51.687}
\emmoveto{124.303}{51.677}
\emlineto{124.489}{51.263}
\emmoveto{124.489}{51.253}
\emlineto{124.661}{50.838}
\emmoveto{124.661}{50.828}
\emlineto{124.820}{50.411}
\emmoveto{124.820}{50.401}
\emlineto{124.966}{49.982}
\emmoveto{124.966}{49.972}
\emlineto{125.097}{49.552}
\emmoveto{125.097}{49.542}
\emlineto{125.216}{49.120}
\emmoveto{125.216}{49.110}
\emlineto{125.320}{48.687}
\emmoveto{125.320}{48.677}
\emlineto{125.411}{48.253}
\emmoveto{125.411}{48.243}
\emlineto{125.489}{47.817}
\emmoveto{125.489}{47.807}
\emlineto{125.552}{47.382}
\emmoveto{125.552}{47.372}
\emlineto{125.602}{46.945}
\emmoveto{125.602}{46.935}
\emlineto{125.637}{46.508}
\emmoveto{125.637}{46.498}
\emlineto{125.659}{46.071}
\emmoveto{125.659}{46.061}
\emlineto{125.668}{45.633}
\emmoveto{125.668}{45.623}
\emlineto{125.668}{45.182}
\emmoveto{125.668}{45.172}
\emlineto{125.665}{44.705}
\emmoveto{125.665}{44.695}
\emlineto{125.647}{44.226}
\emmoveto{125.647}{44.216}
\emlineto{125.615}{43.747}
\emmoveto{125.615}{43.737}
\emlineto{125.568}{43.269}
\emmoveto{125.568}{43.259}
\emlineto{125.505}{42.791}
\emmoveto{125.505}{42.781}
\emlineto{125.427}{42.314}
\emmoveto{125.427}{42.304}
\emlineto{125.334}{41.838}
\emmoveto{125.334}{41.828}
\emlineto{125.227}{41.363}
\emmoveto{125.227}{41.353}
\emlineto{125.104}{40.889}
\emmoveto{125.104}{40.879}
\emlineto{124.966}{40.417}
\emmoveto{124.966}{40.407}
\emlineto{124.814}{39.946}
\emmoveto{124.814}{39.936}
\emlineto{124.646}{39.477}
\emmoveto{124.646}{39.467}
\emlineto{124.464}{39.011}
\emmoveto{124.464}{39.001}
\emlineto{124.267}{38.546}
\emmoveto{124.267}{38.536}
\emlineto{124.055}{38.083}
\emmoveto{124.055}{38.073}
\emlineto{123.829}{37.623}
\emmoveto{123.829}{37.613}
\emlineto{123.588}{37.166}
\emmoveto{123.588}{37.156}
\emlineto{123.333}{36.711}
\emmoveto{123.333}{36.701}
\emlineto{123.064}{36.259}
\emmoveto{123.064}{36.249}
\emlineto{122.780}{35.810}
\emmoveto{122.780}{35.800}
\emlineto{122.482}{35.365}
\emmoveto{122.482}{35.355}
\emlineto{122.171}{34.923}
\emmoveto{122.171}{34.913}
\emlineto{121.845}{34.484}
\emmoveto{121.845}{34.474}
\emlineto{121.505}{34.049}
\emmoveto{121.505}{34.039}
\emlineto{121.152}{33.618}
\emmoveto{121.152}{33.608}
\emlineto{120.785}{33.191}
\emmoveto{120.785}{33.181}
\emlineto{120.405}{32.769}
\emmoveto{120.405}{32.759}
\emlineto{120.011}{32.350}
\emmoveto{120.011}{32.340}
\emlineto{119.605}{31.936}
\emmoveto{119.605}{31.926}
\emlineto{119.185}{31.526}
\emmoveto{119.185}{31.516}
\emlineto{118.752}{31.122}
\emmoveto{118.752}{31.112}
\emlineto{118.307}{30.722}
\emmoveto{118.307}{30.712}
\emlineto{117.849}{30.327}
\emmoveto{117.849}{30.317}
\emlineto{117.379}{29.937}
\emmoveto{117.379}{29.927}
\emlineto{116.896}{29.553}
\emmoveto{116.896}{29.543}
\emlineto{116.402}{29.174}
\emmoveto{116.402}{29.164}
\emlineto{115.895}{28.801}
\emmoveto{115.895}{28.791}
\emlineto{115.377}{28.433}
\emmoveto{115.377}{28.423}
\emlineto{114.848}{28.071}
\emmoveto{114.848}{28.061}
\emlineto{114.306}{27.715}
\emmoveto{114.306}{27.705}
\emlineto{113.754}{27.365}
\emmoveto{113.754}{27.355}
\emlineto{113.191}{27.022}
\emmoveto{113.191}{27.012}
\emlineto{112.617}{26.684}
\emmoveto{112.617}{26.674}
\emlineto{112.033}{26.353}
\emmoveto{112.033}{26.343}
\emlineto{111.438}{26.029}
\emmoveto{111.438}{26.019}
\emlineto{110.833}{25.712}
\emmoveto{110.833}{25.702}
\emlineto{110.218}{25.401}
\emmoveto{110.218}{25.391}
\emlineto{109.593}{25.097}
\emmoveto{109.593}{25.087}
\emlineto{108.959}{24.800}
\emmoveto{108.959}{24.790}
\emlineto{108.316}{24.510}
\emmoveto{108.316}{24.500}
\emlineto{107.663}{24.227}
\emmoveto{107.663}{24.217}
\emlineto{107.002}{23.952}
\emmoveto{107.002}{23.942}
\emlineto{106.332}{23.684}
\emmoveto{106.332}{23.674}
\emlineto{105.654}{23.424}
\emmoveto{105.654}{23.414}
\emlineto{104.968}{23.171}
\emmoveto{104.968}{23.161}
\emlineto{104.274}{22.926}
\emmoveto{104.274}{22.916}
\emlineto{103.572}{22.688}
\emmoveto{103.572}{22.678}
\emlineto{102.863}{22.459}
\emmoveto{102.863}{22.449}
\emlineto{102.146}{22.237}
\emmoveto{102.146}{22.227}
\emlineto{101.423}{22.023}
\emmoveto{101.423}{22.013}
\emlineto{100.694}{21.818}
\emmoveto{100.694}{21.808}
\emlineto{99.957}{21.620}
\emmoveto{99.957}{21.610}
\emlineto{99.215}{21.431}
\emmoveto{99.215}{21.421}
\emlineto{98.467}{21.250}
\emmoveto{98.467}{21.240}
\emlineto{97.713}{21.078}
\emmoveto{97.713}{21.068}
\emlineto{96.954}{20.914}
\emmoveto{96.954}{20.904}
\emlineto{96.190}{20.758}
\emmoveto{96.190}{20.748}
\emlineto{95.421}{20.611}
\emmoveto{95.421}{20.601}
\emlineto{94.648}{20.473}
\emmoveto{94.648}{20.463}
\emlineto{93.870}{20.343}
\emmoveto{93.870}{20.333}
\emlineto{93.089}{20.221}
\emmoveto{93.089}{20.211}
\emlineto{92.303}{20.109}
\emmoveto{92.303}{20.099}
\emlineto{91.515}{20.005}
\emmoveto{91.515}{19.995}
\emlineto{90.723}{19.910}
\emmoveto{90.723}{19.900}
\emlineto{89.928}{19.824}
\emmoveto{89.928}{19.814}
\emlineto{89.131}{19.747}
\emmoveto{89.131}{19.737}
\emlineto{88.331}{19.678}
\emmoveto{88.331}{19.668}
\emlineto{87.529}{19.619}
\emmoveto{87.529}{19.609}
\emlineto{86.726}{19.568}
\emmoveto{86.726}{19.558}
\emlineto{85.921}{19.527}
\emmoveto{85.921}{19.517}
\emlineto{85.115}{19.494}
\emmoveto{85.115}{19.484}
\emlineto{84.308}{19.470}
\emmoveto{84.308}{19.460}
\emlineto{83.501}{19.455}
\emmoveto{83.501}{19.445}
\emlineto{82.693}{19.440}
\emlineto{82.693}{19.450}
\emmoveto{82.693}{19.440}
\emlineto{81.885}{19.453}
\emmoveto{81.885}{19.443}
\emlineto{81.078}{19.465}
\emmoveto{81.078}{19.455}
\emlineto{80.270}{19.486}
\emmoveto{80.270}{19.476}
\emlineto{79.464}{19.517}
\emmoveto{79.464}{19.507}
\emlineto{78.659}{19.556}
\emmoveto{78.659}{19.546}
\emlineto{77.855}{19.604}
\emmoveto{77.855}{19.594}
\emlineto{77.053}{19.661}
\emmoveto{77.053}{19.651}
\emlineto{76.253}{19.727}
\emmoveto{76.253}{19.717}
\emlineto{75.455}{19.802}
\emmoveto{75.455}{19.792}
\emlineto{74.659}{19.885}
\emmoveto{74.659}{19.875}
\emlineto{73.867}{19.978}
\emmoveto{73.867}{19.968}
\emlineto{73.077}{20.079}
\emmoveto{73.077}{20.069}
\emlineto{72.291}{20.189}
\emmoveto{72.291}{20.179}
\emlineto{71.508}{20.308}
\emmoveto{71.508}{20.298}
\emlineto{70.729}{20.435}
\emmoveto{70.729}{20.425}
\emlineto{69.955}{20.572}
\emmoveto{69.955}{20.562}
\emlineto{69.185}{20.716}
\emmoveto{69.185}{20.706}
\emlineto{68.419}{20.870}
\emmoveto{68.419}{20.860}
\emlineto{67.659}{21.031}
\emmoveto{67.659}{21.021}
\emlineto{66.904}{21.201}
\emmoveto{66.904}{21.191}
\emlineto{66.154}{21.380}
\emmoveto{66.154}{21.370}
\emlineto{65.410}{21.567}
\emmoveto{65.410}{21.557}
\emlineto{64.672}{21.762}
\emmoveto{64.672}{21.752}
\emlineto{63.940}{21.965}
\emmoveto{63.940}{21.955}
\emlineto{63.215}{22.177}
\emmoveto{63.215}{22.167}
\emlineto{62.497}{22.396}
\emmoveto{62.497}{22.386}
\emlineto{61.786}{22.623}
\emmoveto{61.786}{22.613}
\emlineto{61.082}{22.859}
\emmoveto{61.082}{22.849}
\emlineto{60.386}{23.102}
\emmoveto{60.386}{23.092}
\emlineto{59.697}{23.352}
\emmoveto{59.697}{23.342}
\emlineto{59.017}{23.611}
\emmoveto{59.017}{23.601}
\emlineto{58.344}{23.876}
\emmoveto{58.344}{23.866}
\emlineto{57.681}{24.150}
\emmoveto{57.681}{24.140}
\emlineto{57.026}{24.430}
\emmoveto{57.026}{24.420}
\emlineto{56.380}{24.718}
\emmoveto{56.380}{24.708}
\emlineto{55.743}{25.013}
\emmoveto{55.743}{25.003}
\emlineto{55.116}{25.315}
\emmoveto{55.116}{25.305}
\emlineto{54.498}{25.624}
\emmoveto{54.498}{25.614}
\emlineto{53.890}{25.940}
\emmoveto{53.890}{25.930}
\emlineto{53.293}{26.262}
\emmoveto{53.293}{26.252}
\emlineto{52.705}{26.591}
\emmoveto{52.705}{26.581}
\emlineto{52.128}{26.927}
\emmoveto{52.128}{26.917}
\emlineto{51.562}{27.269}
\emmoveto{51.562}{27.259}
\emlineto{51.007}{27.617}
\emmoveto{51.007}{27.607}
\emlineto{50.463}{27.971}
\emmoveto{50.463}{27.961}
\emlineto{49.930}{28.331}
\emmoveto{49.930}{28.321}
\emlineto{49.409}{28.697}
\emmoveto{49.409}{28.687}
\emlineto{48.899}{29.069}
\emmoveto{48.899}{29.059}
\emlineto{48.401}{29.446}
\emmoveto{48.401}{29.436}
\emlineto{47.915}{29.829}
\emmoveto{47.915}{29.819}
\emlineto{47.441}{30.217}
\emmoveto{47.441}{30.207}
\emlineto{46.980}{30.611}
\emmoveto{46.980}{30.601}
\emlineto{46.531}{31.009}
\emmoveto{46.531}{30.999}
\emlineto{46.095}{31.413}
\emmoveto{46.095}{31.403}
\emlineto{45.672}{31.821}
\emmoveto{45.672}{31.811}
\emlineto{45.261}{32.234}
\emmoveto{45.261}{32.224}
\emlineto{44.864}{32.651}
\emmoveto{44.864}{32.641}
\emlineto{44.480}{33.073}
\emmoveto{44.480}{33.063}
\emlineto{44.110}{33.499}
\emmoveto{44.110}{33.489}
\emlineto{43.753}{33.929}
\emmoveto{43.753}{33.919}
\emlineto{43.409}{34.362}
\emmoveto{43.409}{34.352}
\emlineto{43.079}{34.800}
\emshow{24.980}{17.700}{}
\emshow{1.000}{10.000}{-6.00e-1}
\emshow{1.000}{17.000}{-4.80e-1}
\emshow{1.000}{24.000}{-3.60e-1}
\emshow{1.000}{31.000}{-2.40e-1}
\emshow{1.000}{38.000}{-1.20e-1}
\emshow{1.000}{45.000}{0.00e0}
\emshow{1.000}{52.000}{1.20e-1}
\emshow{1.000}{59.000}{2.40e-1}
\emshow{1.000}{66.000}{3.60e-1}
\emshow{1.000}{73.000}{4.80e-1}
\emshow{1.000}{80.000}{6.00e-1}
\emshow{12.000}{5.000}{-6.00e-1}
\emshow{23.800}{5.000}{-4.80e-1}
\emshow{35.600}{5.000}{-3.60e-1}
\emshow{47.400}{5.000}{-2.40e-1}
\emshow{59.200}{5.000}{-1.20e-1}
\emshow{71.000}{5.000}{0.00e0}
\emshow{82.800}{5.000}{1.20e-1}
\emshow{94.600}{5.000}{2.40e-1}
\emshow{106.400}{5.000}{3.60e-1}
\emshow{118.200}{5.000}{4.80e-1}
\emshow{130.000}{5.000}{6.00e-1}
{\centerline {\bf Fig. B.2}}

 
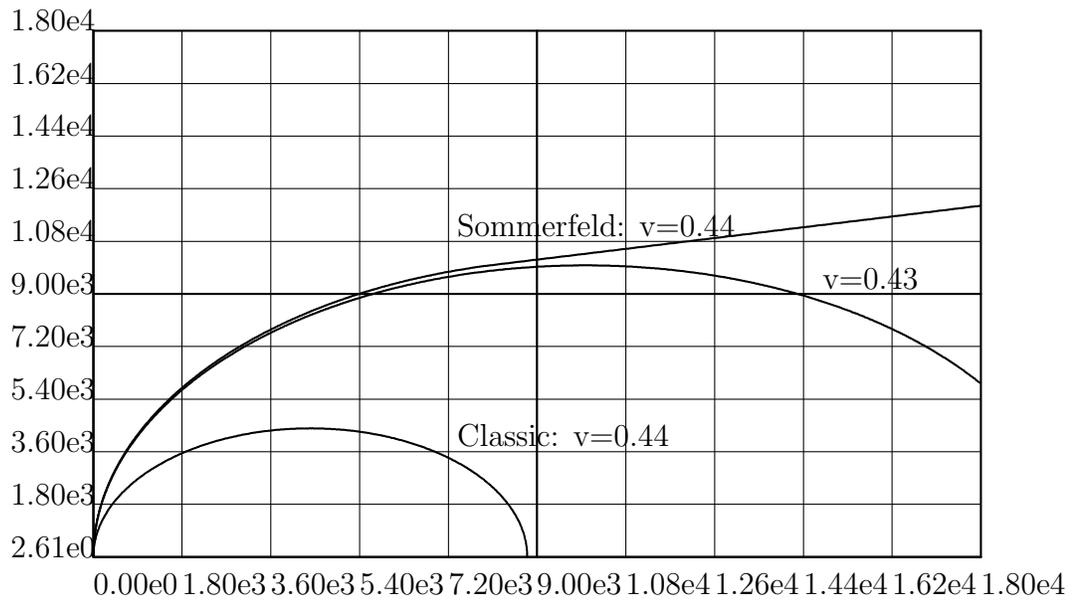
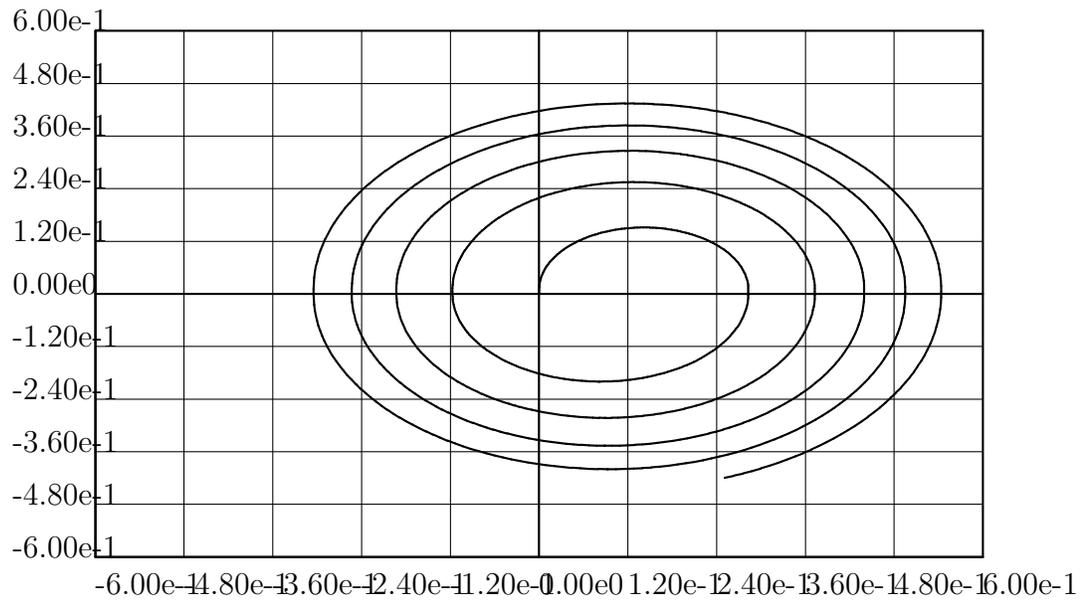
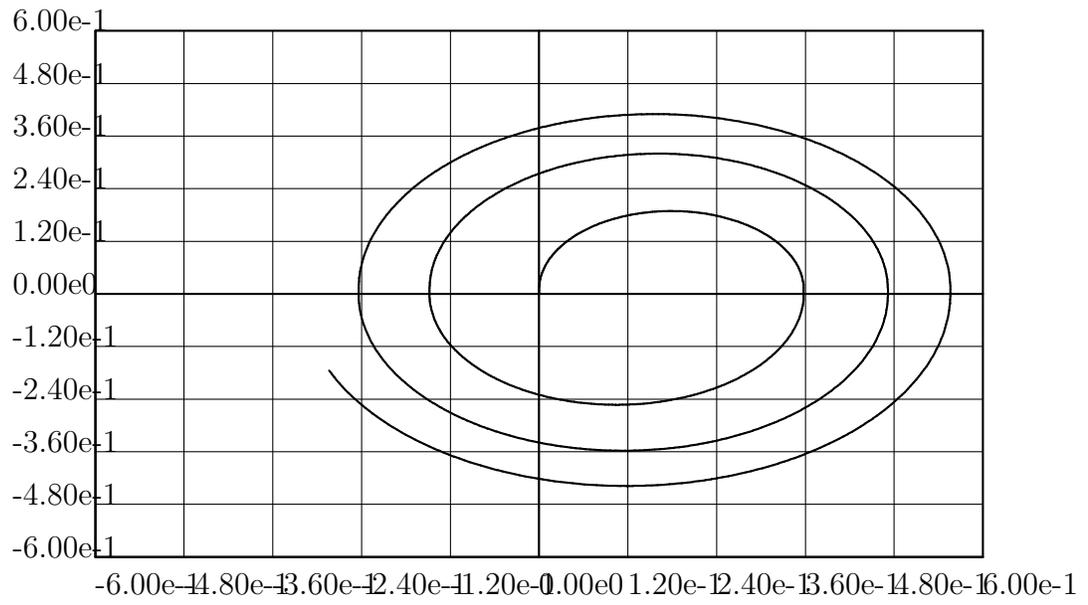
\end{document}